\documentclass{emulateapj}
\usepackage{lscape}

\def\LIR{\hbox{$L_{\rm IR}$}}

\def\HII{\hbox{H\,{\sc ii}}}

\def\ArIII{\hbox{[Ar\,{\sc iii}]8.99\,\micron}}

\def\NeIIion{\hbox{Ne\,$^{+}$}}
\def\NeII{\hbox{[Ne\,{\sc ii}]12.81\,\micron}}
\def\NeIIno{\hbox{[Ne\,{\sc ii}]}}
\def\NeIIIion{\hbox{Ne\,$^{++}$}}
\def\NeIII{\hbox{[Ne\,{\sc iii}]15.55\,\micron}}
\def\NeIIIno{\hbox{[Ne\,{\sc iii}]}}
\def\NeV{\hbox{[Ne\,{\sc v}]14.32\,\micron}}

\def\SIII{\hbox{[S\,{\sc iii}]18.71\,\micron}}

\def\SIV{\hbox{[S\,{\sc iv}]10.51\,\micron}}

\def\Siabs{\hbox{Si\,$_{\rm 9.7\,\mu m}$}}
\def\SSi{\hbox{$S_{\rm Si\,9.7\,\mu m}$}}
\def\H2{\hbox{H$_{2}$}}
\def\PAHa{\hbox{11.3\,\micron\,PAH}}
\def\PAHas{\hbox{11.3\,\micron\,PAHs}}
\def\PAHb{\hbox{8.6\,\micron\,PAH}}
\def\PAHbs{\hbox{8.6\,\micron\,PAHs}}
\def\tauSi{\hbox{$\tau_{\rm Si9.7\,\mu m}$}}
\def\tauV{\hbox{$\tau_V$}}
\def\AV{\hbox{$A_V$}}

\def\Bralpha{\hbox{\rm Br$\alpha$}}

\def\Paalpha{\hbox{\rm Pa$\alpha$}}

\def\Lsun{\hbox{$L_\odot$}}

\def\LIR{\hbox{$L_{\rm IR}$}}

\def\lsd{\hbox{erg$\,$s$^{-1}\,$kpc$^{-2}$}}

\shorttitle{High Spatial Resolution Spectroscopic Study of SF in LIRGs}
\shortauthors{D\'{\i}az-Santos et al.}

\begin{document}

\title{A High Spatial Resolution Mid-Infrared Spectroscopic Study of the Nuclei and Star-Forming Regions in Luminous Infrared Galaxies}


\author{Tanio~D\'{\i}az-Santos\altaffilmark{1,2,\dag},
        Almudena~Alonso-Herrero\altaffilmark{1},
        Luis~Colina\altaffilmark{1},
        Christopher~Packham\altaffilmark{3},
        N.~A.~Levenson\altaffilmark{4,5},
 	Miguel Pereira-Santaella\altaffilmark{1},
 	Patrick F. Roche\altaffilmark{6}
 	and Charles~M.~Telesco\altaffilmark{3}
}
%\altaffiltext{1}{Based on observations obtained with T-ReCS instrument at the
%        Gemini South Observatory, which is operated by AURA, Inc., under a
%        cooperative agreement with the NSF on behalf of the Gemini
%        partnership: NSF (United States), PPARC (UK), NRC (Canada), CONICYT
%        (Chile) ARC (Australia), CNPq (Brazil) and CONICET (Argentina).}
\altaffiltext{1}{Departamento de Astrof\'{\i}sica Molecular e Infrarroja,
        Instituto de Estructura de la Materia (IEM), CSIC, Serrano 121, 
E-28006 Madrid, Spain}
\altaffiltext{2}{Actual address: University of Crete, Department of Physics,
GR-71003, Heraklion, Greece}
\altaffiltext{3}{Department of Astronomy, University of Florida, 211 Bryant Science Center, P.O. Box 112055, Gainesville, FL 32611-2055}
\altaffiltext{4}{Department of Physics and Astronomy, University of Kentucky, Lexington, KY 40506-0055.}
\altaffiltext{5}{Gemini Observatory, Casilla 603, La Serena, Chile}
\altaffiltext{6}{University of Oxford, UK}
\altaffiltext{\dag}{Contact email: tanio@physics.uoc.gr}
%\altaffiltext{7}{Gemini Observatory, Northern Operations Center, Hilo, HI 96720.}

\begin{abstract}

We present a high spatial (diffraction-limited) resolution ($\sim\,$0.3\arcsec)
mid-infrared (MIR) spectroscopic study of the nuclei and star-forming regions
% and some regions of interest
of 4 local luminous infrared galaxies (LIRGs) using T-ReCS on the Gemini
South telescope.
%Although the overall MIR emission traces well the sites of star formation, differences between the MIR and \Paalpha\ luminosities arise on scales of a few hundreds of pc.
We investigate the spatial variations of the features seen in the
$N$-band spectra of LIRGs on scales of $\sim\,$100\,pc, which allow us
to resolve their nuclear regions and separate the AGN emission from
that of the star formation (SF). We compare (qualitatively
and quantitatively) our Gemini T-ReCS nuclear and integrated spectra
of LIRGs with those obtained with \textit{Spitzer} IRS.
Star-forming regions and AGNs show distinct features in the MIR spectra,
and we spatially
separate these, which is not possible using the \textit{Spitzer} data.
%Depending on the source dominating the MIR emission (SF, AGN, or a combination of both) our spectra show distinct features, some of which the \textit{Spitzer} spectra cannot separate.
%% Each galaxy show their own characteristic spectral features depending on the source that is dominating the emission.
The 9.7$\,\micron$ silicate absorption feature is weaker in the
nuclei of the LIRGs than in the surrounding regions. This is probably
due to the either clumpy or compact environment of
%presence of an additional component of hot dust continuum emission produced by
the central AGN or young, nuclear starburst.
% that is contaminating and ``filling-in'' the absorption feature.
We find that the \NeII\ luminosity surface density is tightly and
directly correlated with that of \Paalpha\ for the LIRG star-forming
regions (slope of 1.00$\pm$0.02).
Although the \PAHa\ feature shows also a trend with \Paalpha, this
is not common for all the regions and the slope is significantly lower.
We also find that the \NeII/\Paalpha\ ratio does not depend on the
\Paalpha\ equivalent width (EW), i.e., on the age of the ionizing stellar
populations,
suggesting that, on the scales probed here, the \NeII\ emission line
is a good tracer of the SF activity in LIRGs. On the other hand, the
\PAHa/\Paalpha\ ratio increases for smaller values of the \Paalpha\ EW
(increasing ages), indicating that the \PAHa\ feature can also be
excited by older stars than those responsible for the \Paalpha\ emission.
Finally, more data are needed in order to address the different
physical processes (age of the stellar populations, hardness
and intensity of the radiation field, mass of the star-forming regions)
affecting the energetics of the PAH features in a statistical way.
%with the current amount data it not straightforward to
%we cannot easily
%disentangle the different physical processes (age of the stellar population, hardness and intensity of the radiation field, mass of the star-forming regions) affecting the energetics of the \PAHa\ feature.
Additional high spatial resolution observations are essential to
investigate the star formation in local LIRGs at the smallest scales and
to probe ultimately whether they share the same physical properties as
high-$z$ LIRGs, ULIRGs and submillimiter galaxies and therefore belong
to the same galaxy population.

%This implies that either the PAH carriers are being created (or PAH emission is being enhanced) at a rapid rate during this period of the starburst evolution, or they were being destroyed on the first place at younger ages. We also suggest that the reduction of the \PAHa\ EW with increasing the intensity of the radiation field is probably caused by a mass enhancement of the starburst, which increases the \textit{intensity} of the radiation field and dilutes the PAH emission.

%% That is, this would suggest that the extinction is higher towards the outer regions.
%%However, this result should be taken with caution as the 8.6 and \PAHa\ features are located at both sides of the 9.7$\,\micron$ Silicate absorption feature and could be artifically increasing the depth of the feature.
%We have used AGN clumpy torus models for fitting the T-ReCS nuclear spectra of the two Sy2 LIRGs in the subsample, and of other Sy2 spectra from the literature. We have derived the physical parameters of the torus and found that the ratio between its outer and inner radius needs to be higher for those nuclei showing deeper silicate absorption features.
%% As said above, both the \PAHa\ feature and the \NeII\ emission line are correlated with the \Paalpha\ line.

\end{abstract}


%________________________________________________________________
\section{Introduction}\label{s:intro}

%Now that \textit{Spitzer} is entering its warm phase, MIR observations will be soon restricted to ground-based facilities until the launch of the \textit{James Webb Space Telescope}. However, although currently ground-based telescopes are not able to achieve the sensitivity of \textit{Spitzer}, they can play the other master card in the astronomy: the spatial resolution. $8-10\,$m-class telescopes can achieve resolutions an order of magnitude better than \textit{Spitzer}, down to sub-arcsecond scales. In fact, sometimes this is the essential requisite for studying and disentangling the physical processes occurring in the nuclei of galaxies.

%The Mid-infrared (MIR) emission of luminous and ultra-luminous
%  infrared galaxies (LIRGs: 
%$L_{\rm IR}=10^{11-12}\,$\Lsun; and ULIRGs: $>$10$^{12}\,$\Lsun, respectively)
%is known to be related with the global
%(integrated) star formation processes. 

%Another important feature present in the MIR
%spectra of (U)LIRGs is the 9.7$\,\micron$ silicate feature, which results
%from the absorption (or emission) of the MIR continuum due to cold
%(or hot) dust composed of amorphous silicate grains.

The monochromatic mid-infrared (MIR) emission  (8 and 24$\,\micron$:
\citealt{Wu05}; \citealt{Calzetti07}; \citealt{AAH06b}), as well as 
several MIR features such as the neon forbidden emission lines
(\NeII\ and/or \NeIII: \citealt{Roche91}; \citealt{Ho07}) 
are known to be related to the global (integrated) star formation in
starburst and luminous infrared galaxies (LIRGs,
10$^{11}\,\Lsun\,\leq\,L_{\rm IR [8-1000\,\mu m]}\,<\,$10$^{12}\,\Lsun$).
%%The \NeII\ emission line has been also commonly identified with
%the presence of star formation.
% However, the fact that a fraction of its flux might come from the contribution of an AGN cannot be ruled out. Indeed, the \NeII\ emission line can also be produced in the vicinity (few tens of pc) of an AGN or NLR (e.g., \citealt{Roche06}).
The \NeII\ line emission is also correlated with the number of ionizing photons
in starburst galaxies, LIRGs and star-forming regions and therefore
there is also a direct relation between the \NeII\ flux and the star
formation rate (SFR) (\citealt{Roche91}), although the relative abundance
of \NeIIion\ and \NeIIIion\ ions should be taken into account (\citealt{Ho07}).
%if an accurate relation want to be obtained.
% Thus, a measure of the SFR based only on the \NeII\ line luminosity, although valid at first order, should be taken with caution.

The presence of polycyclic aromatic hydrocarbons (PAHs,
for example the 6.2, 7.7 and 11.3$\,\micron$ features)
has largely been associated to regions of star formation
(from ground-based studies: \citealt{Roche91}; to studies based on
\textit{ISO}: \citealt{Genzel98}; \citealt{Lutz98b}; \citealt{Rigopoulou99};
and \textit{Spitzer} observations: \citealt{Brandl06}; \citealt{Beirao06};
\citealt{Smith07}; \citealt{Desai07}; \citealt{Houck07}; \citealt{Farrah07})
and, in particular, to their
photo-dissociation regions (PDRs) (\citealt{Peeters04}; \citealt{Povich07}).
However, recent works have suggested that the infrared (IR) luminosities
of star-forming galaxies, as estimated using their PAH emission, should
be taken with caution. For instance, the PAH ratios can vary by
up to an order of magnitude among Galactic
star-forming regions, Magellanic \HII\ regions, and galaxies 
(\citealt{Galliano08}),
and the PAH emission relative to the total IR emission
may change from galaxy to galaxy (\citealt{Smith07}). Additionally, the
PAH emission appears to better trace \textit{recent} rather than
\textit{current}, massive star formation (\citealt{Peeters04}).
PAHs also have an interstellar origin and it has been demonstrated that
their emission is more extended than that of hydrogen line emission,
which is more compact and probe mainly the ionizing
stellar populations (\citealt{AAH06b}; \citealt{DS08}). Furthermore,
recently PAH features have been detected in the spectra of a significant
number of local dusty elliptical galaxies (\citealt{Kaneda08}).

The local physical environment in which the star formation is taking
place can also affect the PAH emission. The metallicity, the properties
of the dust, and the intensity and hardness of the radiation field can 
alter the ratios among the PAH features and modify the
absolute amount of PAH emission generated by the star-forming
regions (\citealt{Wu06}; \citealt{Engelbracht08}; \citealt{Gordon08}).
%The emission of an AGN, when it is present, has also to be taken into
%account. 
Moreover, there is some evidence that the PAH molecules
might be destroyed not only by the harsh radiation field of an AGN
(\citealt{Wu07}) but also by that of a young star-forming region if the
PAH carriers are sufficiently close to the source (\citealt{Mason07}).
On the other hand, PAH emission has been detected in the vicinity of
AGNs (e.g., Circinus, \citealt{Roche07}; NGC~1068, \citealt{Mason06}).

MIR observations of high redshift LIRGs, ULIRGs and submillimiter
galaxies (SMGs) suggest that the star formation in these galaxies
is extended over several kpcs (\citealt{Farrah08}; \citealt{MD09}).
In turn, a significant fraction of the star formation taking place
in local (U)LIRGs is, however, mostly confined to their nuclei,
within the inner few kpcs (e.g., \citealt{Gallais04}; \citealt{Egami06};
\citealt{GM06}; \citealt{AAH06a}; \citealt{DS07b}, 2008; \citealt{AAH09a}).
Besides, the fact that the star-forming regions from where the MIR emission
arises are very compact (hundred of pc or less), prevents us from spatially
resolving them in individual sources using current space-based observations
(\textit{IRAS}, \textit{ISO} or even \textit{Spitzer}).
%In fact,
%%While high sensitivity MIR observations from space have been provided by the \textit{IRAS}, \textit{ISO}, \textit{Akari} and \textit{Spitzer} telescopes,
%the scientific community has not exploited to the same degree
%the high spatial resolution capabilities of ground-based MIR facilities.
%MIR observations obtained with $8-10\,$m class telescopes, although
%limited to few number of atmospheric windows (principally $N$-banFd:
%$\sim\,8-13\,\micron$ and $Q$-band: $\sim\,16-23\,\micron$) can achieve
%resolutions of $\sim\,$100\,pc at distances $\lesssim\,70\,$Mpc, and
%allow to study the physical properties of LIRGs and ULIRGs in detail
%and at the smallest scales. Moreover, these studies are essential to
%understand the processes taking place in their counterparts at high-$z$,
%when this class of galaxies were dominating the SFR of the Universe
%(\citealt{PG05}; \citealt{Caputi07}).
%Therefore it is necessary to find reliable tracers of both components to separate out their different contributions to the energy output of these galaxies. %First, to demonstrate the presence of the sources of emission and second, to characterize their properties such as the age, extinction and mass of the stellar populations in the case of star-forming regions, and the optical depth and/or ionization conditions in the case of the existence of an AGN.
In this paper we present high spatial (sub-arcsecond, FWHM$\,\sim\,$0.3\arcsec)
resolution ground-based MIR spectroscopy of 4 local LIRGs (NGC~3256,
IC~4518W, NGC~5135 and NGC~7130), in which star formation and AGN
activity are isolated or mixed on different spatial scales.
Using the spectral features found in the $N$-band
($\sim\,8-13\,\micron$) Gemini T-ReCS spectra of these LIRGs,
we characterize the nature and properties of their energy sources
and perform a detailed study of the star formation in these galaxies
on scales of a few hundreds of pc.
%and explore the reliability of some MIR features tracers of the SFR on LIRGs
%at scales of a few hundreds of pc.
The organization of this
paper is as follows. In \S\ref{s:spec} we present the MIR observations
and the data reduction; in \S\ref{s:adddata} we present the
complementary data at MIR and other wavelengths; \S\ref{s:analysis}
regards the analysis of the data; in \S\ref{s:spitzdiffs} we compare
the MIR spectra of the LIRGs at different spatial scales; in
\S\ref{s:siabsfeat} we examine the strength and spatial variations
of the 9.7$\,\micron$ silicate feature in the nuclear region of the
galaxies; in \S\ref{s:neii} we explore the use of
the \NeII\ emission line as
reliable tracer of the SFR at scales of a few hundreds of pc
and in \S\ref{s:pah} we study the effect of the intensity and
hardness of the radiation field in the emission of the \PAHa\
feature; finally, in \S\ref{s:summary} we
summarize the main results obtained in this paper. Additionally,
in the Appendix we analyze the spatial profiles of the fluxes
of the main $N$-band spectral features and give detailed information
about the specific results found for each LIRG.

% and show that there is not a unique $N$-band reliable indicator to trace them.

\section{MIR Spectroscopy at Sub-arcsecond Scales}\label{s:spec}
%\section{Observations and Data Reduction}\label{s:obs}

We selected
4 galaxies from the sample of local LIRGs of \cite{AAH06b}:
NGC~3256, IC~4518W, NGC~5135 and NGC~7130.
The galaxies were chosen to cover a variety of morphologies
(compact/extended) and to include star-forming processes and AGN
activity in various environments and mixed on different
physical scales. In Table~\ref{t:sample} we summarize the
main properties of these LIRGs.

\begin{deluxetable}{lccccc}
\tabletypesize{\scriptsize}
%\rotate
\tablewidth{0pc}
%\tablenum{}
%\tablecolumns{9}
\tablecaption{\scriptsize The Sample}
\tablehead{\colhead{Galaxy} & \colhead{z} & \colhead{Dist} & \colhead{log
$\LIR$} & \colhead{Type}  & \colhead{12+log(O/H)} \\ 
%\cline{6-8}
\colhead{name} & \colhead{} & \colhead{[Mpc]} & \colhead{[\Lsun]} & \\
\colhead{(1)} & \colhead{(2)} & \colhead{(3)} & \colhead{(4)} & \colhead{(5)} & \colhead{(6)}}
\startdata 
NGC 3256        & 0.00935 & 40.4 & 11.67 & \HII  & 8.8  \\ % 2804 
IC 4518W        & 0.01573 & 68.2 & 11.09 & Sy2   & 8.6  \\ % 4715 
NGC 5135        & 0.01369 & 59.3 & 11.27 & Sy2   & 8.7  \\ % 4105 
NGC 7130        & 0.01615 & 70.1 & 11.39 & L/Sy  & 8.8     % 4842 
\enddata
\tablecomments{\scriptsize (1) Galaxy name; (2) Redshift (from NED); (3) Distance as obtained with the cosmology: $H_0 = 70\,{\rm km\,s}^{-1}\,{\rm Mpc}^{-1}$, $\Omega_{\rm M} = 0.27$, $\Omega_\Lambda = 0.73$; (4) Infrared luminosity ($8-1000\,\micron$) as computed from \textit{IRAS} fluxes (\citealt{Sanders03}) and using the prospect given in (\citealt{Sanders96}); their Table 1); (5) Nuclear activity of the galaxy; (6) Oxygen abundances (see \citealt{DS08} and references therein).}\label{t:sample}
\vspace{0.5cm}
\end{deluxetable}

%\begin{figure*}[!h]
%\epsscale{1.1}
%%\vspace{1.cm}
%\plotone{./f1.ps}
%\vspace{.25cm}
%\caption{\footnotesize $N$-band images of the nuclear (inner $\sim\,2\,$kpc) of NGC~3256, NGC~5135, IC~4518W and NGC~7130. The parallel lines represent (on scale) the position and angle of the slit with which the spectroscopy observations were obtained. The orange cross marks the nucleus of each galaxy while the orange dot marks other regions of interest: in NGC~3256, the southern nucleus of the galaxy; and in NGC~5135 and NGC~7130, a MIR bright \HII\ region. [\textit{See the electronic edition of the Journal for a color version of this figure.}]}\label{f:slits}
%\vspace{.25cm}
%\end{figure*}

\subsection{High spatial Resolution T-ReCS $N$-band Observations}\label{ss:specobs}

\begin{figure*}
\epsscale{1.}
%\plotone{./figures/slits_panel.ps}
\plotone{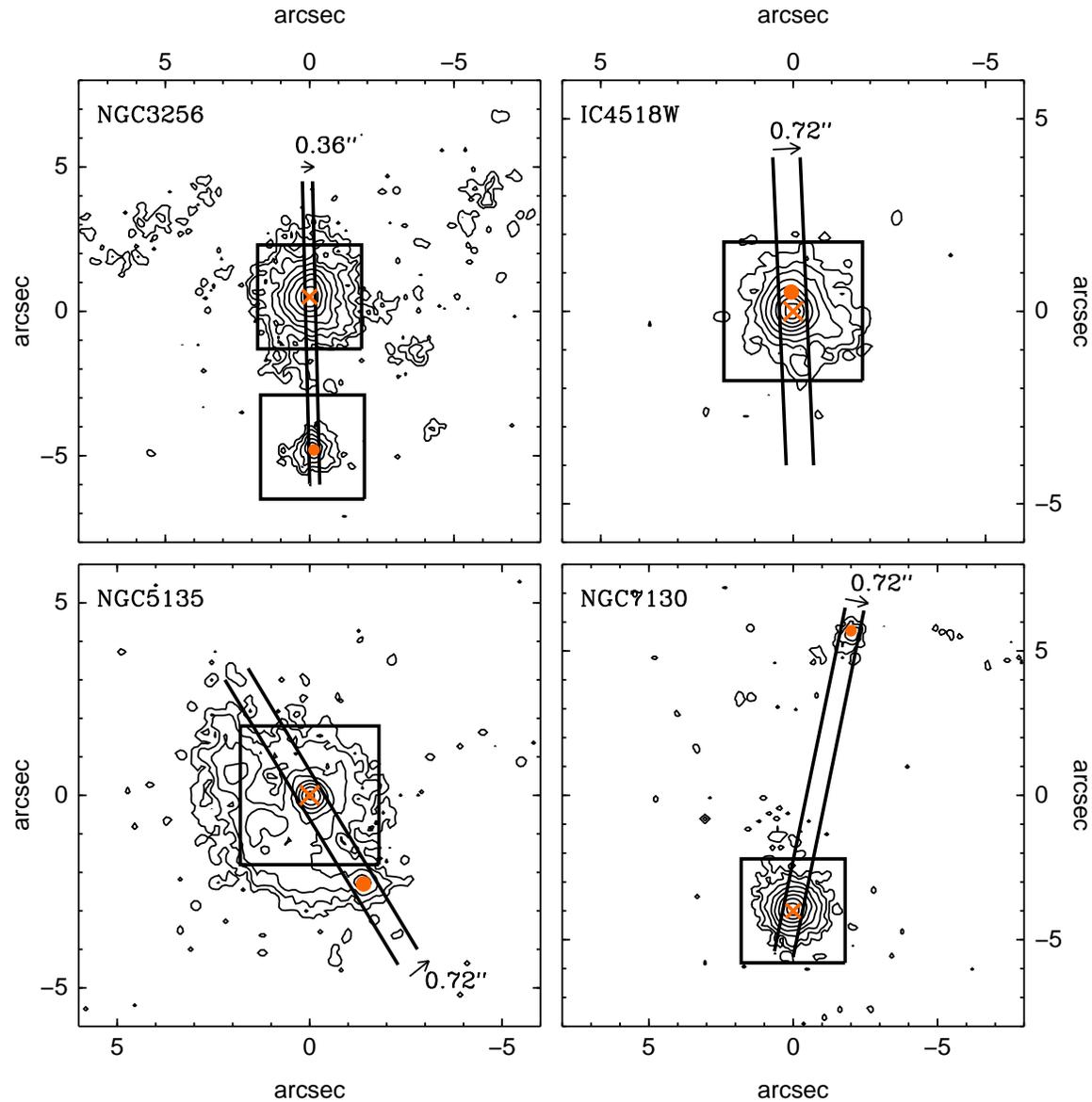}
\vspace{.5cm}
\caption{\footnotesize Si-2 (NGC~3256) and $N$-band (NGC~5135,
  IC~4518W and NGC~7130) images (from \citealt{DS08}) 
  of the central (inner $\sim\,2\,$kpc)
  regions of the LIRGs. The parallel lines represent (true scale) the
  width and orientation of the T-ReCS long slit. 
  The (orange) cross marks the nucleus of each galaxy,
  while the orange dots mark other regions of interest: in NGC~3256,
  the southern nucleus of the galaxy; in IC~4518W, an extended region
  detected in \SIV\ line emission; and in NGC~5135 and NGC~7130, two MIR
  bright \HII\ regions. The squares represent approximately the
  \textit{minimum} aperture size that can be used for extracting the
  low-resolution (SL module) IRS spectra, although their plotted
  position angles do not match those of the \textit{Spitzer} observations.
  [\textit{See the electronic edition of the Journal for a color version
  of this figure}].
}\label{f:slits}
\vspace{.5cm}
\end{figure*}

We obtained $N$-band, low-resolution (from
$R\,\sim\,80$ at $8\,\micron$ to $R\,\sim\,150$ at $13\,\micron$), 
long-slit spectroscopy of the 4 LIRGs mentioned above with T-ReCS
on the 8.1\,m Gemini South telescope. The observations were carried out
during 2005 and 2006 under programs GS-2005B-Q-10,
GS-2006A-Q-7 and GS-2006A-DD-15 (PI: Packham). T-ReCS has a
$320\,\times\,240$ pixel detector with a plate scale of 0.09\arcsec,
which provides a field of view of $\sim 29\arcsec\,\times\,21.5\arcsec$.
We used a slit width of 0.72\arcsec\ for NGC~5135,
IC~4518W, and NGC~7130, and of 0.36\arcsec\ for NGC~3256.
The slits were placed to cover the nucleus of the galaxies
as well as some regions of interest (see Figure~\ref{f:slits},
orange crosses and dots, respectively).

The observations were obtained in a standard chop-nod strategy
to remove the time-variable sky background, telescope thermal emission,
and the 1/f detector noise (see also \citealt{Packham05}).
The chop throw was 15\arcsec\ and perpendicular to the slit.
%The telescope nodding was performed every 30\,s in all cases.
The observations were scheduled to be done in a single night but
they were divided in various datasets to avoid observing problems
or a sudden change in weather conditions.

We observed two Cohen standard stars (\citealt{Cohen99})
for each galaxy to obtain
the wavelength and absolute-flux calibration of the spectra.
The standard star observations were taken with the same
instrument configuration and immediately before and after the
target to minimize the uncertainties in the photometric calibration.
The standard stars were chosen to have similar
air-masses as the galaxies at the time of the observations.
In order to place the slit along the
regions of interest, the instrument was rotated accordingly. The
standard stars were observed with the same configuration.

%The on-source integration time requested for each LIRG was 31 minutes. As mentioned above, the runs were scheduled to be executed in one go but due to several unforeseen circumstances some of the observations were split into different nights and stored in separate datasets.
Details about the integration times and the final useful
data available for each galaxy after discarding chop/nod pairs
affected by noise are given in Table~\ref{t:specobs}.

\begin{deluxetable}{ccccc}
\tabletypesize{\scriptsize}
%\rotate
\tablewidth{0pc}
%\tablenum{}
%\tablecolumns{9}
\tablecaption{\scriptsize Log of the T-ReCS Spectroscopic Observations}
\tablehead{\colhead{Galaxy} & \colhead{Slit width} & \colhead{$t_{\rm int}$} & \colhead{Date} & \colhead{Seeing} \\ 
%\cline{6-8}
\colhead{name} & \colhead{[\arcsec]} & \colhead{[s]} &  & \colhead{[\arcsec]} \\
\colhead{(1)} & \colhead{(2)} & \colhead{(3)} & \colhead{(4)} & \colhead{(5)}}
\startdata 
NGC 3256  & 0.36\arcsec\ &  1800   &  2006/03/07  &  0.35\arcsec\ \\
IC 4518W  & 0.72\arcsec\ &  1900   &  2006/04/17  &  0.36\arcsec\ \\
NGC 5135  & 0.72\arcsec\ &   633   &  2006/03/06  &  0.30\arcsec\ \\
          & 0.72\arcsec\ &  1267   &  2006/03/10  &  0.29\arcsec\ \\
NGC 7130  & 0.72\arcsec\ &   490   &  2005/09/18  &  0.37\arcsec\ \\
          & 0.72\arcsec\ &   760*  &  2006/06/04  &  0.30\arcsec\ \\
          & 0.72\arcsec\ &   633   &  2006/08/29  &  0.36\arcsec\ \\
          & 0.72\arcsec\ &  1267*  &  2006/09/16  &  0.33\arcsec\ \\
          & 0.72\arcsec\ &  1267   &  2006/09/25  &  0.40\arcsec\ 
\enddata
\tablecomments{\scriptsize Note.-- (1) Galaxy; (2) Slit width; (3) Useful on-source integration time; in the observations marked with an asterisk, only the spectrum of the nucleus (i.e., not including the \HII\ region) was obtained; (4) Date(s) of the observations (YYYY/MM/DD); (5) Mean seeing (FWHM of the reference standard star(s) observed right before and/or after the target) throughout each night of observations.}\label{t:specobs}
\end{deluxetable}

\subsection{Data Reduction And Photometric Calibration}\label{ss:specdatared}

\subsubsection{Obtaining the 2-D Spectra}\label{sss:2dspec}

\begin{figure*}
\epsscale{1.}
%\plotone{./figures/2D_panel.ps}
\plotone{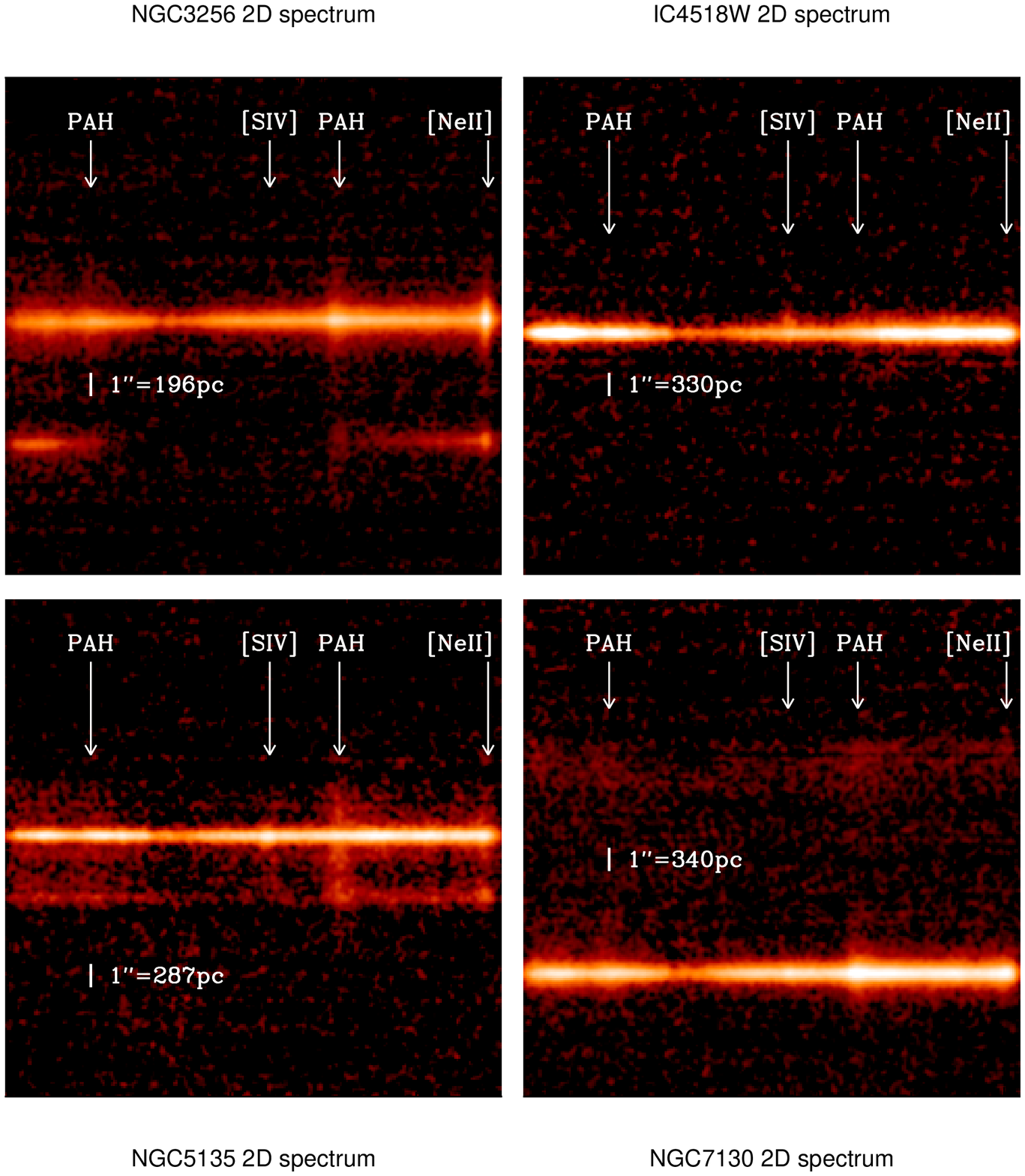}
\vspace{0.5cm}
\caption{\footnotesize Averaged T-ReCS 
2D-spectra of the LIRGs. These images have been partially reduced (see
text). The horizontal axis is the wavelength dispersion direction
(from $\sim\,$8 to 13$\,\micron$), while the vertical axis is the
spatial direction. The physical scale for each galaxy is marked on the
images, as well as the most prominent spectral features. Note that
even in these partially-reduced 2D-spectra, some features are
clearly extended. See Figure~\ref{f:slits} for
details about the orientation and position of the slits used for
obtaining these spectra.
[\textit{See the electronic edition of the Journal
for a color version of this figure}].
}\label{f:2Dspecs}.
\vspace{0.5cm}
\end{figure*}

The T-ReCS data are stored automatically in \textit{savesets},
which contain information on
the on-source (object + sky) and off-source
(sky) frames for each chop/nod position.
%They contain an image of the 2D-spectrum of the target (galaxy or standard star), an image of its associated sky 2D-spectrum, and an image of the sky-subtracted 2D-spectrum of the target.
The savesets can be accessed directly to
discard images affected by any type of instrumental noise pattern.
% (e.g., narrow diagonal stripes of increased signal across the detector or over-variations of the background flux along the array or in between chop sets).
A given number of savesets forms a dataset.
Once the bad savesets of a dataset are discarded, the sky-subtracted
images were stacked to obtain a single image of the 2D-spectrum of
the target. Because some of the datasets were obtained in different
nights and under different atmospheric conditions (see above), the
following procedures were applied to each galaxy-standard pair of
datasets individually as they contained the full information to perform
their own calibration. Figure~\ref{f:2Dspecs} shows the 2D-spectrum
of each LIRG (the different savesets were averaged for obtaining
high S/N images but these were not used for scientific purposes).
Note that even in these partially reduced T-ReCS spectra (not
corrected from the shape of the atmospheric transmission; see below)
we can detect the 8.6 and \PAHas\ and the \SIV, and \NeII\ emission
lines, together with the 9.7$\,\micron$ silicate absorption feature
(see Figure~\ref{f:2Dspecs}).
%However, since our spectra have a limited S/N ratio and spectral resolution, we are not able to see the faintest features as e.g., the \H2 emission lines or some of the weakest PAHs.

After applying a flat-field to the datasets
%We applied a flat-field image to the final 2D-spectrum image (dataset) which was constructed using the average of all the sky pointings taken at each chop-nod integration of a short observation of the target made in imaging mode (a pre-visualization image obtained before the spectroscopy).
we made use of
Gemini-based {\sc iraf}\footnote{{\sc iraf} is written and supported
by the {\sc iraf} programming group at the National Optical Astronomy
Observatories (NOAO), which are operated by the Association of Universities
for Research in Astronomy, Inc., under cooperative agreement with the
National Science Foundation. http://iraf.noao.edu} tasks (\textit{nswavelength}
and \textit{nstransform}) to calibrate in wavelength the 2D-spectra.
%To do so, we measured the positions on the detector of the main absorption bands seen in the reference (sky) stacked spectra of the galaxy dataset and compared them with a reference file containing the wavelengths of the main atmospheric absorption features that can be found in a $N$-band spectrum. Then, we calculated
We used the sky emission features in the reference (sky) stacked spectra to
%($\lambda = f$[pixel])
calibrate in wavelength both the the galaxy and standard star
2D-spectra. For the next final steps of the data reduction,
we used our own in-house developed {\sc idl} routines.

%\begin{figure}[!h]
%\begin{center}
%\plotone{./figures/ngc5135/ast_S20060310S0106.ps}
%\caption{\footnotesize Example of a stacked reference sky spectrum used for wavelength calibrating the entire 2D-spectrum of the associated source (from $\sim\,$8 to 13$\,\micron$). The emission of the sky depends on atmospheric conditions such as the air mass at the position of the source and the water vapour column density at the time of the observations.}\label{f:refsky} 
%\end{center}
%\end{figure}

In order to subtract the residual background from
the 2D-spectrum image, it was fitted to a 2D plane with a fitting
algorithm that iteratively rejects outliers 2.25\,$\sigma$ above/below
this plane to ensure that only ``sky'' pixels are used for the fit.

\subsubsection{Extraction and Calibration of the 1-D Spectra}\label{sss:1dspec}

Once the 2D-spectrum images of the galaxy and the standard star were
wavelength-calibrated and background-subtracted, we used the high
S/N ratio of the spectrum of the standard star to trace the
apertures of the 1-D spectra.

%We extracted the nucleus of each LIRG using fixed and variable aperture sizes (see below). In addition to the nuclei,
We extracted the nuclear spectrum of each LIRG as well as 1-D
spectra at regular positions (1\,pixel\,$\simeq\,$\,0.09\arcsec\ step)
along the slit in the spatial direction using a fixed aperture
of 4\,pixels in length. We then used this grid of spectra to construct
spatial profiles of the features detected in the spectra of each LIRG
(see Appendix). All the regions (nuclei and star-forming regions)
were assumed to be extended. We also extracted the integrated spectra
of the galaxies with a fixed aperture of 40\,pixel ($\sim\,$3.6\arcsec),
as well as the spectra of the regions of interest.
%with a fixed aperture of 4\,pixel.

After the spectrum of a given position was extracted from every dataset,
all the spectra were interpolated to a common wavelength array, with a
spacing approximately similar to that provided by the wavelength-calibration
function ($\sim\,0.022\,\micron\,$pixel$^{-1}$). This step was needed
since the observations of some galaxies were carried out on different nights.
%As a consequence, the wavelength calibration of each dataset yielded slightly different solutions.

Each interpolated spectrum was then flux-calibrated using its
associated Cohen standard star while taking into account the slit
losses in each dataset.
%Each extracted spectrum was then flux-calibrated using its associated Cohen standard star. % associated to its dataset. First, the galaxy spectrum was divided by the observed standard star spectrum as extracted with a wide aperture to include all its flux and then interpolated to the same wavelength array as the object. The result was then multiplied by the synthetic, absolute-flux calibrated spectrum of the standard star and by a constant to account for the slit losses in each dataset.
%The slit losses were calculated independently for each individual dataset and were in the range of $1.05-2.2$ depending on the seeing and the width of the slit used for the observations (see Table~\ref{t:specobs}). The largest corrections were made to the spectra of NGC~3256, obtained with the 0.36\arcsec\ slit.
By comparing the flux-calibration
%conversion factors used for calibrating in flux
of the different datasets
% of the different standard stars,
we determined that this is accurate to within $\sim\,15\%$
uncertainty. Finally, the spectrum of each dataset was weighted by the number of
savesets used for constructing the dataset (i.e., by the exposure time)
and then averaged.
The uncertainties of the background emission were
obtained by calculating the standard deviation of the ``sky'' spectrum
of the grid (see above) defined as those spectra having a $N$-band flux
density below 3\,mJy. The extracted nuclear and integrated spectra of
the LIRGs can be seen in Figure~\ref{f:trecsspitzerspecs}.
%, for the extraction apertures of 1 and 4\,pixel respectively.

\section{Complementary Data}\label{s:adddata}

\subsection{NIR \textit{HST} NICMOS Imaging}\label{ss:imagadddata}

The 4  galaxies selected for this work are from the HST/NICMOS
survey of a volume-limited sample of local LIRGs of \cite{AAH06b}.
The observations were taken
with the NIC2 camera (pixel size of
0.075\arcsec\ and FOV of 19.3\arcsec$\,\times\,$19.3\arcsec)
using two broad-band filters (F110W and F160W) for obtaining
NIR continuum imaging, and two narrow-band filters
(F190N and F187N) for observing the \Paalpha\ line emission
and its associated continuum. We refer the reader to \cite{AAH06b}
for  details on the observations and data reduction.
Using the fully-reduced images 
we constructed continuum-subtracted \Paalpha\ emission,
and color maps of the nuclear regions of the galaxies.

%We also
%calculated the ages of the star-forming regions in the LIRGs
%by means of their \Paalpha\ EWs (see \citealt{DS08} for details).
A few extra steps were needed to make a meaningful
comparison between the NICMOS and the T-ReCS images.
Most importantly, rescaling the pixel size
and match the resolution of the images (see \citealt{DS08} for details).
%{\bf These included ???????}.
In addition, we simulated the positions of the T-ReCS slits 
(see \S\ref{ss:specobs}) over the \textit{HST}
NICMOS \Paalpha\ images and extracted the spatial profiles from them.
These profiles were then compared with those obtained for the MIR
spectral features (see Appendix).
The \Paalpha\ and NIR continuum images where also used to estimate
the ages and extinctions for each of the star-forming regions studied
in this work. The ages were inferred from their \Paalpha\ equivalent
widths (EWs) using
Starburst99 models (\citealt{Leitherer99}) and are upper limits to
the real ages of the regions. The extinctions were calculated using
the NIR colors and are lower limits to the real values.
For more details about the approach used, see \cite{DS08}.

\subsection{MIR \textit{Spitzer} IRS Spectroscopy}\label{ss:specadddata}

For NGC~3256, NGC~5135 and NGC~7130, we compare our T-ReCS data to the
\textit{Spitzer} IRS spectra from \cite{PS09a}. The spectra were obtained
with the short-wavelength, low resolution module (SL), which
%We have additional spectra of NGC~3256, NGC~5135 and NGC~7130 obtained using the spectral mapping capability of the IRS on-board \textit{Spitzer}. The observations were made under program 30577 (PI: G. Rieke) and used all the IRS modules available: SL, SH, LL, LH (see \citealt{AAH09b} and \citealt{PS09a} for a full description of the observing strategies, data reduction and analysis of the data cubes). Here we use data only from the short-wavelength, low resolution module (SL). The SL module
ranges from $\sim\,5.5\,\micron$ to $\sim\,13.5\,\micron$ and has a
spectral resolution of R$\,\sim\,60-130$, similar to that of T-ReCS
spectra.
% at the shortest wavelengths of the $N$-band window.
%The SH module ranges from $\sim\,10\,\micron$ to $\sim\,19.5\,\micron$ with R$\,\sim\,$600.
The extraction of the \textit{Spitzer} IRS 1D spectra for each LIRG
was made using various fixed apertures, from 2$\,\times\,$2\,pixel
to 4$\,\times\,$4\,pixel. The smaller aperture is approximately the
size (FWHM) of a point source at the end of the wavelength range
covered by the module, which is about 3.7\arcsec$\,\times\,$3.7\arcsec\
for the SL module. The flux-calibration was performed assuming
the sources were extended. The extracted spectra of the galaxies
are shown in Figure~\ref{f:trecsspitzerspecs}.

\section{Analysis}\label{s:analysis}

\subsection{Obtaining Sub-arcsecond Spatially Resolved Profiles}\label{ss:spatprof}

\begin{figure}%[!h]
\epsscale{1.1}
%\plotone{./figures/ngc3256_compare_spat_pos.ps}
\plotone{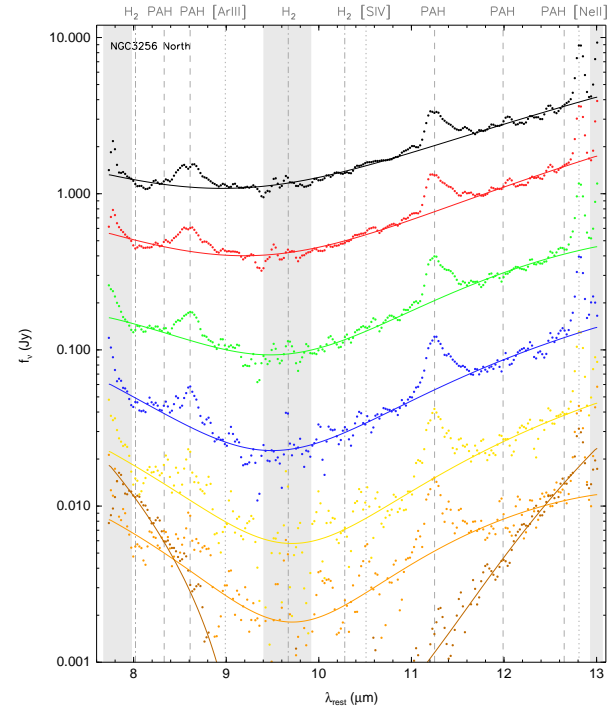}
\vspace{0.25cm}
\caption{\footnotesize T-ReCS spectra of NGC~3256 extracted with a
  fixed aperture of 0.36\arcsec\ at different spatial positions in
  steps of 2\,pixels (0.18\arcsec) from the northern nucleus (top
  black dots) down to 0.9\arcsec\ to the south (orange dots). The
  spectrum of the southern nucleus (at $\sim\,$5\arcsec\ to the south)
  is shown for comparison (bottom brown dots). For clarity, the
  spectra (from top to bottom) have been multiplied by the following
  factors: 64, 32, 16, 8, 4, 2, and 1. The lines are the fits to the
  spectra   (see text for details). The most prominent
  features are marked at the top of the figure. PAH features are
  marked with dashed lines, the unresolved forbidden lines are marked
  with dotted lines, and the location of the \H2 emission lines are
  marked with dotted-dashed lines. The shaded regions correspond to
  wavelength ranges where the atmospheric transmission is very poor,
  causing the spectra to be very noisy.
[\textit{See the electronic edition of the Journal
for a color version of this figure}].
}\label{f:ngc3256spatpos}.
\vspace{0.25cm}
\end{figure}

%To obtain the spatial profiles of the MIR features, we extracted spectra at regular positions (1\,pixel\,$\simeq\,$\,0.09\arcsec\ step) along the spatial direction (parallel to the slit) employing an aperture of 4\,pixel ($\simeq\,$\,0.36\arcsec)
%These spatial profiles were obtained for the fluxes and EWs of the different spectral features and their ratios (see Figures~\ref{f:slits} and \ref{f:2Dspecs}).
Figure~\ref{f:ngc3256spatpos} shows an example of some individual spectra
of NGC~3256 extracted at different positions along the slit
(see \S\ref{sss:1dspec}). The main emission features seen are
the 8.6 and \PAHas, and the \SIV\ and \NeII\ lines.
% and the 9.7$\,\micron$ silicate absorption feature.

To measure each feature, we first fitted
% the local continuum of
each 1-D spectrum with a polynomial function using a $\chi^2$-minimization
method (with a weight $\propto$ 1/$\sigma^2$, where $\sigma$ is the total
uncertainty at each wavelength) after masking out the most prominent
features in the spectra.
% The continuum reference wavelengths were  set to 8.2 and 12.2$\,\micron$.
A few examples of these fits are shown in
%sss:1dspec}
Figure~\ref{f:ngc3256spatpos}.
%Once the local continuum was fitted
Next, we: (1) measured the fluxes
of the PAHs and emission lines;
%(see Tables~\ref{t:nucfluxes} and~\ref{t:intfluxes})
(2) calculated their EWs
%(see Tables~\ref{t:nucews} and~\ref{t:intews});
and (3) obtained the continuum flux at two reference wavelengths:
8.2 and 12.2$\,\micron$. The fluxes of the features
were calculated by fitting them with Gaussian functions with fixed
widths and varying intensities and positions. Although a Gaussian
profile does not represent perfectly the shape of the PAH bands,
it is still a reasonable choice and a good approximation to their
enclosed flux. We considered a feature as detected when its peak
was 2.25$\sigma$ above the continuum.
%, where $\sigma$ is the standard deviation of the fitted local continuum.
This threshold was chosen as a good compromise
between being able to detect low surface brightness emission in
diffuse regions and not including unreliable measurements.

We note that the \NeII\ line may be contaminated with emission
from the 12.7$\,\micron$ PAH feature, since the T-ReCS spectral
resolution is not sufficient to separate both components. In addition,
due to the distance to IC~4518W, NGC~5135 and NGC~7130, the \NeII\
emission line of these LIRGs is red-shifted almost outside of the
T-ReCS $N$-band filter. This causes the red wing of the line to be
significantly affected by the filter and atmospheric transmissions; the
latter starts decreasing significantly at $\lambda\,\gtrsim\,�13\,\micron$.
We therefore doubled
the uncertainties in the fluxes measured of \NeII\ line for these
galaxies because the fits to this line were made using only the half
of the line not affected by the filter and atmospheric transmissions.
%added manually an additional uncertainty to the uncertainties obtained from the fit of this line in the galaxies mentioned above. Indeed, we have 
The uncertainties of the ratios and other quantities obtained using
this line were calculated accordingly, including this additional
uncertainty.

%The fluxes were calculated as: $F = \sqrt{2\pi\sigma}\times I$, where F is the flux, $\sigma$ is the width of the Gaussian, and I is the peak of intensity. Figs.~\ref{f:ngc3256spatproffluxes}, \ref{f:ic4518wspatproffluxes}, \ref{f:ngc5135spatproffluxes}, and \ref{f:ngc7130spatproffluxes} show the flux spatial profiles of the detected features in each LIRG. The tables with the tabulated data of each feature are given in the electronic version of the paper.

\begin{table*}
\begin{center}
\caption{T-ReCS Nuclear Fluxes}\label{t:nucfluxes}
\begin{tabular}{cccccc}
\hline
\hline
Nucleus & \multicolumn{5}{c}{Feature} \\
\cline{2-6}
& \PAHb\ & [S\,{\sc iv}] & \PAHa\ & [Ne\,{\sc ii}] & \SSi\ \\
(1) & (2) & (3) & (4) & (5) & (6)  \\
\hline
NGC~3256 (N)  &   7.5$\pm$3.5   &  $<$1.0         &  11.9$\pm$1.3  &  12.9$\pm$0.5  &  $-$0.39$\pm$0.10    \\
NGC~3256 (S)  &  $<$1.7         &  $<$0.5         &  $<$1.0        &   2.1$\pm$0.2  &  $\dots$            \\
IC~4518W$^b$  &  $<$6.4         &   2.8$\pm$0.9   &  $<$3.7        &   6.9$\pm$1.3  &  $-$1.42$\pm$0.19$^a$   \\
NGC~5135$^b$  &  $<$5.5         &   5.2$\pm$0.9   &  $<$3.2        &   5.1$\pm$1.0  &  $-$0.46$\pm$0.18    \\
NGC~7130      &   5.5$\pm$4.4   &   2.5$\pm$1.8   &  11.4$\pm$1.3  &  10.3$\pm$1.6  &  $-$0.61$\pm$0.08    \\
\hline
\hline
\end{tabular}
\end{center}
\footnotesize{Note.-- (1) Name. (2)--(5) Fluxes of the features and 2.25$\sigma$ upper limits (see text) in units of $\times$\,10$^{-14}$\,erg\,s$^{-1}$\,cm$^{-2}$. (6) Silicate strength of the LIRG nuclei (see \S\ref{ss:siabsstren} and Equation~\ref{e:siabs}).\\ $^a$ The \SSi\ of IC~4518W was calculated using the alternative method explained in \S\ref{ss:siabsstren}.\\
$^b$ The nuclei of IC~4518W and NGC~5135 are unresolved. The values given here have not been corrected for aperture effects.\\
The values correspond to the 1-D spectra extracted with a fixed aperture of 0.36\arcsec$\,\times\,$[0.36\arcsec\ (NGC~3256), and 0.72\arcsec\ (IC~4518W, NGC~5135, NGC~7130)] centered at the nuclei of the galaxies.}
\end{table*}

\begin{table*}
\begin{center}
\caption{T-ReCS Integrated Fluxes}\label{t:intfluxes}
\begin{tabular}{cccccc}
\hline
\hline
Nucleus & \multicolumn{5}{c}{Feature} \\
\cline{2-6}
& \PAHb\ & [S\,{\sc iv}] & \PAHa\ & [Ne\,{\sc ii}] & \SSi\ \\
(1) & (2) & (3) & (4) & (5) & (6)  \\
\hline
NGC~3256 (N) &  23.0$\pm$9.3   &  $<$3.0         &  40.0$\pm$3.9  &  42.8$\pm$2.0  &  $-$0.62$\pm$0.13    \\
NGC~3256 (S) &  $<$5.4         &  $<$1.          &  $<$1.0        &   3.2$\pm$1.7  &        \dots        \\
IC~4518W     &  $<$17.2        &  12.5$\pm$2.5   &  $<$10.1       &  22.9$\pm$4.1  &  $-$1.45$\pm$0.28$^a$   \\
NGC~5135     &  19.2$\pm$13.3  &  13.1$\pm$6.3   &  36.7$\pm$7.1  &  19.2$\pm$4.8  &  $-$0.63$\pm$0.23    \\
NGC~7130     &  17.5$\pm$10.9  &   5.2$\pm$5.2   &  45.0$\pm$4.4  &  25.4$\pm$4.2  &  $-$0.75$\pm$0.10    \\
\hline
\hline
\end{tabular}
\end{center}
\footnotesize{Note.-- (1) Name. (2)--(5) Fluxes of the features and 2.25$\sigma$ upper limits (see text) in units of $\times$\,10$^{-14}$\,erg\,s$^{-1}$\,cm$^{-2}$. (6) Silicate strength of the LIRG nuclei (see \S\ref{ss:siabsstren} and Equation~\ref{e:siabs}).\\
$^a$ The \SSi\ of IC~4518W was calculated using the alternative method explained in \S\ref{ss:siabsstren}.\\
The values were calculated for the 1-D spectra as extracted using a fixed aperture of 3.6\arcsec$\,\times\,$[0.36\arcsec\ (NGC~3256), 0.72\arcsec\ (IC~4518W, NGC~5135, NGC~7130)] centered at the nuclei of the galaxies.}
\end{table*}

\begin{table*}
\begin{center}
\caption{T-ReCS Nuclear EWs}\label{t:nucews}
\begin{tabular}{ccccc}
\hline
\hline
Nucleus & \multicolumn{4}{c}{Feature} \\
\cline{2-5}
& \PAHb\ & [S\,{\sc iv}] & \PAHa\ & [Ne\,{\sc ii}] \\
(1) & (2) & (3) & (4) & (5) \\
\hline
NGC~3256 (N) &   0.11$\pm$0.05   &       \dots      &   0.16$\pm$0.02   &   0.12$\pm$0.01   \\%$\pm$0.006
NGC~3256 (S) &       \dots       &       \dots      &       \dots       &   0.07$\pm$0.01   \\
IC~4518W     &       \dots       &   0.05$\pm$0.02  &       \dots       &   0.03$\pm$0.01   \\%$\pm$0.006  \\
NGC~5135     &       \dots       &   0.05$\pm$0.01  &       \dots       &   0.04$\pm$0.01   \\%$\pm$0.007  \\
NGC~7130     &   0.05$\pm$0.04   &   0.03$\pm$0.02  &   0.18$\pm$0.02   &   0.06$\pm$0.01   \\
\hline
\hline
\end{tabular}
\end{center}
\footnotesize{Note.-- (1) Name. (2)--(5) EWs of the features in $\micron$. See also Table~\ref{t:nucfluxes} for notes about the extraction apertures.}
\end{table*}

\begin{table*}
\begin{center}
\caption{T-ReCS Integrated EWs}\label{t:intews}
\begin{tabular}{ccccc}
\hline
\hline
Nucleus & \multicolumn{4}{c}{Feature} \\
\cline{2-5}
& \PAHb\ & [S\,{\sc iv}] & \PAHa\ & [Ne\,{\sc ii}] \\
(1) & (2) & (3) & (4) & (5)  \\
\hline
NGC~3256 (N) &   0.13$\pm$0.05   &       \dots      &   0.23$\pm$0.03   &   0.16$\pm$0.01   \\
NGC~3256 (S) &       \dots       &       \dots      &   0.38$\pm$0.33   &   0.10$\pm$0.01   \\
IC~4518W     &       \dots       &   0.11$\pm$0.03  &       \dots       &   0.06$\pm$0.01   \\
NGC~5135     &   0.06$\pm$0.05   &   0.07$\pm$0.04  &   0.15$\pm$0.03   &   0.06$\pm$0.01   \\
NGC~7130     &   0.07$\pm$0.04   &   0.03$\pm$0.03  &   0.18$\pm$0.02   &   0.07$\pm$0.01   \\
\hline
\hline
\end{tabular}
\end{center}
\footnotesize{Note.-- (1) Name. (2)--(5) EWs of the features in $\micron$. See also Table~\ref{t:intfluxes} for notes about the extraction apertures.}
\end{table*}

%Figs.~\ref{f:ngc3256spatproffluxes}--\ref{f:ngc3256spatprofews} are for NGC~3256. Figs.~\ref{f:ic4518wspatproffluxes}--\ref{f:ic4518wspatprofews} are for IC~3256W.  Figs.~\ref{f:ngc5135spatproffluxes}--\ref{f:ngc5135spatprofews} are for NGC~5135. Figs.~\ref{f:ngc7130spatproffluxes}--\ref{f:ngc7130spatprofews} are for NGC~7130.

%Figs.~\ref{f:ngc3256spatproffluxes}, \ref{f:ic4518wspatproffluxes}, \ref{f:ngc5135spatproffluxes}, and \ref{f:ngc7130spatproffluxes} show the flux spatial profiles of 8.6 and \PAHas, and the \SIV\ and \NeII\ emission lines. Figs.~\ref{f:ngc3256spatprofconts}, \ref{f:ic4518wspatprofconts}, \ref{f:ngc5135spatprofconts}, and \ref{f:ngc7130spatprofconts} shows the spatial profile of the dust continuum emission at 8.2 and 12.2$\,\micron$ (almost outside of the wings of the \Siabs\ absorption feature).  Figs.~\ref{f:ngc3256spatprofratios}, \ref{f:ic4518wspatprofratios}, \ref{f:ngc5135spatprofratios}, and \ref{f:ngc7130spatprofratios} show the ratios between the different spectral features.  Figs.~\ref{f:ngc3256spatprofews}, \ref{f:ic4518wspatprofews}, \ref{f:ngc5135spatprofews}, and \ref{f:ngc7130spatprofews} show their EWs.

The fluxes and EWs of the different features measured in the nuclear
and integrated T-ReCS spectra are given in
Tables~\ref{t:nucfluxes}, \ref{t:intfluxes}, \ref{t:nucews}, and
\ref{t:intews}.

\subsection{Measuring the Strength of the \Siabs\ Feature}\label{ss:siabsstren}

An important quantity related to the \Siabs\ feature is its depth or
strength. This is calculated as:

\begin{equation}\label{e:siabs}
S_{\rm Si\,9.7\,\mu m}=ln\frac{F^{obs}_{\lambda}}{F^{cont}_{\lambda}}
\end{equation}

\noindent
where $F^{obs}_{\lambda}$ is the observed flux density of the feature
and $F^{cont}_{\lambda}$ is the continuum, both evaluated at
wavelength $\lambda$
(usually, 9.7$\,\micron$). A negative value indicates absorption, while
a positive value indicates emission. The \SSi\ can be associated
with an optical depth (and therefore with a
visual extinction) via an extinction law and a dust obscuration model
or geometry. For a dust screen configuration, the relation
between the \SSi\ (or equivalently in this case: apparent \tauSi\ $= -$\SSi)
and the apparent \tauV\ is unique and it is given by the shape of
the absorption curve of the adopted extinction law.

%However,
%in order to measure accurately the \Siabs\ strength one needs to know
%where the MIR continuum ($F^{cont}_{\lambda}$) at the peak of the
%absorption is. I.e., one needs to be able to measure the continuum
%outside the feature. 

The limited spectral range afforded by ground-based 
$N$-band  observations  in conjunction with the broad silicate
feature, and  the prominent PAH features in our
galaxies make it difficult to measure the MIR continuum. 
%We took two different approaches to estimate the MIR continuum and thus the strength of the \Siabs\ feature:
%(a) The first approach
To circumvent this problem, we made use of the
SL \textit{Spitzer} IRS spectra which has a larger spectral range.
We stress that the \textit{Spitzer} spectra are extracted with much
larger apertures (see \S\ref{ss:specadddata}) than the T-ReCS spectra,
and thus the main assumption here is that the shape of the continuum
over the \textit{Spitzer} (kpc) spatial scales is the same as that
over the T-ReCS (hundred pc) scales. That is, for a given galaxy,
we used the same
\textit{Spitzer} continuum for all the T-ReCS spectra extracted along
all spatial steps. This may have some implications in cases were the
\textit{Spitzer} spatial resolution cannot separate AGN and star
formation sites.

\begin{figure}%[!h]
\epsscale{1.1}
%\plotone{./figures/SL_sp_nuclear_3x3pix_Spitzerfit.ps}\plotone{./figures/SL_sp_nuclear_3x3pix_TReCSfit.ps} 
\plotone{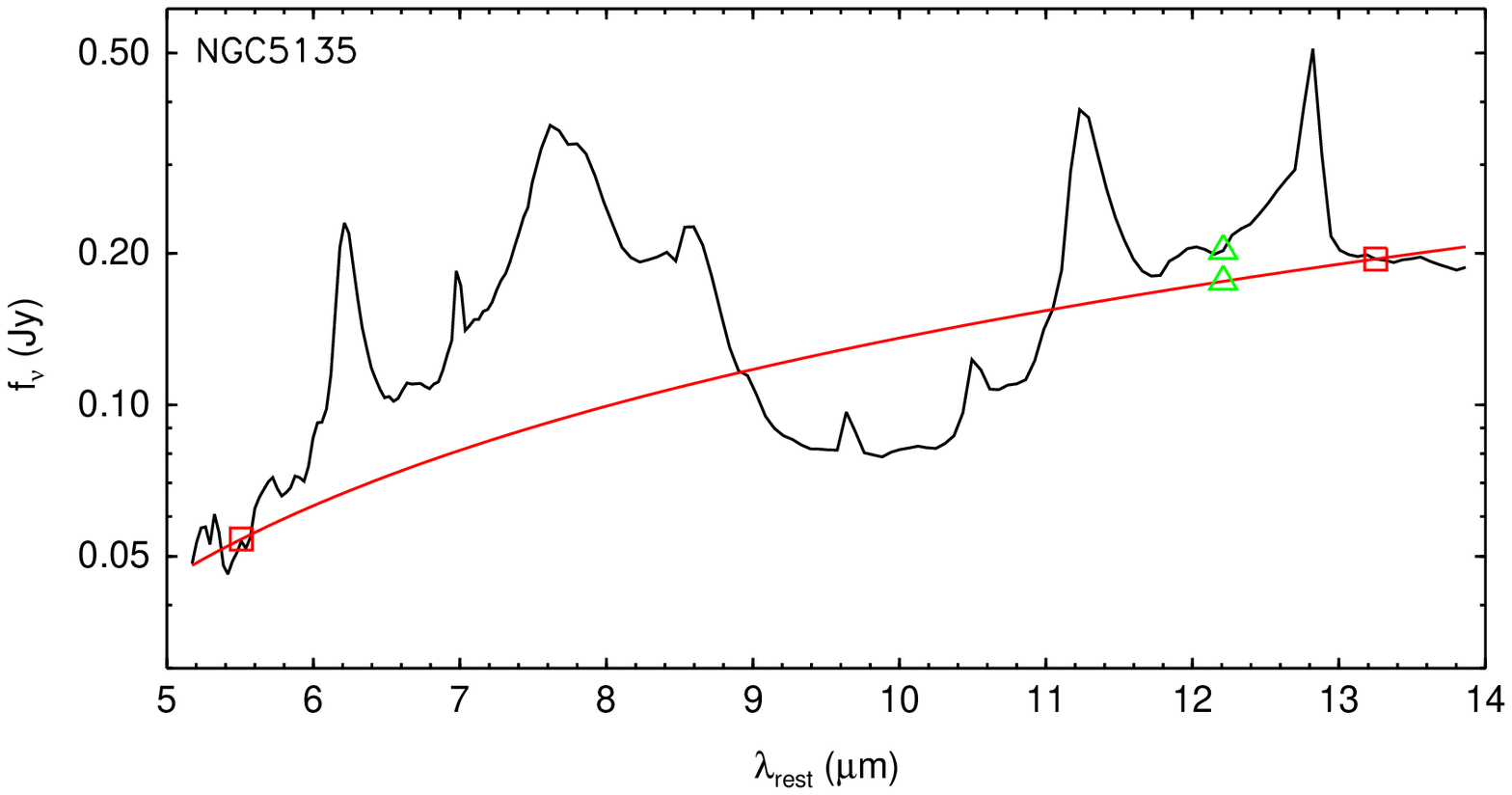}\plotone{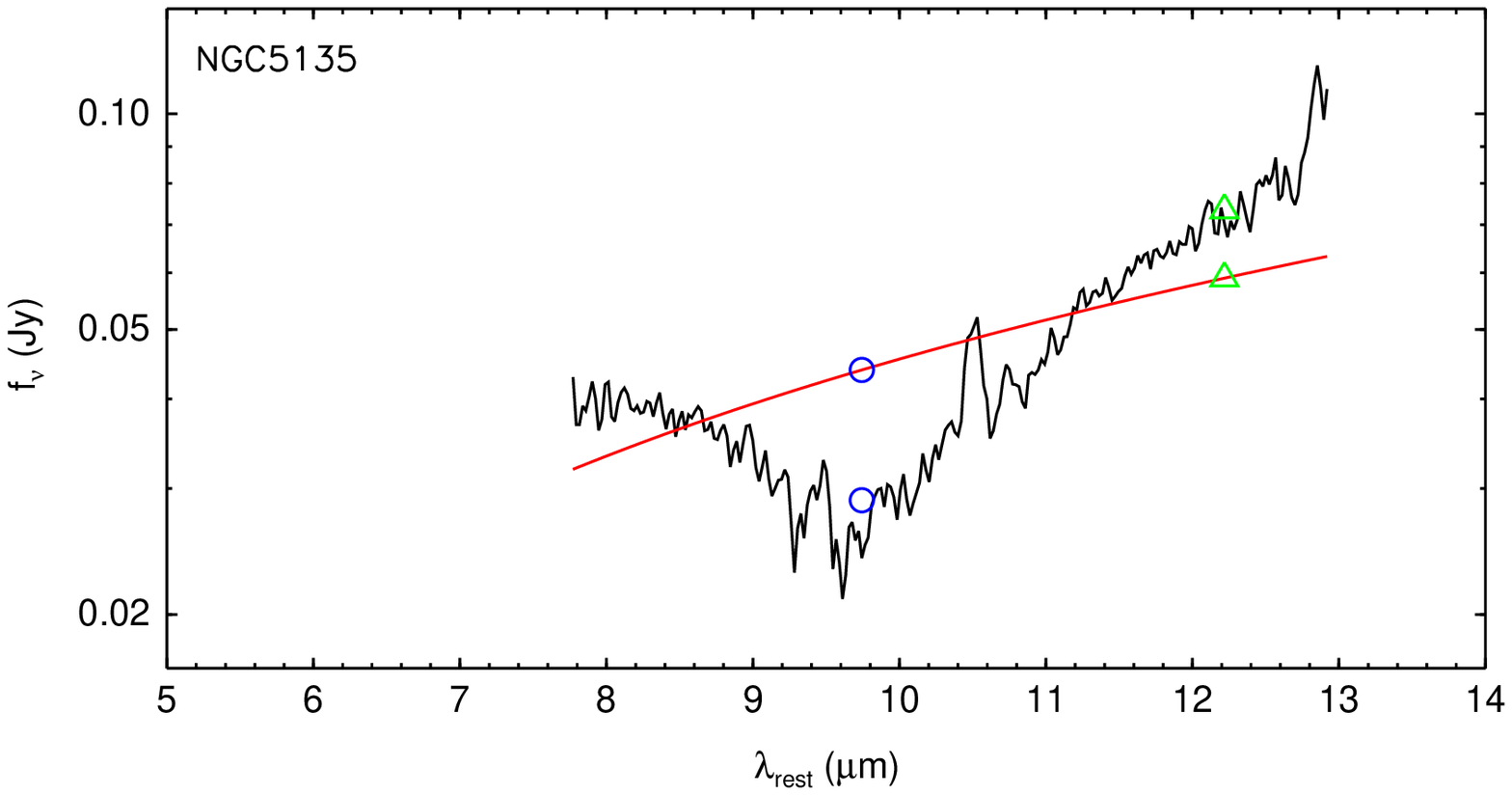} 
\vspace{0.25cm}
\caption{\footnotesize Upper panel: Example of the fitted continuum  
(red line) to
  the SL IRS spectrum (black line) of NGC~5135  to measure the strength
  of the silicate feature. The spectrum was extracted with an aperture
  size of 5.4\arcsec$\,\times\,$5.4\arcsec. The red squares are the
  anchors used for the fit. The green triangles are the spectrum and
  the fitted continua at 12.2$\,\micron$ , chosen as the reference
  wavelength for the normalization to the T-ReCS spectra (see text for
  details). Bottom panel: The red line is the IRS continuum fit scaled
  to the nuclear T-ReCS spectrum of NGC~5135 (black line), 
as extracted with an
  aperture size of 0.72\arcsec$\,\times\,$0.36\arcsec. 
 The ratio between the green
  triangles is the offset applied to the normalization of the fit. 
 The blue circles show the spectrum and fitted continua
  fluxes at the maximum of the \Siabs\ absorption
  feature. [\textit{See the electronic edition of the Journal
for a color version of this figure}].}\label{f:siabsfit}
\vspace{0.25cm}
\end{figure}

We fitted the IRS spectrum to a simple
linear function with anchors at 5.5 and 13.2$\,\micron$ (see
Figure~\ref{f:siabsfit}, top). This method yielded similar
results to the method of \cite{Sirocky08}, which uses a spline
interpolation. We then measured the
ratio between the measured flux at 12.2$\,\micron$
in the IRS spectrum, and the fitted value at the same wavelength.
We used this ratio to scale the fit to the T-ReCS spectra at
12.2$\,\micron$. We have to note that because of the wavelength
selected to scale the spectra is still inside the absorption feature
(though in the more outer part of it), the measured silicate strengths
might be slightly underestimated. In any case, relative comparisons
are not affected by this.
%This method is illustrated in Figure~\ref{f:siabsfit}}.

%When fitting the \textit{local} continuum of the T-ReCS spectra
We found that the
maximum of the \Siabs\ absorption feature was not always centered
at 9.7$\,\micron$ (see \citealt{Sirocky08} and reference therein).
It varies between
9.4 and 10$\,\micron$ and sometimes is even located at shorter
wavelengths down to $\sim\,9\,\micron$ (see, e.g., the spectra
of NGC~3256 in Figure~\ref{f:ngc3256spatpos}).
%This is probably due to the low S/N ratio of the spectra in that wavelength range that makes the fit very uncertain there.
Because of these displacements, we computed the \SSi\
of the T-ReCS spectra using Equation~(\ref{e:siabs}) but
evaluating it at the maximum of the absorption, rather than exactly
at $\lambda_{rest}\,=\,$9.7$\,\micron$.

\begin{figure}%[!h]
\epsscale{1.1}
%\plotone{./figures/ngc3256_compare_trecsspitz.ps}
\plotone{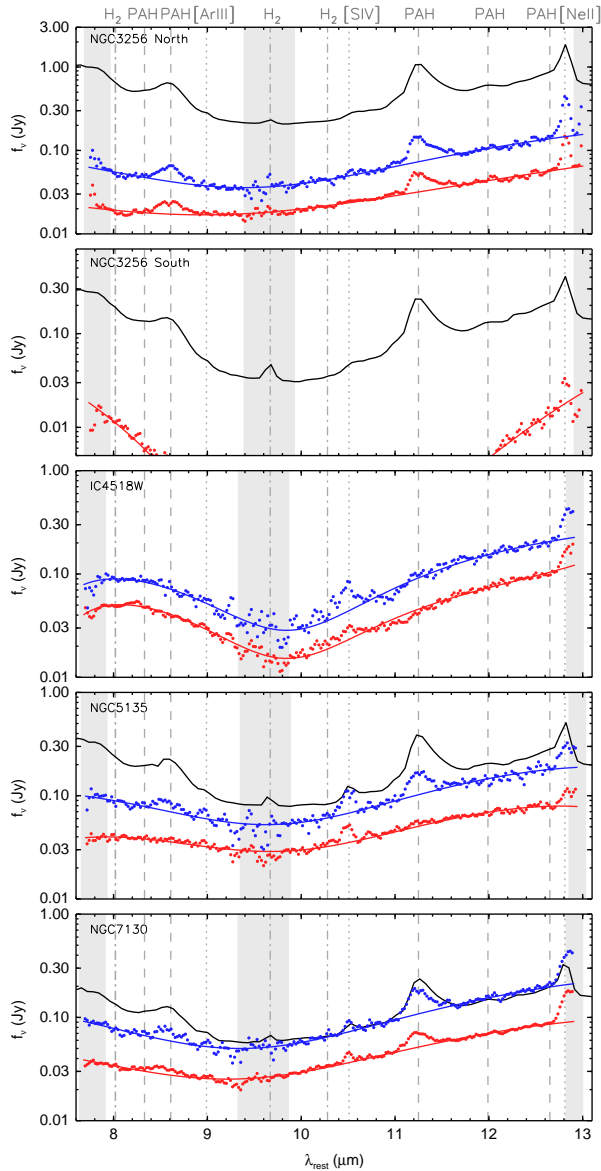}
\vspace{0.25cm}
\caption[Comparison between T-ReCS and IRS spectra]{\footnotesize
  T-ReCS and IRS spectra (where available) of the four LIRG under
  analysis. The nuclear and integrated T-ReCS spectra of the LIRGs
  are shown as red and blue dots, respectively.
  The low-resolution (SL module) IRS spectrum is shown as a black line
  (for IC~4518W an IRS spectrum is not publicly available yet).
  %Comparison between T-ReCS and IRS spectra. The nuclear and
  %integrated T-ReCS spectra of the LIRGs are shown as red and blue
  %dots, respectively.
%% These spectra are the same as in Figures~\ref{f:ngc3256intnucroispecs}, \ref{f:ic4518wintnucroispecs}, \ref{f:ngc5135intnucroispecs}, and \ref{f:ngc7130intnucroispecs}.
%The low-resolution (SL module) IRS spectrum is shown as a black line.
%\textbf{For completeness, we also show the T-ReCS spectra of IC~4518W although
%the IRS spectrum has not been released yet.}
%while the high-resolution (SH module) is shown as a gray line. Both 
The \textit{Spitzer} spectra were extracted with CUBISM from
mapping-mode data (see \citealt{PS09a}, for more details). We
used an aperture of 5.4\arcsec$\,\times\,$5.4\arcsec\ for the
extraction, except for the southern nucleus of NGC~3256 for which
we used a 3.7\arcsec$\,\times\,$3.7\arcsec\ aperture. The labels and
fits are as in
Figure~\ref{f:ngc3256spatpos}.
[\textit{See the electronic edition of the Journal
for a color version of this figure}].}\label{f:trecsspitzerspecs} 
\vspace{0.25cm}
\end{figure}

Since no IRS spectrum is available (to the date) for IC~4518W,
we used a second approach to measure the \SSi.
%The second approach was used for IC~4518W since no IRS spectrum is available for this galaxy. 
In this case we measured the continuum
emission above the \Siabs\ feature using directly the T-ReCS spectra.
Note, however, that the PAHs and in particular the 8.6$\,\micron$
feature, in this galaxy are not very strong (see
Figure~\ref{f:trecsspitzerspecs}).
For each step of the spatial profiles we fitted the spectra between
8.2 and 12.2$\,\micron$ (at the edges of the silicate feature)
with a linear function and interpolated the
fitted continuum to the maximum of the absorption.
We also applied this alternative method to the other
LIRGs to compare these values with those obtained using the 
IRS spectra. We found that both methods yielded similar results, 
as can be seen for  NGC~3256, NGC~5135 and NGC~7130
in Figure~\ref{f:spatprofsiabs}.

Tables~\ref{t:nucfluxes} and \ref{t:intfluxes} give the \Siabs\
strengths measured in the nuclear and integrated spectra of the
galaxies, respectively. In \S\ref{s:siabsfeat} we explore in detail
the spatial profile of the $9.7\,\micron$ silicate feature in each LIRG.

\section{Crucial Differences Between T-ReCS and IRS Spectra}\label{s:spitzdiffs}

We can compare the nuclear and integrated T-ReCS spectra of NGC~3256,
NGC~5135 and NGC~7130 with the \textit{Spitzer} IRS spectra (there is
no \textit{Spitzer} spectrum for IC~4518W) to highlight
the different physical regions and processes probed by the two instruments.
While the \textit{Spitzer} IRS spectra
are representative of the emission  over 
scales of $\sim\,$1\,kpc, our ground-based observations improve this
resolution by almost one order of magnitude. Thus with the T-ReCS data
we can disentangle the emission arising from the nuclear regions of the
galaxies on scales of $\sim\,$100\,pc.

%One of the first things we can notice when comparing the T-ReCS integrated and IRS spectra is the difference in the shape of the continuum emission. While in the $\sim\,11-13\,\micron$ wavelength range the ratio between the flux of both spectra remains constant, when approaching to lower wavelengths, the IRS spectrum show an excess of emission with respect to the integrated T-ReCS spectrum. This difference (seen in all galaxies) may be produced by the way in which the IRS spectra have been extracted and calibrated. The limited spatial resolution of \textit{Spitzer} does not allow to resolve the sub-structure of these LIRGs so that they appear as unresolved sources (nuclei) relying over a more diffuse emission (extended star formation, when present). Thus, the emission cannot be considered point-like nor totally extended. This makes the the extraction and associated flux-calibration of the sources to be somewhat uncertain.

%Keeping in mind these considerations,

\subsection{NGC~3256}

NGC~3256 has
%is a young merger ($\sim\,50\,$Myr; \citealt{Lipari00}) with
two nuclei separated by $\sim\,$5\arcsec\
($\sim\,$1\,kpc) along the north-south
direction. The northern nucleus (see Figure~\ref{f:slits}, orange cross)
has been clearly classified as a star-forming region (\citealt{Lipari00};
\citealt{Lira02}) and is resolved in the T-ReCS MIR image (FWHM of
0.50\arcsec\,$\simeq\,$100\,pc; \citealt{DS08}). The southern nucleus
is heavily obscured (\AV$\,\gtrsim\,12-15\,$mag;
\citealt{Kotilainen96}; \citealt{Lira02}; \citealt{AAH06a}; \citealt{DS08}),
and it is classified as \HII-like.
The T-ReCS (nuclear and integrated) and IRS spectra of the
northern nucleus of NGC~3256 are all very similar
(see Figure~\ref{f:trecsspitzerspecs}). All of them show
prominent 8.6 and \PAHa\ features together with a conspicuous
\NeII\ emission line. Therefore, despite the nucleus is resolved,
the nuclear spectrum is representative of the whole region,
even at kpc scales (see IRS spectrum). 

The IRS spectrum of the southern \HII-like nucleus of NGC~3256 shows
a larger absorption (\SSi\,$\simeq\,-1.4$) than that of the northern
nucleus (\SSi\,$\simeq\,-0.5$; \citealt{AAH09b};
\citealt{PS09a}).
However, our T-ReCS spectrum of the southern nucleus shows that
%the value measured from the IRS spectra is only a lower limit to the true depth of the silicate feature.
the IRS spectrum underestimates the true depth of the silicate feature.
In fact, the T-ReCS spectrum is so absorbed that even the intense
8.6 and \PAHas\
(clearly seen also in the spectrum of \citealt{MH06})
cannot be measured because they are totally extinguished. The \NeII\
emission line is also likely to be somewhat affected by the absorption.
All these features are however clearly detected in the IRS
spectrum. Thus, the T-ReCS spectrum effectively separates the
emission arising from the heavily absorbed
southern nucleus of NGC~3256 from that of
the surrounding regions. In turn, the IRS spectrum
is a combination of both.

%For example, if we suppose that the medium surrounding the
%nucleus has a typical spectrum that is a combination of the spectra
%of the \HII\ and the diffuse regions seen in NGC~5135 (see Appendix),
%the hole created by the \Siabs\ absorption feature in the ``real''
%T-ReCS spectrum of the southern nucleus would be ``filled'' by this
%emission thus giving rise to the spectrum that is observed
%with \textit{Spitzer}.

%The most impressive differences between the T-ReCS and IRS spectra are seen in the southern nucleus (see Figure~\ref{f:trecsspitzerspecs}). While a large absorption is clearly seen in the T-ReCS spectrum, the IRS spectrum shows a shape very similar to that of the northern nucleus with the exception of the \Siabs\ feature, {\bf THIS IS NOT TRUE: which is slightly deeper. CHECK THE ASR PAPER, FIG2, NGC3256-N S-SI=-0.5 AND NGC3256-S S-SI=-1.4, AND ADD REFERENCES TO BOTH MIGUEL AND MY PAPER.} {\bf It is clear that our T-ReCS spectrum shows that the value measured from the IRS spectra is only a lower limit to the true depth of the silicate feature}  {\bf would be infinite as there is  I DONT UNDERSTAND THIS no continuum emission at 9.7$\,\micron$}.

%Therefore, the high spatial resolution afforded by Gemini T-ReCS and,
%in general, by $8-10\,$m class telescopes is crucial to disentangle
%the sources of emission producing the MIR spectra that is seen at
%kpc scales in this type of galaxies.

\subsection{NGC~5135}

The central $5\arcsec \times 5\arcsec$ region of this galaxy is 
known to host a Compton-thick Seyfert 2 nucleus as well as a number of bright
\HII\ regions (\citealt{Levenson04}; \citealt{GD01}; \citealt{Bedregal09}).
Our MIR imaging data showed that the AGN is only contributing
about 25\% of the MIR emission in this region (\citealt{AAH06a}).
The IRS spectrum (Figure~\ref{f:trecsspitzerspecs})
shows intense 8.6 and \PAHa\ features,
indicating the presence of star formation, as well as 
\SIV\ and \NeII\ line emission, and a hint of the \ArIII\
emission line. Outside the ground-based $N$-band spectral range
the IRS spectra display
evidence for the presence of the AGN in the form of high-excitation
emission lines and a strong dust continuum at $6\,\micron$ (see
\citealt{AAH09b}; \citealt{PS09a}).  From the IRS spectrum alone it is not 
clear whether the spectral
%AGN and star-formation
features are being emitted by the same source (see Figure~\ref{f:slits}). 
The T-ReCS nuclear spectrum is almost featureless, except for the 
\SIV\ and the \NeII\ emission lines. The comparison 
between the line fluxes from the
IRS and the T-ReCS spectra shows that
the AGN of NGC~5135 is responsible for a large fraction, 
at least 50\% (this is a lower limit, since we are not
correcting for aperture),
% and slit losses),
of the \SIV\ line emission measured
from the IRS spectrum (see Table~\ref{t:spitzintfluxes}).
The \NeII\ emission on the other hand is coming mostly
from extra-nuclear regions. It is also
worth noting that the T-ReCS nuclear spectrum does not
display PAH emission (although the \PAHa\ may be marginally
present) or \ArIII\ line emission. We can 
conclude from the T-ReCS spectroscopy that the star formation
in this galaxy is mostly circumnuclear ($\gtrsim\,$150\,pc)
rather than nuclear.

In contrast, both the integrated T-ReCS spectrum of NGC~5135
and the kpc-scale IRS spectrum show the same features, including
the PAHs.
% (with the exception of the faint 12.7$\,\micron$ PAH).
The
similarity between both spectra seems to indicate that the PAH emission
does not vary strongly from region to region within the central kpc of
NGC~5135. The T-ReCS 
nuclear spectrum of NGC~5135 has a shallower
\Siabs\ feature than that of the integrated spectrum (see below).
%, implying that it might be suffering a much lower extinction than the \HII\ and diffuse emission regions (but see next section).
%Figure~\ref{f:ngc5135intnucroispecs} and \S\ref{s:siabsfeat}).

\begin{table*}
\begin{center}
\caption{Integrated Fluxes of \textit{Spitzer} IRS Spectra}\label{t:spitzintfluxes}
\begin{tabular}{ccccccccc}
\hline
\hline
Nucleus & %\multicolumn{5}{c}{Feature} \\
%\cline{2-9}
 \PAHb\ & $f$ & [S\,{\sc iv}] & $f$ & \PAHa\ & $f$ & [Ne\,{\sc ii}] & $f$ \\
(1) & (2) & (3) & (4) & (5) & (6) & (7) & (8) & (9)  \\
\hline
NGC~3256 (N) & 342 &   2\% & \dots & \dots & 458 &   3\% & 267 &  5\%  \\
NGC~3256 (S) &  53 & \dots & \dots & \dots & 103 & \dots &  60 &  4\%  \\
NGC~5135     &  66 & \dots &    11 &  47\% & 172 & \dots &  67 &  8\%  \\
NGC~7130     &  37 &  15\% &   5.7 &  44\% &  87 &  13\% &  41 & 25\%  \\
\hline
\hline
\end{tabular}
\end{center}
\footnotesize{Note.-- (1) Name. (2), (4), (6), and (8) Fluxes of the features in units of $\times$\,10$^{-14}$\,erg\,s$^{-1}$\,cm$^{-2}$. (3), (5), (7) and (9) Percentage of flux contained in the nuclear T-ReCS spectra. Note that the T-ReCS spectra have not been corrected for aperture, so these fractions are lower limits.\\
The values were measured from IRS spectra (SL module) and extracted with CUBISM from mapping-mode data and using an extraction aperture of 6.8\arcsec$\,\times\,$6.8\arcsec\ centered at the nuclei of the galaxies, except for the southern nucleus of NGC~3256 that was extracted with an aperture of 4.5\arcsec$\,\times\,$4.5\arcsec\ to avoid overlapping with the northern nucleus.}
\vspace{0.5cm}
\end{table*}

\subsection{NGC~7130}

NGC~7130 also hosts a Compton-thick Seyfert 2 nucleus, as well
as star formation within the central $\sim\,$150\,pc
(\citealt{Levenson05}; \citealt{GD98}). The IRS spectrum
%of NGC~7130 is very similar to that of NGC~5135, with 
shows both AGN and star formation features (\citealt{PS09a}).
However, unlike NGC~5135, our T-ReCS
imaging data of NGC~7130 show a resolved nucleus
(FWHM $\sim\,$0.45\arcsec) suggesting that
in this region the star formation and the AGN emissions are
still mixed (\citealt{AAH06a}). Indeed,
Figure~\ref{f:trecsspitzerspecs} shows that the T-ReCS
nuclear spectrum of NGC~7130 displays the \SIV\ and \NeII\ emission lines
as well as the 8.6 and \PAHa\ features. The presence of PAH
features in the nuclear spectrum clearly
indicates that there is  star formation in the very nuclear 
region ($\lesssim\,$100\,pc) of NGC~7130, as well as in regions surrounding it
(at distances of several hundreds of pc). Nevertheless, at least half of the \SIV\ 
emission measured in the \textit{Spitzer} spectrum is contained within
the inner 0.36\arcsec$\,\times\,$0.72\arcsec\ probed by the T-ReCS
nuclear spectroscopy (see Table~\ref{t:spitzintfluxes}).
In contrast, most of the \NeII\ emission is steming
from regions outside the central 
0.36\arcsec$\,\times\,$0.72\arcsec\ region of this galaxy.

\section{The Strength of the 9.7$\,\micron$ Silicate Feature in Star-Forming Regions and AGNs}\label{s:siabsfeat}

\begin{figure}%[!h]
\epsscale{1.1}
%\plotone{./figures/ngc3256_ap4pix_fix2_cal_SpitzSiabs_spat_dist_panel.ps}
\plotone{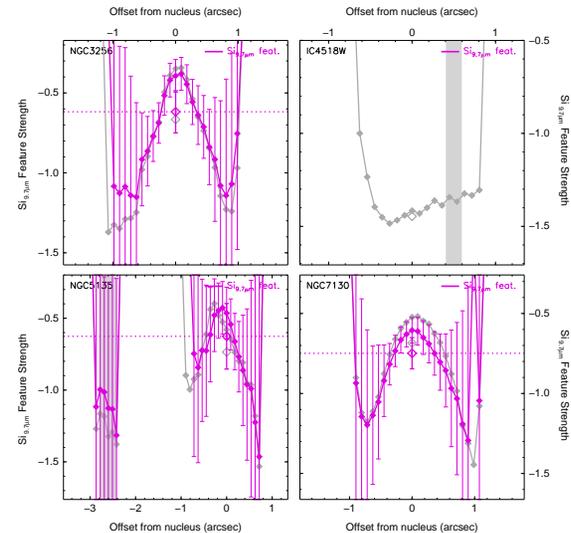}
\vspace{0.25cm}
\caption{\footnotesize Spatial profiles of the \Siabs\ feature strength
  for each LIRG (pink profiles; see \S\ref{ss:siabsstren} for details
  on the calculation). Lower values of the strength imply higher absorptions,
  and viceversa. The open diamond symbol is the \Siabs\ strength as
  measured from the integrated T-ReCS spectrum of the galaxy using
  its IRS spectrum to calculate the continuum emission outside the feature.
  The pink dashed line also mark this value.
  The gray profiles were derived using the alternative (see
  \S\ref{ss:siabsstren}). The results from both
  methods agree within the uncertainties. The shaded regions mark the
  positions of the regions of interest (see Figure~\ref{f:slits},
  orange dots).
[\textit{See the electronic edition of the Journal
for a color version of this figure}].
}\label{f:spatprofsiabs}.
\vspace{0.25cm}
\end{figure}

The spatially-resolved information afforded by the 
T-ReCS spectroscopy allows us to study in detail the
behavior of the $9.7\,\micron$ silicate feature in the 
nuclear and circumnuclear regions of our LIRGs. The most
intriguing result here is that the \SSi, in general, is shallower
towards the nuclei of the LIRGs than in the surrounding regions 
(see Figure~\ref{f:spatprofsiabs}). We measure
\Siabs\ strength values for the innermost regions of the LIRG
nuclei in the range of $-$0.5 (NGC~3256) to $-$0.7 (NGC~7130), while
the extra-nuclear regions show values ranging from $-$1 to up to $-$1.5.
The \HII\ region in NGC~5135 also shows a mean value of $\simeq\,-1.2$.
The only galaxy that presents a different behavior is IC~4518W, which
shows an almost constant value of the \SSi. This is not surprising
however, as the AGN totally dominates the MIR nuclear emission
of this LIRG and there is no hint for star formation in the
innermost regions of the galaxy. Interestingly, the \Siabs\ strength in
the high-excitation line region of IC~4518W
(see Appendix~\ref{s:ic4518wsp}) is
slightly lower than but comparable to that of the nucleus.

%%%%%%Using \textit{Spitzer} data (an with the caveat in mind about the very different spatiall resolutions achieved by \textit{Spitzer} and T-ReCS), \cite{Hao07} have found that a value of $\simeq -0.5$ is typical of Sy2 nuclei, which is in agreement with the values we have measured in the integrated spectra of NGC~5135 and NGC~7130 (see Figure~\ref{f:spatprofsiabs}). However, this is also in agreement for the nucleus of NGC~3256 which is starburst-like.

%%%%%%Interestingly, on the other hand, the mean value for integrated spectra of ULIRGs is $\simeq -1.5$ (from the sample of \citealt{Hao07}), which is more closer to the values found for the outer, star-forming regions of our LIRGs than to those measured in their nuclei where the AGN domains the emission (in the case of IC~4518W and NGC~5135).
%%%%%%%Thus, our results would indicate that low values of the \Siabs\ strength are typical of star-forming regions. Furthermore, this would suggest that the ULIRGs from the Hao et. al. sample would be dominated by star formation.
%%%%%%%(whose existence is statistically more likely than in LIRGs; \citealt{Veilleux95} and more recent REFERENCES).
%%%%%%%However, this does not mean that such star formation is young but, otherwise, is the one traced by the extended PAH emission, i.e., the one associated to the diffuse medium (reach in PAHs; see the spatial profiles of our LIRGs, e.g.,Figs.~\ref{f:ngc3256spatproffluxes} or \ref{f:ngc5135spatproffluxes}).

%One possible explanation for the behavior of the \SSi\ in the rest of the LIRGs could be
One may think that intense
PAH emission could be ``artificially'' increasing
the continuum emission at the wings of the \Siabs\
feature\footnote{Note that the 7.6 and 8.6$\,\micron$ PAH
features are on the blue side of the T-ReCS spectrum,
and the 11.3 and 12.7$\,\micron$ PAH complexes on the red side.}.
%This would make the silicate feature appear deeper in the star-forming regions surrounding the LIRG nuclei.
We used, however, the \textit{Spitzer} IRS spectrum of the galaxies for
estimating the real continuum at two wavelengths
% (5.5 and 13.2$\,\micron$)
where the PAHs do not dominate the emission. In addition, the two
methods used to calculate the \SSi\ (see \S\ref{ss:siabsstren} and
Figure~\ref{f:spatprofsiabs}) yielded similar values, indicating that the
tendencies seen in the spatial profiles of the \Siabs\ strength in our
LIRGs are real.

The nuclei of all these galaxies heat the nearby dust effectively.
Because the same dust produces both the observed continuum and the
silicate feature, radiative transfer effects determine the resulting
feature strength. If AGNs are indeed located within clumpy environments,
they can produce only shallow silicate absorptions (\citealt{Nenkova08b}).
A clumpy geometry allows direct views of some hot, directly-illuminated
cloud surfaces whose emission fills in the absorption created by cold
clouds. That is, even when the clumpy dust distribution is extended,
the temperature gradient along a clumpy torus is never high due to
the probability of viewing hot, directly-illuminated clouds at all
distances (\citealt{Nenkova08a}). Therefore, the silicate feature
remains always weak (\SSi$\,\gtrsim\,- 1$; \citealt{Nenkova08b}). A deeper
absorption feature requires a steeper temperature gradient, as radiative
transfer computations demonstrate (\citealt{Levenson07}). An optically and
geometrically thick smooth dusty medium can provide this temperature
gradient.
%, in contrast to the shallower gradient of a clumpy medium.}

In the case of NGC~3256, where the nuclear starburst is compact,
the immediate dusty surroundings remain hot. Without a strong temperature
gradient, the nuclear silicate absorption is weak. The dust farther
($\gtrsim\,100\,$pc) from the nucleus is somewhat cooler, without
producing such a hot continuum. More importantly, the dust distribution
in the off-nuclear extended star-forming regions of NGC~3256,
NGC~5135, and NGC~7130 can be optically and geometrically thick,
therefore showing deeper absorptions (\citealt{Levenson07}).
Thus, we interpret these cases as the
transition from either the clumpy or compact environment of an AGN or
a nuclear starburst to the more smooth, extended dust distributions
typical of circumnuclear star formation. The AGN in NGC~5506 offers
another similar example, with a weaker silicate absorption in the
nucleus than in the surrounding extended regions (\citealt{Roche07}).

On the other hand, the silicate strength remains roughly constant
and large across the
central region of IC~4518W. This is not characteristic
of a clumpy environment but instead suggests foreground extinction
by cool dust, possibly by a dust lane in this highly inclined galaxy.
%In any case, both explanations are in agreement with the fact that the EWs of the $N$-band features (PAHs and sometimes the \NeII\ and \SIV\ emission lines) have always a minimum in the nuclei of the galaxies, probably due to the dilution caused by this ``extra'' continuum of hot dust emission.

In general however, using a simple foreground dust screen
geometry to estimate the optical depth across the nuclei of these
LIRGs is not correct because of the existence of multiple dust
components at different temperatures (as explained above).
%Therefore, using a foreground dust screen geometry and an extinction law to estimate the extinction may not be correct for these LIRGs because of the existence of multiple components of  dust at different temperatures.
In fact and due to this, it is because if we used a simple
foreground dust screen model for calculating the
extinction (i.e., the apparent \tauV, see \S\ref{ss:siabsstren})
we would obtain lower extinctions for the LIRG nuclei than
for the surrounding regions (except in IC4518W, where the \SSi\
is nearly constant). That is, we now know that the apparent
\tauSi\ ($= -$\SSi) values found for the nuclei are only lower limits to the
real ones because of the re-emission of the hot dust emission component
(associated either to a clumpy geometry or a compact
environment) near $\sim\,$10$\,\micron$ that is not accounted for
by the screen model.
% when compared to the surrounding regions.

%%%%%%%This is partially in agreement with what has been found by other authors. For example, from \textit{Spitzer} data \cite{Farrah07} suggest that ULIRGs with \SSi\,$\gtrsim$--0.8 contain an IR-luminous AGN. In addition, they also suggest that those ULIRGs with \SSi\,$\lesssim$--2.4 harbours an AGN too. In this case we cannot verify it as we do not have any region with such a deep \SSi.

The \SSi\ measured over the integrated spectra of
galaxies (Figure~\ref{f:spatprofsiabs}, open diamonds)
indicate that the absorption is never very deep.
Moreover, all LIRGs show almost the same strength (from
$\simeq\,-0.6$ to $-0.8$, except for IC~4518W which presents
a deeper \SSi\ of $-$1.45), closer
to those values found in their nuclei rather than to those measured in
the surrounding regions. Indeed, this is an expected result if we realize
that these values are ``luminosity-weighted'' and that the flux included
within our slits arising from these surrounding regions represents a low
fraction of the MIR luminosity when compared with which is arising
from the nucleus itself. The strengths measured in the T-ReCS integrated
spectra are in agreement with those obtained from the IRS spectroscopy.

% (or, in other words, our LIRG nuclei generate the bulk of the MIR luminosity of the galaxies).

%%%%%%%Taken this into account, it seems that it is the star formation surrounding the LIRG AGN that that is ``artificially'' increasing the \Siabs\ strength when it is measured in spectra where both physical processes cannot be effectively isolated, as is the case for \textit{Spitzer} spectra of LIRGs and ULIRGs which commonly include large amounts of emission not directly associated with the AGN.

%%%%%%%Another important result inferred from Figure~\ref{f:spatprofsiabs} is that the \Siabs\ strength, and therefore the deepen of the absorption suffered by the hot dust continuum, is not correlated with the absorption due to cold dust as measured from NIR colors. We can see that the shape of the spatial profiles of both measurements do not coincide even in relative magnitudes. For a cold-``slab'' (cold-screen) dust configuration, that is, with the slab being far from the heating source and where the emission of the dust continuum is decoupled from the absorption, the \Siabs\ strength is, by definition, equal to the 9.7$\,\micron$ optical depth (\tauSi), which in turn is proportional to \tauV. Since \AV\,=\,1.068\,\tauV, the \Siabs\ strength and the extinction calculated from NIR colors are in the same relative scale. In fact, this is very interesting since \textbf{the cold dust causing the extinction in the NIR is not the same cold dust that is creating the \Siabs\ feature through the absorption of the hot dust continuum.}

%\section{Discussion}\label{s:discussion}

\section{The [NeII] emission}\label{s:neii}

%A number of works have shown that  the \NeII\ emission line is related to star formation processes
The \NeII\ emission line luminosity scales with SFR in galaxies
(\citealt{Roche91};  \citealt{Ho07}). \cite{Ho07} found a good empirical
correlation between the sum of the \NeII\ and the \NeIII\ emission lines
and the \Bralpha\ emission line for \HII\ regions in the Galaxy,
the Small and Large Magellanic Clouds, and M33. They also calibrated
theoretically the SFR in terms of the \NeIIno+\NeIIIno\ luminosity and
found a good agreement between observations and theory.

\begin{figure}
\epsscale{1.}
%\plotone{./figures/all_ap4pix_fix2_cal_0_step4pix_NeII_flux-__vs_paa_flux-__extc.ps}\vspace{0.25cm}
\plotone{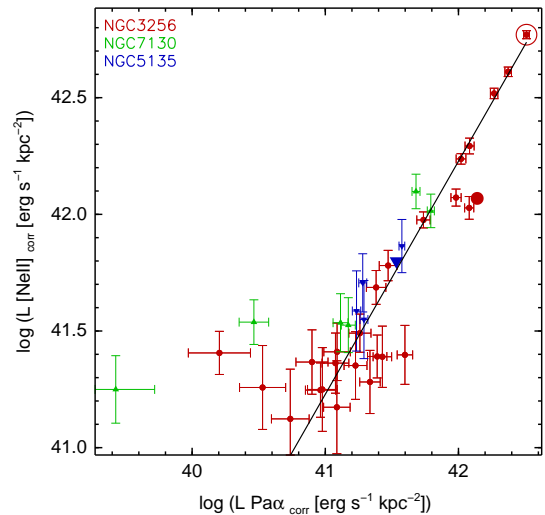}\vspace{0.25cm}
\caption{\footnotesize \NeII\ LSD vs. \Paalpha\ LSD (corrected for extinction) for the star-forming regions in our LIRGs. Each datapoint represents a spatial position along the slit in steps of 0.36\arcsec.
%(see \S\ref{ss:spatprof}, e.g., Figures~\ref{f:ngc3256spatproffluxes}--\ref{f:ngc3256spatprofews}).
The nucleus of NGC~3256 is marked with a big open symbol. The Seyfert nuclei of the remaining LIRGs have been excluded from the study. The regions of interest (when the data are available, see Figure~\ref{f:slits}) are marked with big filled symbols.
[\textit{See the electronic edition of the Journal
for a color version of this figure}].}\label{f:neiivspaaflux}
\vspace{0.5cm}
\end{figure}

We support these conclusions showing here that \NeII\ scales with \Paalpha,
a direct tracer of the youngest (ionizing) stellar populations, on the physical
scales probed by our data ($\sim\,$100\,pc)\footnote{We excluded
the regions of IC~4518W from this section, since we do not find any
clear evidence for star formation, as well as the Seyfert nuclei of
NGC~5135 and NGC~7130.}.
%In this section we compare the behavior of the \NeII\ emission  in star-forming regions in our sample of LIRGs
%with a direct tracer of the youngest (ionizing) stellar populations like \Paalpha,
%To make this comparison
Here and in the next section, we chose to use
units of luminosity surface density (LSDs) to represent the luminosities
of the regions because our galaxies are at different distances and were
observed with different slit widths.
%If we used luminosities we would be probing different physical areas depending on the galaxy.

Figure~\ref{f:neiivspaaflux} compares the \NeII\ and \Paalpha\ LSDs
%(corrected for extinction)
for the star-forming regions in our sample of LIRGs. Each point in these
figures represents a spatial position along the slit in 0.36\arcsec\ bins.
%\footnote{Although the spectra were extracted in 1 pixel (0.09\arcsec) step, here we show (and use) only the datapoints in steps of 4\,pixel (0.36\arcsec, the aperture size used for the extraction, which is also approximately the spatial resolution of our observations) to avoid the redundancy of the data (i.e., to overlap regions) when the fits of the trends are obtained.}.
We estimated
the extinction to each star-forming regions as explained by \cite{DS08}
and corrected the \NeII\ and \Paalpha\ LSDs for it using the
\cite{Calzetti00} extinction law (see \S\ref{ss:imagadddata}).
There is a tight correlation between the \NeII\ and \Paalpha\
LSDs (Figure~\ref{f:neiivspaaflux}).
Furthermore, this correlation is common to the regions of all LIRGs
and spans about 2 orders of magnitude. We fitted the
datapoints above log(L\,\Paalpha\,[\lsd])\,=\,40.5.
The 3 outliers with log(L\,\Paalpha\,[\lsd]) below 40.5 are
from regions just above the detection limit of the \NeII\ line
and therefore the measurements are very uncertain.
We found:

\begin{equation}\label{e:neiivspaa}
%{\rm log(} [NeII]{\rm )} = (0.276\pm0.904) + (1.000\pm0.021) {\rm log(} Pa\alpha {\rm )}
{\rm log(} [NeII]{\rm )} = (0.28\pm0.90) + (1.00\pm0.02) {\rm log(} Pa\alpha {\rm )}
\end{equation}

\noindent
where the \NeII\ and \Paalpha\ lines are in LSD units (\lsd).
The scatter around the trend is $\sim\,0.2\,$dex. The slope of the fit
is 1.00$\pm$0.02 and therefore the correlation is linear,
suggesting that the \NeII\ emission is effectively tracing the
youngest ionizing stellar populations in LIRGs.

This correlation is similar to that
found by \cite{Ho07} but without including the \NeIII\ line contribution.
The fact that the \NeII\ luminosity is directly proportional
to the \Paalpha\ luminosity with a unity slope is
%well understood in terms of
in agreement with
the relatively low
\NeIII/\NeII\ nuclear and integrated ratios observed in most
of our LIRGs from our kpc-scale \textit{Spitzer} IRS observations
(\citealt{AAH09b}; \citealt{PS09a}).
This may be explained in terms of the 
ages probed by our regions: $\simeq\,5.6-6.7\,$Myr.
Indeed, photo-ionization models predict
for starbursts up to $\sim\,$6\,Myr that: (a) the neon atoms are no
longer double ionized (\NeIIIion); and (b) the
\NeII/\Paalpha\ ratio does not vary significantly (\citealt{Dopita06};
\citealt{Groves08}; \citealt{Thornley00}; \citealt{MH05};
\citealt{Rigby04}; \citealt{Snijders07}).
This is in agreement with our star-forming regions,
%Figure~\ref{f:neiitopaafluxvspaaew}, which shows that most regions in our LIRGs
which present a rather constant \NeII/\Paalpha\ ratio
as a function of their age (log(\NeII/\Paalpha)$\,\simeq\,0.3\pm0.2$).

Therefore, the \NeII\ emission line alone can be used as a good
tracer of the SFR,
provided that the regions are not dominated by extremely young stellar
populations ($<<\,5-6$Myr) and/or have low metallicity, which
is in agreement with previous results (\citealt{Roche91}; \citealt{Ho07}).
For example, this is the case for the observations obtained with
\textit{Spitzer} of local LIRGs, whose integrated properties are
very similar to the physical conditions in which the \NeII\ emission line
can be used as a SFR indicator (\citealt{PS09a}). Moreover, if the
population of LIRGs found by \textit{Spitzer} at cosmological distances
(\citealt{LeFloch05}; \citealt{PG05}) turn out to be the high-redshift
counterparts of those detected in the
local Universe (not only in luminosity but also in their physical
properties), this relation could be also applied to them, allowing
to calculate the SFR of this important population of LIRGs in a simple
manner.

\section{The PAH emission}\label{s:pah}

The good correlation between the integrated PAH emission and the IR
luminosity of high metallicity starbursts and ULIRGs seems to indicate
that the PAH emission is tracing the star formation processes at least
on large scales (\citealt{Brandl06}; \citealt{Farrah07};
\citealt{Smith07}; \citealt{Weedman08}).
Indeed, \cite{Farrah07}
derived from integrated MIR spectra of ULIRGs a SFR calibration based
on the luminosity of the 6.2 plus the \PAHa\ features through the
correlation of these with the luminosity of the \NeII\ and the \NeIII\
emission lines.

\begin{figure}
\epsscale{1.}
%\plotone{./figures/all_ap4pix_fix2_cal_0_step4pix_11.3um_PAH_flux-__vs_paa_flux-__extc.ps}\vspace{0.25cm}
\plotone{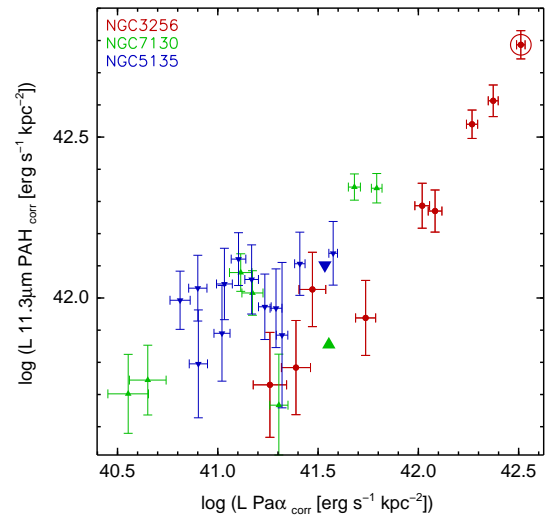}\vspace{0.25cm}
\caption{\footnotesize \PAHa\ LSD vs. \Paalpha\ LSD (corrected for extinction) for the star-forming regions in our LIRGs. Each datapoint represents a spatial position along the slit in steps of 0.36\arcsec.
%(see \S\ref{ss:spatprof}, e.g., Figures~\ref{f:ngc3256spatproffluxes}--\ref{f:ngc3256spatprofews}).
The nucleus of NGC~3256 is marked with a big open symbol. The Seyfert nuclei of the remaining LIRGs have been excluded from the study. The regions of interest (when the data are available, see Figure~\ref{f:slits}) are marked with big filled symbols.
[\textit{See the electronic edition of the Journal
for a color version of this figure}].}\label{f:113pahvspaaflux}
\vspace{0.5cm}
\end{figure}

As can be seen in Figure~\ref{f:113pahvspaaflux},
the \PAHa\ and \Paalpha\ LSDs seem to be broadly correlated
on the scales of a few hundred parsecs probed for our LIRG
star-forming regions. However, this trend is not as tight as the
\NeII\ vs. \Paalpha\ relation and the slope is considerably lower
(0.48$\pm$0.03). Similarly, \cite{Peeters04}
found that Galactic \HII\ regions do not show a very tight
correlation between their $6.2\,\micron$ PAH luminosities
and their number of ionizing photons. 
Moreover, Figure~\ref{f:113pahvspaaflux} shows that, individually,
the star-forming regions of each LIRG appear to show their own trend.
For example, the star-forming regions of NGC~3256 show quite
low \PAHa\ LSDs for a given \Paalpha\ LSD when compared
with those of NGC~5135.
%(see Figure~\ref{f:neiiand113pahvspaaflux}, right).
In addition, the regions of NGC~5135, unlike
those of NGC~3256 and NGC~7130,
show a very weak dependence of the \PAHa\ on the \Paalpha\ LSD.
Therefore Figure~\ref{f:113pahvspaaflux} suggests that the PAH emission
in the star-forming regions of LIRGs differs from galaxy to galaxy not
only in its total luminosity (see also \citealt{Smith07}) but
also in relation with the ionizing stellar populations.

The metallicity of the star-forming regions, and in particular
low-metallicity environments, is known to have a strong
impact on the observed PAH emission (e.g., \citealt{Madden06};
\citealt{Engelbracht06}; \citealt{Wu06}; \citealt{Calzetti07}).
Metallicity is not likely to have a strong effect on the observed
PAH properties of the LIRG star-forming regions
since all the galaxies have similar oxygen abundances
(see Table~\ref{t:sample}). For a fixed metallicity, the hardness and
the intensity of the radiation field are also believed to play an
important role in the PAH energetics (\citealt{Wu06}; \citealt{Gordon08}).
In the next sections we investigate the effects of both on
the observed PAH emission of star-forming regions in LIRGs.

\subsection{The Age of the Ionizing Stellar Populations}\label{ss:ageeffect}

At a given metallicity, the hardness of the radiation field is a
strong function of the age of the ionizing populations. The radiation
field decreases with age as stars evolve off the main
sequence, but it also shows a temporary increase when
the most massive stars enter the Wolf-Rayet phase (see e.g.,
\citealt{Snijders07}). In this section we use the age of the ionizing
stellar populations, as probed by the Pa$\alpha$ EW, as an approximate
proxy for the hardness of the radiation field.

\begin{figure}
\epsscale{1.}
%\plotone{./figures/all_ap4pix_fix2_cal_0_step4pix_11.3um_PAH_flux-paa_flux_vs_paa_ew-__extc.ps}
\plotone{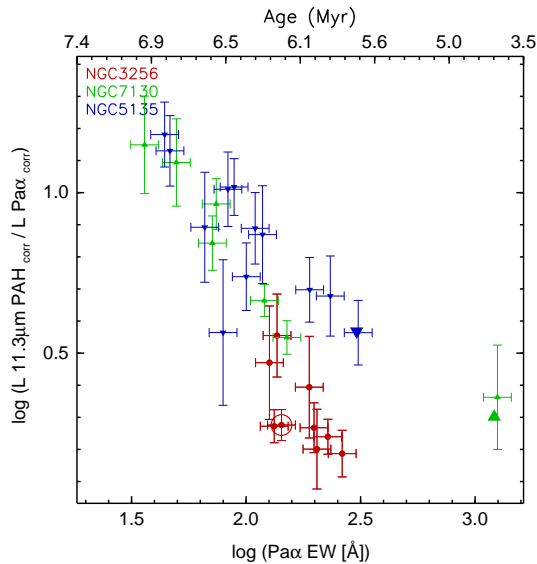}
\vspace{0.25cm}
\caption{\footnotesize \PAHa/\Paalpha\ ratio (corrected for extinction) vs. \Paalpha\ EW for the star-forming regions in our LIRGs. Each datapoint is as in Figure~\ref{f:neiivspaaflux}.
[\textit{See the electronic edition of the Journal
for a color version of this figure}].}\label{f:113pahtopaafluxvspaaew}
\vspace{0.25cm}
\end{figure}

Despite the relatively narrow range of
ages ($\sim 3.5-7\,$Myr) of our star-forming regions,
Figure~\ref{f:113pahtopaafluxvspaaew} clearly shows
that the \PAHa/\Paalpha\ ratio depends on the age of the
ionizing stellar populations in the sense that
\textit{more evolved} regions show higher \PAHa/\Paalpha\ ratios.
This tendency is mainly (but not only) caused by the decreasing
of the \Paalpha\ emission during this period (as seen in \citealt{DS08}).
In addition we can also see that the trend is common to all star-forming
regions in our sample of LIRGs. This is in agreement with our results
in \cite{DS08} that showed that the 8$\,\micron$ luminosity
(\PAHb\ + continuum emission) of LIRG \HII\ regions is correlated
with their age.
This may suggest that: PAHs may trace B stars (i.e., \textit{recent})
rather than O stars (i.e., \textit{current} star formation;
see e.g., \citealt{Peeters04}), and/or that at the earliest
stages of the star formation, PAHs may be destroyed by the hard
radiation fields of the youngest massive stars. However, the later
explanation is unlikely since in our sample of LIRGs we do not see
evidence for radiation fields hard enough (\citealt{PS09a})
to explain the low \PAHa/\Paalpha\ ratios observed in the youngest
regions in terms of the destruction of the PAH carriers (at least
at the spatial scales probed in this study).

%Disentangling the different effects (age of the stellar populations, hardness and intensity of the radiation field, mass of the star-forming regions) that affect the \PAHa\ luminosity is not straightforward.
We conclude then, as we did
for the 8$\,\micron$ monochromatic (PAH + continuum) luminosity
(\citealt{DS08}), that the \PAHa\ luminosity, unlike the \NeII\
emission line, is not a good tracer of the current star formation
when measured on scales of a few hundreds of pc.
% from galaxy to galaxy and maybe from \HII\ region to \HII\ region.
Thus, in order to use the PAH emission as a measure of the global SFR
in LIRGs (or in galaxies in general), the calibration must be done
using integrated emission of galaxies, where all these local physical
properties are averaged out (see also \citealt{Wu05}; \citealt{AAH06b};
\citealt{Brandl06}).

\subsection{The Density of the Radiation Field}\label{ss:denseffect}

\begin{figure}
\epsscale{1.}
%\plotone{./figures/all_ap4pix_fix2_cal_0_step4pix_11.3um_PAH_flux-NeII_flux_vs_NeII_flux-__extc.ps}
\plotone{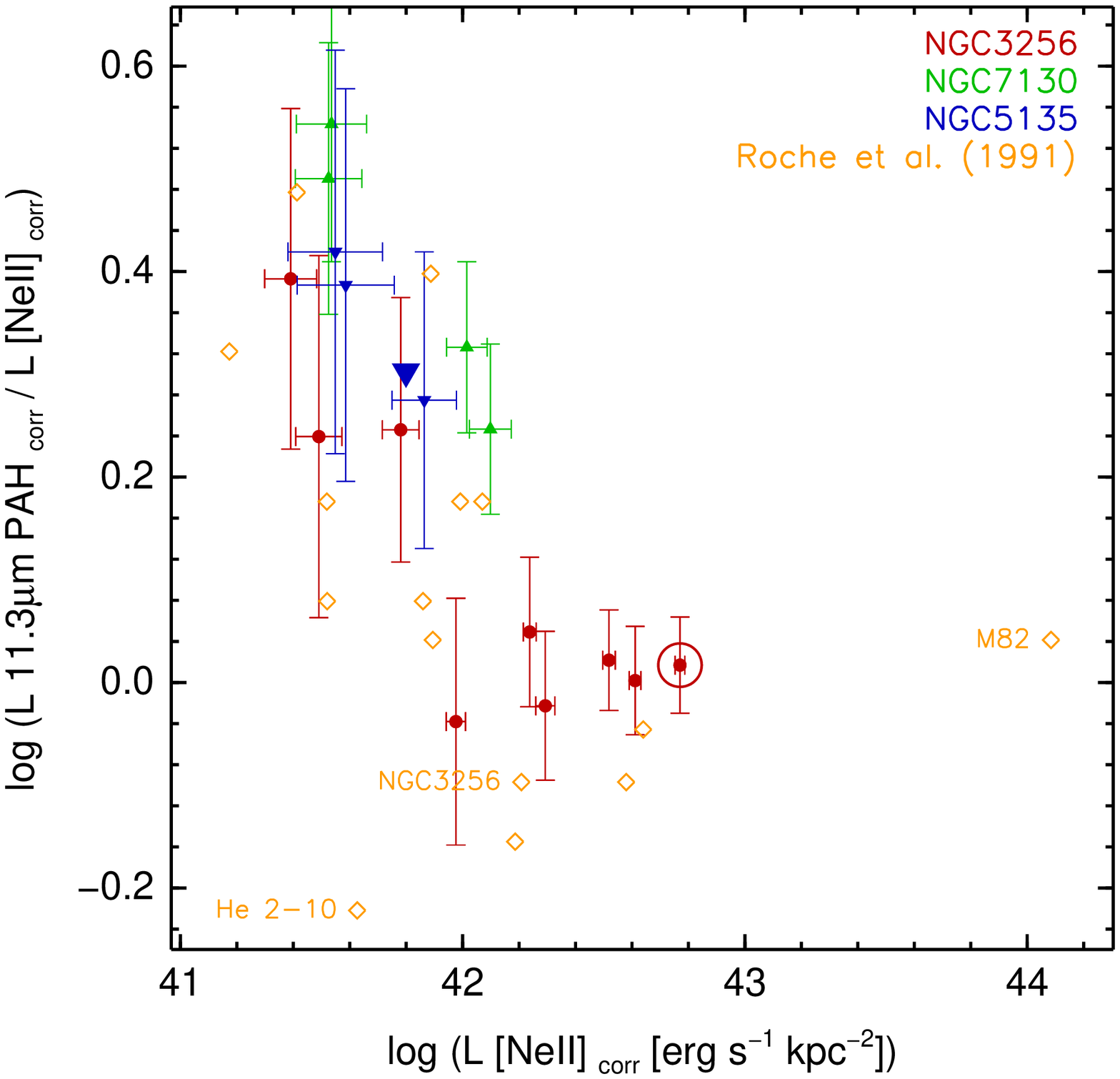}
\vspace{0.25cm}
\caption{\footnotesize \PAHa/\NeII\ ratio vs. \NeII\ LSD (corrected for extinction) for the star-forming regions in our LIRGs. Each datapoint is as in Figure~\ref{f:neiivspaaflux}. The orange open diamonds are the nuclei of star-forming galaxies in the sample of \cite{Roche91} for which measurements of the \NeII\ line and the \PAHa\ (not corrected for extinction) are available. As explained in the text, the \PAHa/\NeII\ ratio increases with the decreasing of the hardness of the radiation field (decreasing of the \NeII\ LSD) due to the ageing of the star-forming regions. On the other hand, the intensity of the radiation field (in the sense of the mass density) of the starburst does not affect the \PAHa/\NeII\ ratio but modifies the \NeII\ LSD and causes the spreading of the data along the x-axis (see the nuclear regions of NGC~3256).
%The arrows are indicating the trend of the data with increasing age and mass of the star-forming regions accordingly with the deductions made in the text.
[\textit{See the electronic edition of the Journal
for a color version of this figure}].}\label{f:113pahtoneiivsneiiflux}
\vspace{0.25cm}
\end{figure}

In Figure~\ref{f:113pahtopaafluxvspaaew} we showed the evolution of the
\PAHa/\Paalpha\ ratio as a function of the hardness (age) of the starburst
\textit{only}. In this section we study the variation of the
\PAHa/\NeII\ ratio (equivalent to the \PAHa/\Paalpha\ ratio) as a
function of the \NeII\ LSD, i.e., as a function of the density
of the radiation field, which takes into account not only the
hardness (age) but also the intensity (mass density) of the starburst.
%Since the the \NeII\ vs. \Paalpha\ correlation has a unity slope, we can use the \NeII\ LSD as a measure of the density of the radiation field, \textbf{which depends on the hardness (age) and the intensity (mass density) of the starburst.}
%As explained in \S\ref{s:neii}, using LSD units is more appropriate for the different star-forming regions in our sample as we are sampling different physical regions.
Figure~\ref{f:113pahtoneiivsneiiflux} shows that
star-forming regions in LIRGs with high
values of the \NeII\ LSD tend to display lower \PAHa/\NeII\ ratios.
This tendency is further supported by the observations of the
\HII-like nuclei in the sample of \cite{Roche91}\footnote{The data of
\cite{Roche91} are not corrected for extinction.}.
%Therefore, since the star-formation in our regions is fully accounted by the \NeII\ luminosity (see previous section), the fact that we are seeing this trend suggests that there is either a deficit in the \PAHa\ luminosity in regions with higher LSDs \textbf{or an excess in regions with lower LSDs.
In fact, the trend seen in Figure~\ref{f:113pahtoneiivsneiiflux}
is connected to the age effect we discussed in \S\ref{ss:ageeffect}
since, as it is said above, the density of the radiation field
depends on the hardness (age) of the starburst.
%can be generated by a more massive star-forming region \textit{and/or} also by a younger starburst.}

%Indeed, there is a correlation between the \NeII\ LSD and the \Paalpha\ EW except for the nuclear ($\lesssim\,$1\arcsec\,$\simeq\,$200\,pc) regions of NGC~3256. For these regions the \NeII\ LSD increases without increasing the \Paalpha\ EW (see also Figure~\ref{f:113pahtopaafluxvspaaew}) suggesting that it is the mass of the starburst that is larger. Interestingly, these star-forming regions are those in  Figure~\ref{f:113pahtoneiivsneiiflux} presenting a rather constant value of the \PAHa/\NeII\ ratio.
On the other hand, despite the star-forming regions in the inner
$\lesssim\,$1\arcsec\,$\simeq\,$200\,pc of NGC~3256 have larger
\NeII\ LSDs than those in NGC~5135 and NGC~7130 (see
Figure~\ref{f:113pahtoneiivsneiiflux}), they do not show
significantly lower \PAHa/\NeII\ ratios but instead rather
constant values. Given that the age range of these regions of
NGC~3256 is quite narrow (see Figure~\ref{f:113pahtopaafluxvspaaew}),
we can attribute the enhancement of the \NeII\ LSD and
the constantness of the \PAHa/\NeII\ ratio to a mass escalation
of the starburst with the \NeII\ LSD
(through its dependecy on the mass density), which increases
as we move towards the nucleus. Therefore, for the case
of these regions, we interpret LSD units as a measure of the intensity
(not hardness) of the radiation field in the sense of luminosity
(or, equivalently, number of ionizing photons) per unit of physical
area, that is, as a measure of the mass density of the starburst.
In fact, there exist a correlation
between the \NeII\ LSD and the \Paalpha\ EW for the star-forming
regions in our LIRGs that holds except for the nuclear regions
of NGC~3256. In turn these present an enhanced \NeII\ LSD without
increasing their \Paalpha\ EW. This is in agreement
with our interpretation and supports the idea of that it is the
mass density of the starburst which is larger towards the nuclei
of NGC~3256, and that the age does not play any role in this case.
This behavior is not unexpected since a more massive (denser)
starburst will have a higher \NeII\ LSD but the same \PAHa/\NeII\
ratio than a less massive one because the hardness of the radiation
field is not modified (only its intensity). Moreover, in general,
the scatter of the data in the horizontal direction in
Figure~\ref{f:113pahtoneiivsneiiflux} could be explained
as an effect of the mass density of the starburst (even for
those regions that presents a dependence of the \PAHa/\NeII\ ratio
with the age). Unfortunately the
nuclear regions of both NGC~5135 and NGC~7130 host an AGN,
and thus cannot be used to explore the effects of higher
density radiation fields in star-forming regions in LIRGs.
In any case, our data suggest that it is only the age
(hardness), but not the mass (intensity) of the starburst that
modifies the \PAHa/\NeII\ ratio as a function of the \NeII\ LSD
(or equivalently \Paalpha\ LSD; see
Figure~\ref{f:113pahtoneiivsneiiflux}).

\begin{figure}
\epsscale{1.}
%\plotone{./figures/all_ap4pix_fix2_cal_0_step4pix_11.3um_PAH_ew-__vs_paa_flux-__extc.ps}
\plotone{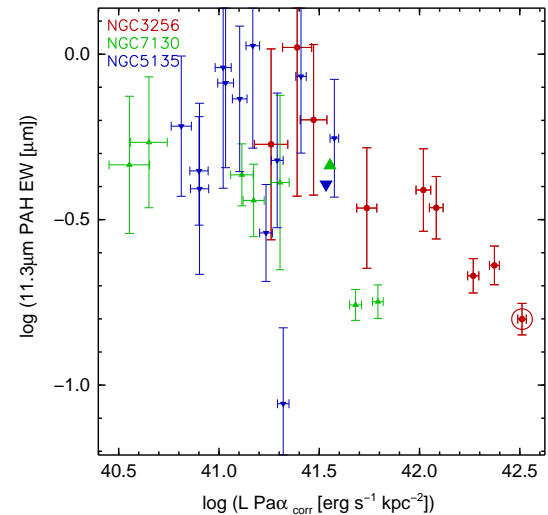}
\vspace{0.25cm}
\caption{\footnotesize \PAHa\ EW vs. \Paalpha\ LSD (corrected for extinction) for the star-forming regions in our LIRGs. Each datapoint is as in Figure~\ref{f:neiivspaaflux}.
[\textit{See the electronic edition of the Journal
for a color version of this figure}].}\label{f:113pahewsvspaaflux} 
\vspace{0.5cm}
\end{figure}

The intensity of the radiation field, however, does have an effect
on the EW of the \PAHa\ feature. Figure~\ref{f:113pahewsvspaaflux}
shows that there is a trend of the \PAHa\ EW to have lower
values for increasing \Paalpha\ LSD (equivalent to \NeII\ LSD).
This tendency is mostly driven by the star-forming regions of
NGC~3256 (and the two \HII\ regions of NGC~7130) and therefore
is related to the mass of the starburst, not the age (see above).
Hence, given that the \PAHa/\NeII\ ratio of these regions
is not varying with the \NeII\ LSD we could argue that,
% it is the continuum
%luminosity at 11.3$\,\micron$ that is increasing faster
%with the \Paalpha\ LSD than that of the \PAHa. That is,
for a given age, the more massive (denser) is a star-forming
region, the stronger is the continuum at 11.3$\,\micron$ with
respect to its \PAHa\ emission (i.e., the lower is the \PAHa\ EW).
Therefore
%, and because of the same reasons,
our data suggest that the PAHs (at least in these regions and at
the spatial scales probed here) are not destroyed by the
fact of being exposed to more massive starbursts but instead
they are being diluted by the enhanced continuum emission.
This dilution could be due to smaller surfaces of the PDRs
(from where the PAH emission arises) when compared to the
volumes of the continuum emitting dust for more massive starbursts.
%({\bf comentario sugerido por Vassilis}).
%This conclusion however, should be taken with caution given the uncertainties of the data.

\cite{Wu06} also established that the density (hardness and intensity)
of the radiation field has an impact on the PAH emission
of low-metallicity blue compact dwarf galaxies, and now our study
reveals that this effect is also present in the relatively high
metallicity star-forming regions of LIRGs.
However, our work also suggest that, at least for the age
and metallicity ranges of our star-forming regions,
a higher radiation field intensity (in the form of a more
massive starburst) causes the dilution of the PAHs
(i.e., a decreasing of the \PAHa\ EW, in this case) but not their
destruction (since the \PAHa/\NeII\ ratio is not modified).
Therefore our data support the
PAH \textit{dilution} scenario also claimed by other studies
of low-redshift ULIRGs (\citealt{Desai07}) and high-redshift
SMGs (\citealt{MD09}).
The \textit{destruction} of the PAHs (if it would be taking place)
would be related to the age of the star-forming regions (hardness
of the radiation field).
%, and would be seen in starbursts younger that those we are studying here (see previous subsection).

%Therefore, when star-forming regions of the same age (and metallicity) are considered, our data support the PAH \textit{dilution} scenario also claimed by other studies of low-redshift ULIRGs (\citealt{Desai07}) and high-redshift submillimiter galaxies (SMGs, \citealt{MD09}). In turn, the PAH \textit{destruction} scenario would be related to age effects.

Moreover, the diffuse regions located in between the \HII, \Paalpha-emitting
regions of our LIRGs show \PAHa\ EWs larger (approximately twice)
and \PAHa/\NeII\ ratios higher than those measured in the \HII\ regions
(see, e.g., Figure~\ref{f:ngc5135spatproffluxes}). This is in agreement
with the idea of these diffuse regions being less massive (have lower
intensities, implying higher \PAHa\ EWs, see above) and older
(have milder radiation
fields, implying higher \PAHa/\NeII\ ratios) than the \HII\ regions,
as it was suggested in \S\ref{ss:ageeffect}. This also reinforces the
idea of the PAH \textit{dilution} scenario in which the \HII\ regions
would be diluted by the enhanced dust continuum emission
that is not otherwise seen in the diffuse regions.

%En este caso NO es supresion, es DILUCION!!!!!!!!!! Por eso he modificado este parrafo:

%A similar behavior of supressed PAH emission in regions with high
%radiation field densities is seen in
%Figure~\ref{f:113pahewsvspaaflux}, but shown in terms
%of the EW of the \PAHa\ feature as a function of the
%\Paalpha\ LSD. This trend is mostly driven by the star-forming
%regions of NGC~3256, and the two \HII\ regions of
%NGC~7130. \cite{Wu06} also established that the intensity of the
%radiation field has an impact on the PAH emission of
%low-metallicity blue compact dwarf galaxies, and now our study
%reveals that this effect is also present in the relatively high
%metallicity star-forming regions of LIRGs. Apart from having a young
%starburst, which would also increase the hardness of the radiation
%field (previous section), a  massive starburst would also increase
%significantly the density of the radiation field.

Sumarizing, we have shown that the PAH and MIR continuum
emitting regions are not always spatially overlapped and
display different properties. For example, the MIR dust continuum
is mainly associated with massive, young, \Paalpha-emitting \HII\
regions, i.e., with \textit{current} star formation.
The PAH emission on the other hand is also characteristic of
more diffuse (less massive) and evolved stellar populations,
i.e., of \textit{recent} star formation, that show large \PAHa\ EWs
and high \PAHa/\NeII\ ratios. This is very important from the
point of view of knowing \textit{how} recent and \textit{where}
is located the star formation
%\textbf{that is taking place}
in LIRGs (either $\lesssim\,8-10\,$Myr or tens of Myr).
%in which the starbursting periods do not last for much more than $\sim\,$100\,Myr (\citealt{Kennicutt98}).
Depending on the
measurement used (MIR dust continuum or PAH emission),
different stages in the process of star formation
and different locations in galaxies will be probed.
%This is even more critical when the SFR is obtained based on the IRAC $8\,\micron$ fluxes. Because this is a wide filter, it contains both dust continuum and PAH emission in different relative amounts. Thus, there exist not only the problem of averaging out the properties of the star-forming regions due to the limited spatial resolution of \textit{Spitzer}, but also because of the PAH and the MIR continuum cannot be separated.
Therefore, high spatial resolution observations of local
and also high-redshift LIRGs are essential to determine: (1) how
recent and (2) where is located the star formation in these galaxies
(compact \HII\ regions and/or diffuse emission?) in a statistical way.
This is also ultimately important in order to know whether local
and high-$z$ LIRGs, ULIRGs and SMGs share the same physical
properties and belong to the same galaxy population.

\section{Summary}\label{s:summary}

%Our sample of LIRGs covers a wide range of physical conditions: from nuclear isolated star formation, to sources whose nuclear emission is a mixture of AGN and star-forming regions, and to nuclei purely dominated by AGN activity.
The high spatial resolution afforded by Gemini T-ReCS has allowed us to
resolve the nuclei and star-forming regions in our sample of LIRGs
on hundred pc scales. In addition, we have been able to separate the
AGN contribution (if any) from that of the star formation.

By comparing the nuclear and integrated T-ReCS spectra with those
obtained with \textit{Spitzer} IRS we find that, for LIRGs containing
an AGN, at least half of the \SIV\ emission arises
from the central $\sim\,$150\,pc. We have shown that the T-ReCS
spectrum of the Southern nucleus of NGC~3256 is very obscured and
is totally absorbed between 9 and 11$\,\micron$, in contrast to the
conclusion reached from the IRS spectrum
since it is contaminated by emission from the
diffuse medium that surrounds it. Therefore, only with the combination
of the exceptional sensitivity of \textit{Spitzer} and the high spatial
resolution of T-ReCS, we can gain a clearer and less confused picture
of the central regions of the LIRGs.

\begin{itemize}
 \item Regarding the individual study of the LIRGs (in the Appendix):

For NGC~3256
we have shown
that the integrated spectrum of the northern nucleus is very similar to
that of the very nuclear region (0.36\arcsec$\,\times\,$0.36\arcsec),
with prominent 8.6 and \PAHa\ features and an intense \NeII\ emission line.
On the other hand, the spectrum of the southern nucleus is extremely
absorbed and no emission is detected between $9-11\,\micron$.
IC~4518W shows extended \SIV\ line emission out to
$\sim\,$0.8\arcsec\ ($\simeq\,$265\,pc) to the north of the nucleus.
We have interpreted this high excitation region as a signature of
the presence of a narrow line region. For NGC~5135 we have separated
the unresolved AGN component from that of the surrounding \HII\ and
diffuse regions. The Sy2 nucleus of NGC~5135 accounts for at least
50\% of the \SIV\ emission of the LIRG. The \HII\ region and the
diffuse medium display intense \PAHa\ emission with the latter showing
a greater EW. We have resolved the nucleus of NGC~7130. However,
the emission from the Sy nucleus and that of the surrounding star-forming
regions are still mixed on scales of $\sim\,$100\,pc. Both, the very
nuclear (0.72\arcsec$\,\times\,$0.36\arcsec) and the integrated
(0.72\arcsec$\,\times\,$3.6\arcsec) spectra show the same features:
faint \SIV\ emission and intense \PAHa\ and \NeII\ emission.
In all LIRGs we find that the PAH emission is always more extended
than the \Paalpha\ or \NeII\ emission suggesting that PAHs are not
only associated with the youngest ionizing stellar populations but
also to a more diffuse medium.

%The innermost nuclear regions of the LIRGs with star formation show a higher \NeII/\PAHa\ ratio ($\sim\,$1.1) than those regions located in the surrounding ($\gtrsim\,$1\arcsec\,$\simeq\,$250\,pc) area ($\sim\,$0.6). This suggests a very rapid change in the physical conditions within the nuclear regions of LIRGs and/or changes in extinction.

\item Regarding the general conclusions:

For NGC~3256, NGC~5135 and NGC~7130, the strength of the silicate
feature is lower in the nucleus than in the surrounding regions.
While we find a \SSi$\,\simeq\,-0.5$ for the Sy2 nuclei of NGC~5135
and NGC~7130, and also for \HII-like nucleus of NGC~3256, the
extra-nuclear regions show values up to $\simeq\,-1.2$. The nucleus
of IC~4518W shows a deeper absorption, with \SSi$\,\simeq\,-1.4$.
Such variations in the depth of the silicate feature have been also
observed in some nearby Sy galaxies. We attribute the fact of observing
lower values of the \SSi\ in the nucleus than in the surrounding regions
to the transition from either the clumpy or compact environment
of an AGN or young, nuclear starburst to the more smooth, extended
distributions of extra-nuclear star-forming regions.
%presence of a hot dust emission component generated by the nuclear AGN or young starburst that is ``filling'' the absorption feature.

Both the \NeII\ line and the \PAHa\ emission are related to
the number of ionizing photons as measured from the \Paalpha\ emission
line. However, while we find that the \NeII\ and \Paalpha\ LSDs are
directly proportional for all the star-forming regions in our LIRGs
(slope of 1.00$\pm$0.02), the trend seen between the \PAHa\ and
\Paalpha\ LSDs, although clear, is different for the regions
of each LIRG.

The \NeII/\Paalpha\
ratio does not depend on the \Paalpha\ EW, that is, on the age of
the ionizing stellar populations, suggesting that the \NeII\ line
can be used as a good tracer of the SFR in star-forming regions
in LIRGs. On the other hand, the \PAHa/\Paalpha\ ratio is strongly
dependent on the \Paalpha\ EW, and increases as the star-forming
regions age, in agreement with the findings obtained by \cite{DS08}.
%However, the slope of the trend is higher than that of the 8$\,\micron$/\Paalpha\ ratio, and indicates that the luminosity of the \PAHa\ is effectively evolving (increasing) with time.
%This implies that either the PAH emission is being enhanced % carriers are being created rapidly %during this epoch of the starburst evolution, and/or the PAH carriers were being suppressed at younger times.
This adds support to the scenario in which PAHs are better
tracers of \textit{recent} (tens of Myr) rather than of
\textit{current} ($\lesssim\,8-10\,$Myr), massive star formation.

The \PAHa/\NeII\ ratio (equivalent to the \PAHa/\Paalpha\ ratio) also
varies with the \NeII\ LSD, which depends not only on the hardness
of the radiation field (age) but also on the intensity (mass density)
of the starburst.
However, the decreasing of the \PAHa/\NeII\ ratio with increasing
\NeII\ LSD is only related to an age effect, in agreement with
the findings above. The intensity of the starburst (in the sense
of its mass density) does not seem to modify the \PAHa/\NeII\ ratio.
On the other hand, the \PAHa\ EW does not present a clear dependence
with the age of the stellar population but it does (mainly for the
nuclear regions of NGC~3256) with the intensity of the radiation field.
We suggest that an increasing of the starburst mass density causes
the PAH to be \textit{diluted} (through the increasing of the
MIR continuum) but not to be \textit{destroyed}.
The MIR dust continuum is mainly associated with massive,
young, compact \HII\ regions, i.e., with \textit{current} star
formation ($\lesssim\,8-10\,$Myr). The \PAHa\ emission is rather more
(but not only) characteristic of more diffuse (less massive)
and evolved (tens of Myr) stellar populations,
i.e., of \textit{recent} star formation, showing large \PAHa\ EWs
and high \PAHa/\NeII\ ratios.

%the decreasing of the \PAHa\ EW with the increasing LSD increasing of the mass of the starburst (through its increasing of the MIR continuum  emission) is causing the dilution of the \PAHa\ EW, as the size of the PDR probably remains almost constant.

Disentangling the different effects on the energetics of the PAH
features in star-forming regions is not straightforward and more
observations are needed to draw statistically significant conclusions.
Therefore we want to stress that high spatial resolution
observations of local and high-$z$ (U)LIRGs are essential to
determine: (1) how recent and (2) where is located the star formation
in these galaxies (compact \HII\ regions and/or diffuse emission?).
This will allow us to investigate whether they share the same physical
properties and therefore belong to the same galaxy population,
providing ultimately the missed link between the local
and high-redshift Universe when LIRGs dominated the SFR.\\

\end{itemize}

\section*{Acknowledgments}

We thank the referee, Konrad Tristram, for his very useful comments
and suggestions which significantly improved the paper.
This work is based on observations obtained with T-ReCS instrument at the
Gemini South Observatory, which is operated by AURA, Inc., under a
cooperative agreement with the NSF on behalf of the Gemini
partnership: NSF (United States), PPARC (UK), NRC (Canada), CONICYT
(Chile) ARC (Australia), CNPq (Brazil) and CONICET (Argentina).
This work has been supported by the Plan Nacional del Espacio under
grant ESP2005-01480 and ESP20076-65475-C02-01.
%TDS wants to thank the Univerity of Kentucky group for their kind hospitality during the spring of 2007, and specially to Robert Nikutta and Matthew Sirocky for many interesting discussions.
TD-S acknowledges support from the Consejo
Superior de Investigaciones Cient\'{\i}ficas under grant I3P-BPD-2004.
NAL acknowledges work supported by the NSF under Grant No. 0237291.
AA-H also acknowledges support from the Spanish Ministry of Science and
Innovation through grant Proyecto Intramural Especial 200850I003.
TD-S wants to thank Brent Groves for providing
us with the photo-ionization models and for interesting discussions.
This research has made use of the NASA/IPAC Extragalactic Database
(NED), which is operated by the Jet Propulsion Laboratory, California
Institute of Technology, under contract with the National Aeronautics
and Space Administration, and of NASA's Astrophysics Data System (ADS)
abstract service.\\

\appendix\label{s:appendix}

\section{-A- NGC~3256: Isolated, Spatially-Resolved Star Formation}\label{s:ngc3256sp}

\subsection{Emission from Star-Forming Regions}\label{ss:ngc3256nucspec}

\begin{figure}%[!h]
\epsscale{0.75}
%\plotone{./figures/ngc3256_trecs_spectra.ps}
\plotone{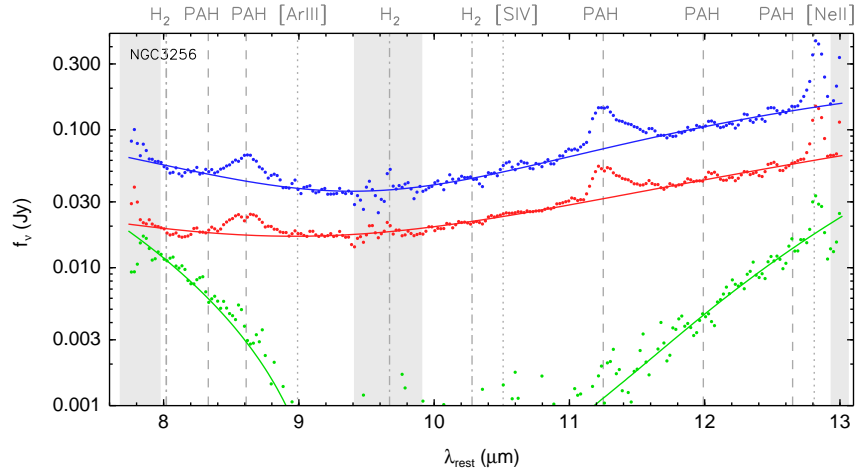}
\vspace{0.25cm}
\caption{\footnotesize Integrated (blue dots) and nuclear (red dots) T-ReCS $N$-band spectra of NGC~3256. The region of interest (green dots) in NGC~3256 is the southern nucleus of the galaxy (as extracted in the same way as the northern one). The integrated spectrum was extracted with a fixed aperture of 3.6\arcsec\ in length centered at the nucleus of the galaxy, while the nuclear spectra were extracted with a fixed aperture of 0.36\arcsec. The labels are as in Figure~\ref{f:ngc3256spatpos}.
[\textit{See the electronic edition of the Journal
for a color version of this figure}].}\label{f:ngc3256intnucroispecs} 
\vspace{0.25cm}
\end{figure}

The northern nuclear spectrum of NGC~3256
(see Figure~\ref{f:ngc3256intnucroispecs}; if no specification is given,
we will refer to the northern nucleus as the main nucleus of this LIRG)
shows prominent PAH features (the \PAHb\ feature and the \PAHa\ complex)
%and even the $11.99\,\micron$ PAH feature seems to be marginally present),
%, although slightly displaced toward longer wavelengths),
which are indicative of intense star formation. This is corroborated
by the presence of conspicuous \NeII\ line emission.
% and is in agreement with the classification of this galaxy as \HII-like.
Furthermore, there is no hint for other lines such as \SIV\ that could
suggest the existence of an AGN (and/or very young star formation).
The northern nucleus of NGC~3256 was previously observed by \cite{MH06}
and \cite{Lira08}. Both works also detected intense PAH features and
the \NeII\ emission line. In an aperture of 1.2\arcsec$\,\times\,$3\arcsec,
\cite{MH06} measured a flux of the \NeII\ emission line in the northern
nucleus of NGC~3256 of 2.1$\,\times\,10^{-12}\,$erg\,s$^{-1}$\,cm$^{-2}$,
%(without correcting for slit losses)
which means that only $\sim\,$6\%
of this flux is contained within our T-ReCS nuclear aperture (about
30 times smaller than theirs). In addition, \cite{MH06} also measured
a \PAHa\ flux from the northern nucleus of
2.5$\,\times\,10^{-12}\,$erg\,s$^{-1}$\,cm$^{-2}$; the T-ReCS nuclear
flux accounts for $\sim\,$5\% of their measurement.
%Although with some uncertainty, this would indicate that the \NeII\ emission is slightly more compact than that of the \PAHa.
%\subsection{Integrated Northern Nuclear Spectrum}\label{ss:ngc3256intspec}
The T-ReCS integrated spectrum (not including the southern nucleus)
displays the same features as its nucleus, but with both the 8.6 and
\PAHas\ and the \NeII\ emission line showing larger EWs
(see Figure~\ref{f:ngc3256intnucroispecs} and Tables~\ref{t:nucews}
and \ref{t:intews}).
%Therefore, despite the nucleus is resolved we find the same signatures of star formation in the nuclear and in the integrated spectra (which cover from the inner hundred of pc to the central kpc of the galaxy, respectively). This is in agreement with this galaxy undergoing a large scale merger that is driving the star-forming processes at all spatial scales (\citealt{Lipari00}), from a few tens of pc to several kpc. Maybe the main difference between both spectra is the slope of the continuum at the shortest wavelengths. While the integrated spectrum from 9 to 13$\,\micron$ is just a scaled-up version of the nuclear spectrum, below 9$\,\micron$ the slopes of both spectra diverge. This could be due to the higher contribution of the 7.6$\,\micron$ PAH to the integrated spectrum. In spite of this feature is outside of the $N$-band window, its red wing can contribute to the emission up to $\sim\,8\,\micron$ and beyond. It is known that PAH emission is more spatially extended than the continuum emission emitted by the hot dust (see results of \citealt{AAH06a} and \cite{DS08}). This result would then imply that this PAH would be more extended than the rest of the MIR emission and therefore the integrated spectrum would include more PAH emission than the nuclear spectrum. This is also in agreement with the fact that the EWs of the 8.6 and \PAHas\ are higher in the integrated spectrum.

%\subsection{Extra-Nuclear Regions: The Southern Nucleus}\label{ss:ngc3256regofintspec}

The southern nucleus of NGC~3256 (the region of interest of this galaxy)
is located $\sim\,$5\arcsec\ to the south of the northern nucleus.
It is behind large amounts of dust even making the MIR wavelengths to
be very affected by a high extinction (\AV$\,\gtrsim\,12-15\,$mag;
\citealt{Kotilainen96}; \citealt{Lira02}; \citealt{AAH06a}; \citealt{DS08}).
This is reflected in the $N$-band T-ReCS spectrum as a deep absorption of
the \Siabs\ feature.
%The most impressive finding among all is the spectrum of the southern nucleus of NGC~3256.
Figure~\ref{f:ngc3256intnucroispecs} shows an extremely absorbed continuum
whose emission is totally suppressed between 9 and 11$\,\micron$.
The \HII\ classification of this nucleus would imply the detection
of prominent 8.6 and \PAHas\ as in the northern one. In fact, \cite{MH06}
and \cite{Lira08} find both features in their spectrum of the
southern nucleus. We do not. The lack of PAH emission
in the T-ReCS spectrum is probably due to that our aperture
is much smaller than theirs implying that either there is not PAH
emission in the very nuclear region or the extinction is that
high (\AV$\,>\,12\,$mag, with $A_{12\,\mu m}\,\simeq\,0.037\,$\AV,
\citealt{Rieke85b}) that prevents us from detecting the PAHs.
In fact, their spatial resolutions and, most important, the
size of their extraction apertures ($\sim\,$2$\,\times\,$2\arcsec\
$\approx$ 390$\,\times\,$390\,pc) are $>$\,5 times larger than ours
(0.36\arcsec$\,\times\,$0.36\arcsec\ $\approx$ 70$\,\times\,$70\,pc).
Therefore, their spectrum contains not only the emission from the nucleus
but also from the surrounding regions, which includes diffuse continuum
and PAH emission. Regarding the \NeII\ emission line, the flux of the
southern nucleus contained in our aperture is only $\sim\,$10\% that
of Martin-Hernandez et al.'s, suggesting that probably much of the
\NeII\ emission they measure is mostly off-nuclear.

%the extinction towards this area may have completely erased any signature from these features. Only the \NeII\ emission line is detected, but it would also be affected by extinction as $A_{12\,\mu m}\,\propto\,27\,$\AV\ (\citealt{Rieke85b}). In fact, the flux contained in our aperture is only $\sim\,$10\% that of Martin-Hernandez et al.'s
%Regarding the southern nucleus and the \NeII\ emission line, the flux contained in our aperture is only $\sim\,$10\% that of Martin-Hernandez et al.'s. In fact, none of the 8.6 or \PAHas\ are detected in our T-ReCS spectrum. This is probably because of our aperture is much smaller than that used by them, and therefore we are looking at the very center of the nucleus where the extinction is very high (\AV$\,\gtrsim\,12\,$mag), which prevent us from detecting the PAHs. However, their spectrum contains not only the emission from the nucleus but also from the surrounding regions, plenty of PAH and \NeII\ emitting regions.
%On the other hand, \cite{MH06} and \cite{Lira08} find both 8.6 and \PAHa\ emission in the spectrum of the southern nucleus. However, their spatial resolutions and, most important, the size of their extraction apertures ($\sim\,$2$\,\times\,$2\arcsec\ $\approx$ 390$\,\times\,$390\,pc) are $>$\,5 times larger than ours (0.36\arcsec$\,\times\,$0.36\arcsec\ $\approx$ 70$\,\times\,$70\,pc), and thus they include diffuse continuum and PAH emission in the spectrum that is not directly related with the nucleus. % (see NGC~5135).

\subsection{Spatial Profiles: Differences Between Nuclear and Extra-Nuclear Star-Forming Regions}\label{ss:ngc3256spatprof}

\begin{figure}%[!h]
\epsscale{0.37}
%\plotone{./figures/ngc3256_ap4pix_fix2_cal_flux_spat_dist_panel_02.ps}\hspace{0.2cm}\plotone{./figures/ngc3256_ap4pix_fix2_cal_flux_spat_dist_panel_12.ps}\hspace{0.5cm}\plotone{./figures/ngc3256_ap4pix_fix2_cal_flux_spat_dist_panel_11.ps}
\plotone{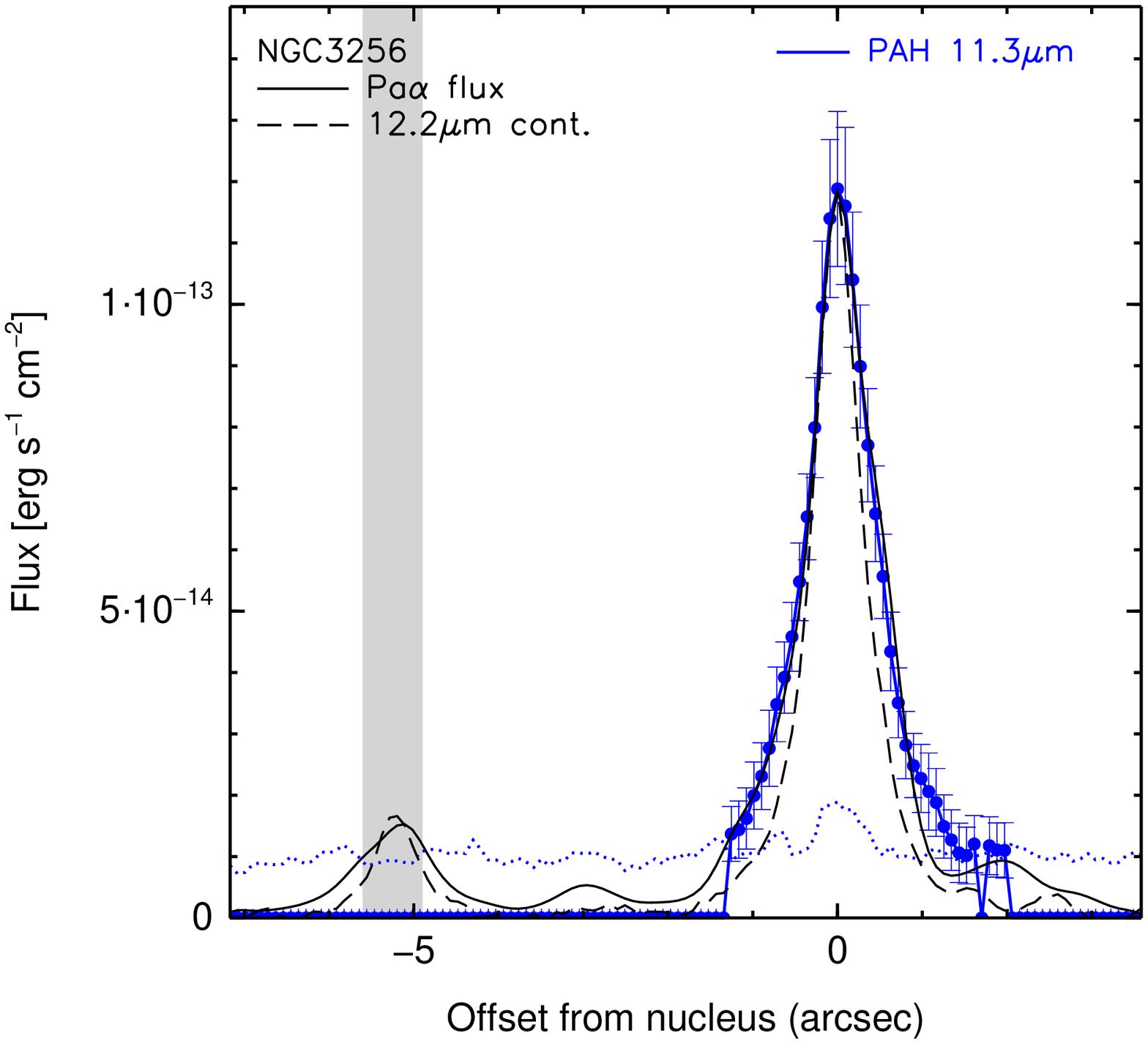}\hspace{0.2cm}\plotone{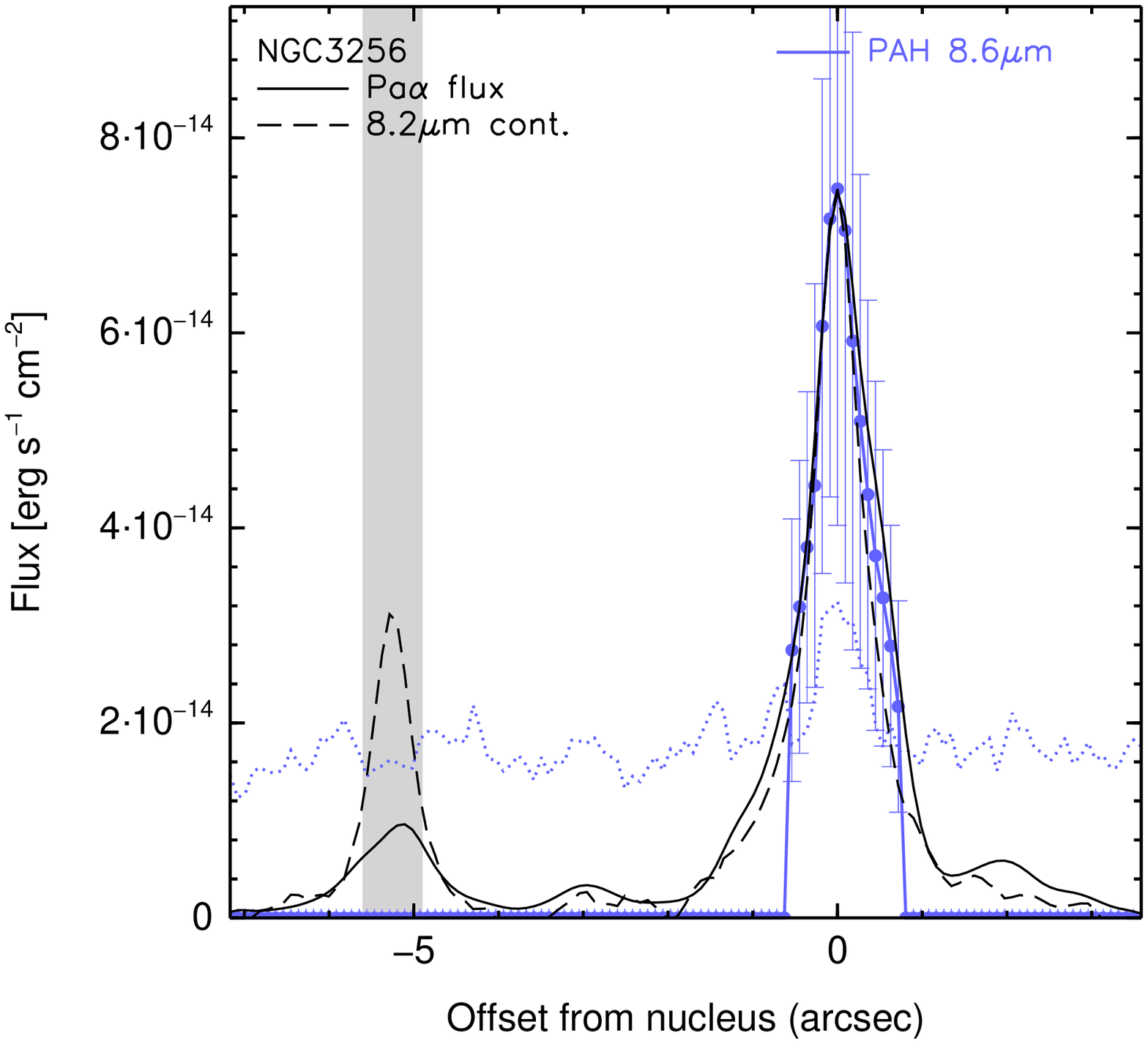}\hspace{0.5cm}\plotone{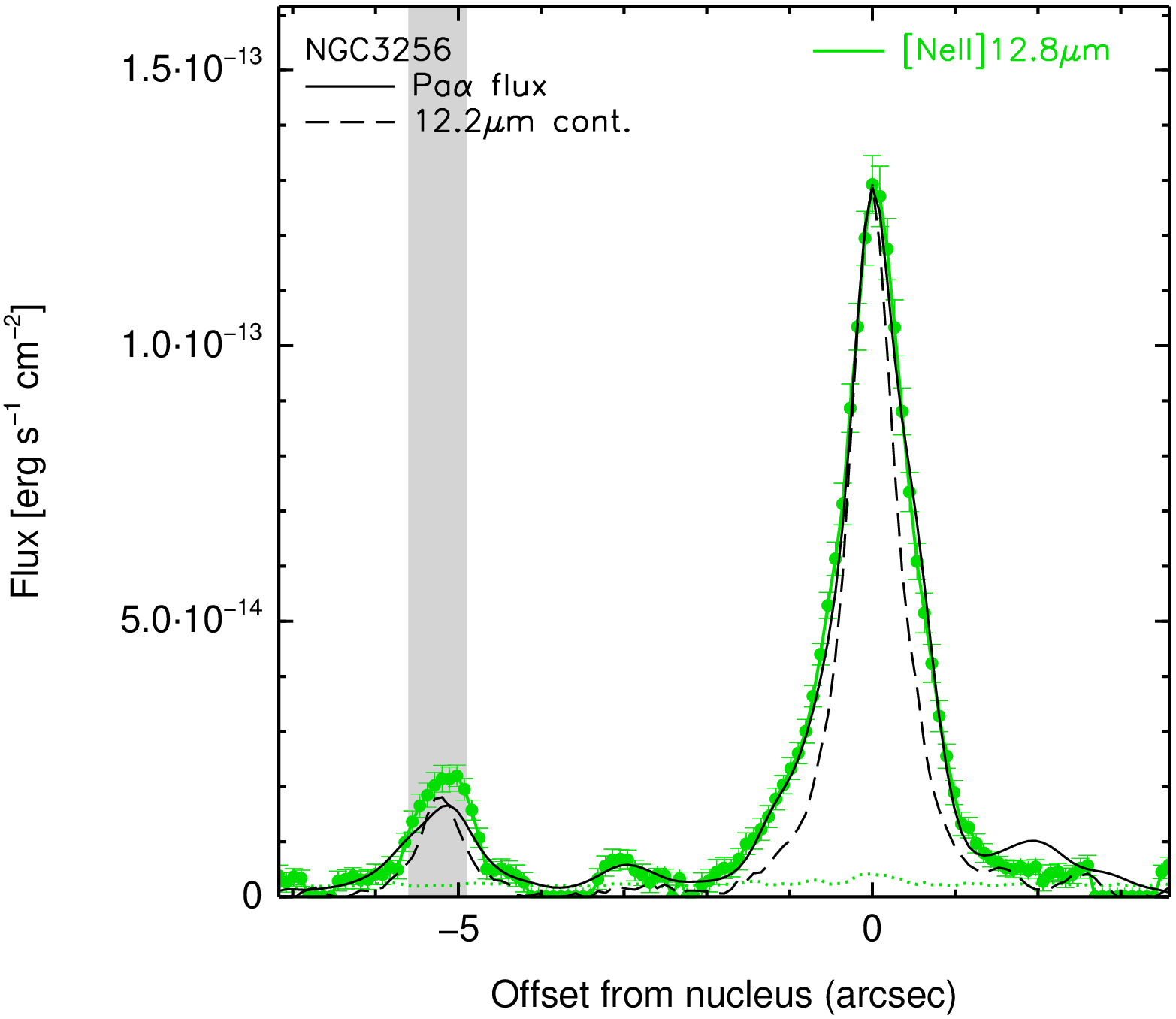}
\vspace{0.25cm}
\caption{\footnotesize Spatial profiles of the flux (not corrected for extinction) of different features detected in the MIR T-ReCS spectra of NGC~3256. On the x-axis, the 0 offset marks the position of the northern nucleus. From left to right and top to bottom: the 11.3 and \PAHbs, and the \NeII\ emission line. The dotted profile in each panel is the threshold limit for the detection of the feature. The black solid line is the \Paalpha\ profile as measured from the \textit{HST} NICMOS images by simulating the same aperture size with which the T-ReCS spectra were extracted. The flux has been scaled to the maximum value of the spatial profile of the given feature. The black dashed line is the continuum emission profile at the given wavelength, scaled in the same manner as the \Paalpha\ profile. The gray shaded column marks the location of the region of interest in each galaxy. For NGC~3256 the region of interest is the southern nucleus, located at $\sim\,$5\arcsec\ to the south (negative offset) of the northern nucleus. Figure~\ref{f:slits} shows how the slit was positioned over the galaxy.
[\textit{See the electronic edition of the Journal
for a color version of this figure}].}\label{f:ngc3256spatproffluxes} 
\vspace{0.25cm}
\end{figure}

We detect \PAHb\ emission at distances of about 1\arcsec\
($\sim\,$196\,pc) north and south from the northern nucleus
(see Figure~\ref{f:ngc3256spatproffluxes}). The \PAHa\
emission is even more extended, out to 2\arcsec\ to the north. Because
the \PAHb\ detection threshold is higher than that of the \PAHa\ we can
only measure it in the central regions. There is no PAH emission in the
southern nucleus. The \NeII\ emission line is detected at almost all
positions along the slit and, as with the PAH features, its nuclear
emission is more extended than the continuum emission.
% (see Figure~\ref{f:ngc3256spatprofconts}).

Moreover, the spatial profile of the \NeII\ emission is remarkably similar
to that of \Paalpha. In particular, the \NeII\ emission is also detected
in the southern nucleus. This agreement is not only qualitative but also
quantitative as their ratio is almost constant along the slit.
% ($\simeq\,$3).
%, see Figure~\ref{f:ngc3256spatprofneiipaafluxratio}).
%Both emission lines show the same extended emission to the south of the northern nucleus and decay at equal rate to the north, unlike the \PAHa\ which shows extended emission towards the north (at $\sim\,$1.5\arcsec, see the profile of the \PAHa/\NeII\ ratio in Figure~\ref{f:ngc3256spatproffluxes}, left panel).
This finding is in agreement with previous results
that suggest that the \NeII\ emission line is directly linked to
the total number of ionizing photons of a region, that is, to its
star-formation rate (see \S\ref{s:neii}, and also \citealt{Roche91};
\citealt{Ho07}) as traced by, e.g., the \Paalpha\ line.
%In Figure~\ref{f:ngc3256spatprofneiipaafluxratio} we can see that
The \NeII/\Paalpha\ ratio is broadly constant
%along the slit
(it ranges between $\sim\,2-5$, not corrected for extinction).
However, small-scale variations (less than an arcsecond, i.e.,
few hundreds of pc) can be found.
%For example, the southern nucleus show a slightly higher ratio than its adjacent regions.
%These changes
%%The fact that the \NeII\ to \Paalpha\ emission ratio at the southern nucleus of the galaxy is sligltly higher than in its surroundings (see Figure~\ref{f:ngc3256spatprofneiipaafluxratio}) %The only location where the \NeII\ and the \Paalpha\ emission line profiles slightly disagree is at the southern nucleus of the galaxy where the \NeII\ to \Paalpha\ emission ratio is higher than in the rest of the regions (the \Paalpha\ profile is normalized to that of the \NeII\ at the peak of the northern nucleus).
%could be partially explained in terms of the extinction. For example, the southern nucleus of NGC~3256 suffers from severe obscuration
%%\AV$\,\sim\,12-15\,$mag (\citealt{Kotilainen96}; \citealt{Lira02}; \citealt{AAH06a}; \citealt{DS08} and Figure~\ref{f:ngc3256intnucroispecs}),
%($\sim\,1.75\,$mag at $1.87\,\micron$ whereas this is only $\sim\,0.75\,$mag at 12.8$\,\micron$, \citealt{Calzetti00}). This differential extinction could roughly explain the variations of the \NeII/\Paalpha\ ratio at the location of the southern nucleus. On the other hand,
Variations in the fraction of
\NeIIion\ ions, %\NeII/\NeIII\ ratio
which in turn depends on the age of the ionizing stellar populations
(\citealt{MH05}), could be responsible for the \NeII\ vs. \Paalpha\
differences. %\textit{It could be also that the [NeII]/[NeIII] ratio varies and there was less [NeII]? Or less Ne abundance in general...}

%Figure~\ref{f:ngc3256spatprofratios} shows that 
The 11.3/\PAHb\ ratio is constant
%($\simeq\,1.7$)
along the slit indicating that the ionization conditions (that are
suggested to be traced by the amount of neutral to ionized PAHs;
\citealt{Galliano08}) do not vary significantly
throughout the nuclear region of the LIRG on the physical scales probed
here ($\sim\,$100\,pc). The \PAHa/\NeII\ ratio (not corrected for
extinction) is also almost constant ($\simeq\,1$) within 1\arcsec\
from the nucleus (see also Figure~\ref{f:113pahtoneiivsneiiflux}).
However, the ratio starts to increase towards the
north (positive offset; between $1-2\arcsec$).
% reaching a value of $\simeq\,1.6$.
A very young population where all the Neon were in the
\NeIIIion\ state would show a high \PAHa/\NeII\ ratio and would
explain this behavior.
However, the \Paalpha\ EW map of the galaxy (see \citealt{DS08})
does not display any enhancement of the \Paalpha\ EW in this area.
In addition, we see a very good correlation (slope of 1.00$\pm$0.02,
see \S\ref{s:neii}) between the \NeII\ and the \Paalpha\
lines, indicating that all the star formation is accounted by the
\NeII\ emission. We refer the reader to \S\ref{ss:denseffect}
with regards to the discussion on the \PAHa/\NeII\ ratio.
%the other explanation would be that this region of low \NeII/\PAHa\ ratio is not as young as the nuclear regions but the stars are sufficiently young ($\lesssim\,30-50\,$Myr) to produce UV radiation and thus to produce the excess of PAH emission.
% due to the extended emission of the \PAHa\ seen in Fig~\ref{f:ngc3256spatproffluxes} but that is not seen in the \NeII\ line profile. \textbf{This would suggest that a low ratio of \NeII\ to \PAHa\ emission ($\simeq 0.5$) is representative of a region that is not longer dominated by the youngest ionizing stars but is rather characteristic of the diffuse ISM formed by intermediate-age stellar populations, which provides the UV photons able to excite the PAHs but not the hard radiation field necessary to ionize the Ne.}

%\textit{Esto se puede dejar para el articulo especifico de NGC~3256 -----$>$}
%The spatial profiles of the EW of the different spectral features are shown in Figure~\ref{f:ngc3256spatprofews}.
The northern nucleus of the galaxy
shows lower EWs of the \NeII\ emission line and 8.6 and \PAHas\
than its surrounding regions.
%Furthermore, in general, the features seem to show a lower EW where they are more intense (see Figure~\ref{f:ngc3256spatproffluxes}). This is only apparent, though. The fact is that the emission of the features is more extended than that of the continuum, making their EWs to increase outwards the nucleus (away form where the continuum emission peaks). In other words,
This is probably because the features are diluted by the more compact
%(see Figure~\ref{f:ngc3256spatprofconts})
continuum emitted by the hot dust that is concentrated towards the nucleus
(see also \S\ref{s:siabsfeat} and \S\ref{ss:denseffect}).

\section{-B- IC~4518W: Isolated AGN Activity}\label{s:ic4518wsp}

\subsection{Signatures of the Sy2 Nucleus and a High-Excitation Region}\label{ss:ic4518wnucspec}

\begin{figure}%[!h]
\epsscale{0.75}
%\plotone{./figures/ic4518w_trecs_spectra.ps}
\plotone{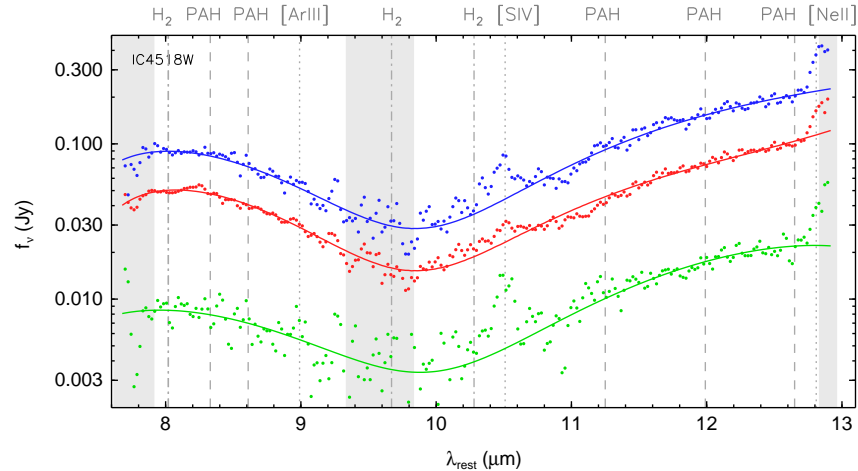}
\vspace{0.25cm}
\caption{\footnotesize Integrated (blue dots) and nuclear (red dots) T-ReCS $N$-band spectra of IC~4518W. The region of interest (green dots) of IC~4518W is an extended region seen in \SIV\ emission located 0.5\arcsec\ to the north of the nucleus (the spectrum was extracted in the same way as the nucleus). The integrated spectrum was extracted with a fixed aperture of 3.6\arcsec\ in length centered at the nucleus of the galaxy, while the nuclear and the high-excitation region spectra were extracted with a fixed aperture of 0.36\arcsec. The labels are as in Figure~\ref{f:ngc3256spatpos}.
[\textit{See the electronic edition of the Journal
for a color version of this figure}].}\label{f:ic4518wintnucroispecs} 
\vspace{0.25cm}
\end{figure}

%On the other extreme to NGC~3256 is IC~4518W, whose nuclear spectrum
Unlike NGC~3256, the nuclear spectrum of IC~4518W displays a
rather featureless continuum with the exception of the barely
detected \SIV\ emission line and the \NeII\ emission line
(see Figure~\ref{f:ic4518wintnucroispecs}). It is not clear whether the
lack of PAH emission at both 8.6 and 11.3$\,\micron$ can be interpreted
in terms of absence of star formation, since the PAH molecules could have
been destroyed by the Sy2 nucleus of the LIRG.
% Moreover, taken into account this assumption, we cannot assure that all the measured \NeII\ flux is related to the AGN emission.
%\subsection{Integrated Spectrum}\label{ss:ic5418wintspec}
As for NGC~3256, the nuclear and integrated spectra of IC~4518W are
very similar. The integrated spectrum of the galaxy (see
Figure~\ref{f:ic4518wintnucroispecs}) also shows an almost featureless
continuum as the nuclear one, as well as the presence of the \SIV\ and
the \NeII\ emission lines.
An important remark here is that the flux of the \SIV\ line is
larger for the integrated emission of the galaxy when it is subtracted
from that of the nucleus, than for the nucleus itself. That is,
for the rest of the LIRGs in which the \SIV\ line is detected,
the nuclear emission accounts for at least or around half of the flux
seen in their integrated spectra (compare Tables~\ref{t:nucfluxes} and
\ref{t:intfluxes}; see also Appendix \ref{s:ngc5135sp} and \ref{s:ngc7130sp};
the nuclear fluxes of IC~4518W and NGC~5135 have not been corrected for
aperture effects). In other words, the \SIV\ emission line usually stems
from the very nuclear region ($\lesssim\,100\,$pc) of the galaxies
except for IC~4518W (see below).
%In fact, the EW of the \SIV\ emission line measured in the integrated spectrum is about twice the value obtained from the nuclear one.

%\subsection{Extra-Nuclear Regions}\label{ss:ic4518wregofintspec}

The region of interest of IC~4518W is a region located
0.5\arcsec\ ($\simeq\,$165\,pc) to the north of the galaxy. Its
spectrum is very similar to that of the nucleus,
not only in the shape of the continuum but also in its spectral
features (see Figure~\ref{f:ic4518wintnucroispecs}).
In fact, the \SIV\ emission line is as intense as in the nucleus.
This is interesting because the nucleus of IC~4518W is unresolved
(FWHM $\lesssim\,0.4\arcsec$; see Table~\ref{t:specobs})
in continuum emission whereas the \SIV\ emission is clearly
detected out to $\sim\,$0.8\arcsec\ ($\simeq\,$265\,pc)
north from the nucleus. Indeed, this extended \SIV\ emission
can explain the difference seen above between the nuclear and
integrated spectrum of the galaxy.

%of this extra-nuclear region is practically not mixed with that of the AGN. Thus, such intense emission of the \SIV\ line must be associated with processes involving large-scale heating from the central BH (see below).

\subsection{Spatial Profiles: Line Emission in Hard Radiation Fields}\label{ss:ic4518wspatprof}

\begin{figure}%[!h]
\epsscale{0.37}
%\plotone{./figures/ic4518w_ap4pix_fix2_cal_flux_spat_dist_panel_01.ps}\hspace{0.2cm}\plotone{./figures/ic4518w_ap4pix_fix2_cal_flux_spat_dist_panel_11.ps}
\plotone{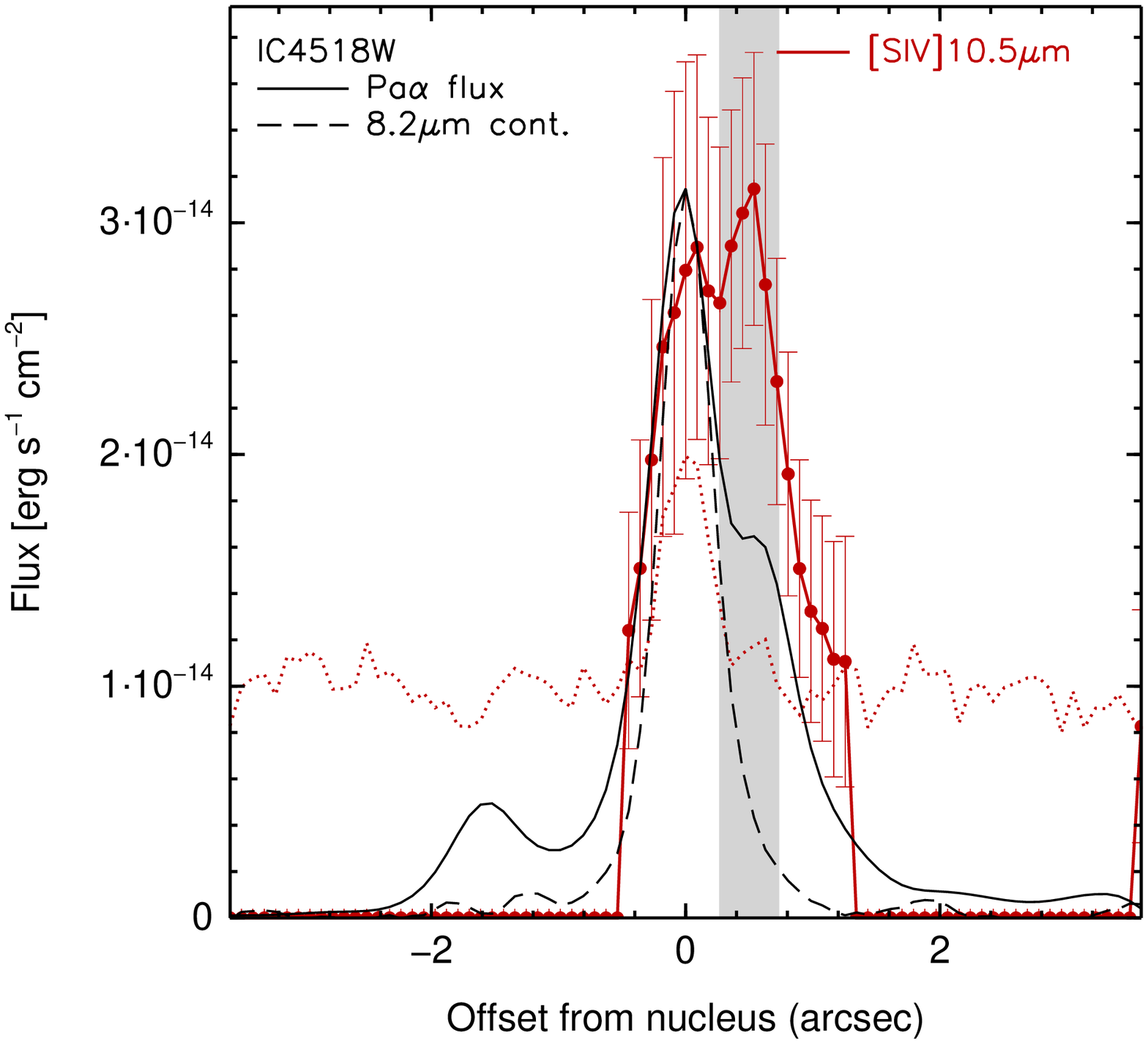}\hspace{0.2cm}\plotone{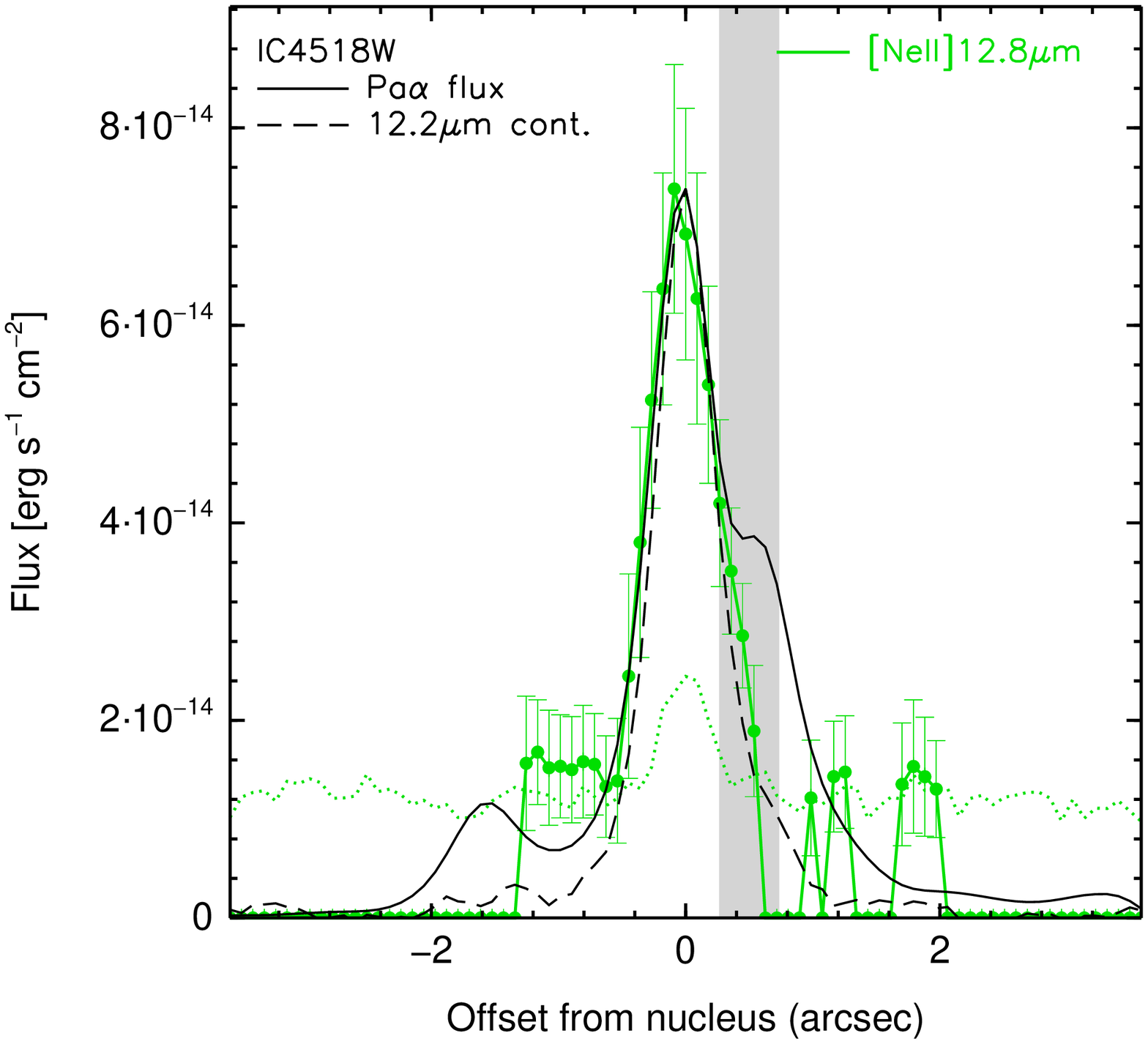}
\vspace{0.25cm}
\caption{\footnotesize Same as Figure~\ref{f:ngc3256spatproffluxes} but for IC~4518W. In this case, the flux spatial profiles of the \SIV\ (left) and \NeII\ (right) emission lines are shown. No PAH emission was detected.
[\textit{See the electronic edition of the Journal
for a color version of this figure}].}\label{f:ic4518wspatproffluxes}
\vspace{0.25cm}
\end{figure}

%The case of IC~4518W is very interesting too.
%This would be in agreement with the destruction of the molecules in the closest environment of an AGN, but also with the lack of nuclear star formation. In addition, it was also shown that a region located $\sim\,0.5\arcsec\,$ to the north of the nucleus (positive offset) exhibits an important emission of the \SIV\ line. This is confirmed in
Fig~\ref{f:ic4518wspatproffluxes} shows quantitatively that the \SIV\
line emission of the region located $\sim\,0.5\arcsec\,$ to the north
of the nucleus (positive offset) is comparable to that of the nucleus.
Although the \SIV\ line can be produced in very young star-forming
regions, we do not detect significant \Paalpha\ nor \NeII\ emissions
spatially coincident with the \SIV\ extended emission.
%There is only a little bump of \Paalpha\ emission towards this region but its intensity is 3 times fainter than the flux measured in the nucleus.
Moreover, we do not detect any of the two dominant 8.6 and \PAHas\
%of the $N$-band spectra
in the spectrum of the Sy2 nucleus of this LIRG. Thus we
conclude that the extended \SIV\ emission is most likely associated to the
central AGN and produced in an extended narrow line region (NLR) excited
by the Seyfert nucleus.
%%This extended emission extends out to 0.8\arcsec\ ($\simeq\,$265\,pc) to the north of the nucleus.
%We do not detect the same extended \SIV\ emission to the south of the nucleus (see Fig~\ref{f:ic4518wspatproffluxes}; negative offset).
%In \S\ref{s:specfeat} we explained that it is not clear whether this line is ``activated'' by AGN emission or by star formation. Here we show that the extended emission seen in IC~4518W is not associated, at least clearly, with star formation. Although there is a little bump of \Paalpha\ emission that coincides with the \SIV\ extended emission, its intensity is 3 times fainter than the flux measured in the nucleus. In addition, there is no such a bump seen in the \NeII\ line emission profile (although it is somewhat extended to this region) and there is no PAHs within and away from the nucleus. \textbf{Thus, the different profiles suggest that the \SIV\ extended emission is associated with the AGN activity. We propose that this could be the signature of a ionization cone arising from the Sy2 nucleus.}
Because of the intermediate ionization potential of the
\SIV\ line, it can be ionized by the AGN emission far away from
the nucleus.
% and therefore be detected as extended emission (see Fig~\ref{f:ic4518wspatprofconts}).
This has already been found in other Sy2 galaxies, such as Circinus
(\citealt{Roche06}) or NGC~5506 (\citealt{Roche07}), where the
\SIV\ line is more extended
%(detected up to almost a hundred of pc)
than the dust continuum emission
% (see also Fig~\ref{f:ic4518wspatprofconts})
and has been associated to high-excitation (coronal or narrow line) regions.
%Moreover, we \textit{do} probably detect the extension of the \SIV\ line to the north of the nucleus due to the decreasing of the extinction towards that direction (see the F110W/F160W color-like map in \citealt{DS08}). In contrast, the region to the south suffers from an obscuration similar to that of the nucleus, and this might be preventing us from detecting the \SIV\ emission of the counter high-ionization zone (the \SIV\ line is located almost at the maximum of the \Siabs\ absorption feature).

%\begin{figure}%[!h]
%\epsscale{0.37}
%\plotone{./figures/ic4518w_ap4pix_fix2_cal_flux_spat_dist_panel_00.ps}\hspace{0.2cm}\plotone{./figures/ic4518w_ap4pix_fix2_cal_flux_spat_dist_panel_10.ps}
%\vspace{0.25cm}
%\caption{\footnotesize Same as Figure~\ref{f:ngc3256spatprofconts} but for IC~4518W.}\label{f:ic4518wspatprofconts}
%\vspace{0.25cm}
%\end{figure}

%\begin{figure}%[!h]
%\epsscale{0.37}
%\plotone{./figures/ic4518w_ap4pix_fix2_cal_ratio_spat_dist_panel_13.ps}
%\vspace{0.25cm}
%\caption{\footnotesize Same as Figure~\ref{f:ngc3256spatprofneiipaafluxratio} but for IC~4518W.}\label{f:ic4518wspatprofneiipaafluxratio}
%\vspace{0.25cm}
%\end{figure}

%\begin{figure}%[!h]
%\epsscale{0.37}
%\plotone{./figures/ic4518w_ap4pix_fix2_cal_ratio_spat_dist_panel_11.ps}
%\vspace{0.25cm}
%\caption{\footnotesize Same as Figure~\ref{f:ngc3256spatprofratios} but for IC~4518W.}\label{f:ic4518wspatprofratios}
%\vspace{0.25cm}
%\end{figure}

%\begin{figure}%[!h]
%\epsscale{0.37}
%\plotone{./figures/ic4518w_ap4pix_fix2_cal_ew_spat_dist_panel_01.ps}\hspace{0.2cm}\plotone{./figures/ic4518w_ap4pix_fix2_cal_ew_spat_dist_panel_11.ps}
%\vspace{0.25cm}
%\caption{\footnotesize Same as Figure~\ref{f:ngc3256spatprofews} but for IC~4518W.}\label{f:ic4518wspatprofews}
%\vspace{0.25cm}
%\end{figure}

The change on the physical conditions between the nucleus and the
location identified as the extended narrow line region is also shown by
the
%spatial profile of the
\NeII/\SIV\ ratio.
%(Figure~\ref{f:ic4518wspatprofratios}).
While it is approximately constant
in the nucleus (almost $\simeq\,3$), the ratio steeply diminishes
towards the \SIV\ extended emission region. The value measured
in the nucleus is in agreement with the limit given by \cite{MH06},
\NeII/\SIV$\,\lesssim\,$3, for sources with no PAH detection.
They use the \NeII/\SIV\ ratio as a measure of the
hardness of the radiation field. A higher \NeII/\SIV\ ratio
implies a softer medium, where PAH emission would be detected.
Our values are always below this limit along the slit,
in agreement with the lack of PAH emission in this galaxy.
%Figure~\ref{f:ic4518wspatprofratios} also shows that the extinction is higher in the nucleus than in the outer regions.
%Conversely, we find a higher ratio in the nucleus of IC~4518W than in the outer regions. However, since the \SIV\ emission line is located almost at the wavelength where the absorption peak of the \Siabs\ feature is, it is therefore more affected by extinction than the \NeII\ emission line. Thus, the measured ratio at the nucleus of the galaxy %(see Figure~\ref{f:ic4518wspatprofratios}) is an upper limit to the real value, if the extinction is high there. Indeed, the \Siabs\ absorption feature is shallower to the north (positive offset) of the nucleus (see \S\ref{s:siabsfeat}). This would explain why the \NeII/\SIV\ ratio is higher in the nuclear region, where the radiation field should be harder due to the AGN, i.e., the \NeII/\SIV\ ratio should be lower than in the surrounding regions.
The EWs of the \SIV\ and the \NeII\ lines show
their lowest values in the nucleus, increasing slowly outwards. This is
again an effect of the continuum emission being more spatially compact than
the line emission.
% (compare Figures~\ref{f:ic4518wspatproffluxes} and~\ref{f:ic4518wspatprofconts}).

%The \NeII/\Paalpha\ ratio is almost constant ($\simeq\,4$) in the nuclear region of IC~4518W. This is approximately the same value as measured in the spatial profile of NGC~3256, and it increases rapidly to the south (up to $\sim\,$10) due to the lack of \Paalpha\ emission.
%, which causes the \NeII/\Paalpha\ ratio to increase significantly
%(see Figure~\ref{f:ic4518wspatprofneiipaafluxratio}).

%Although our values are in agreement with what they find, the \NeII\ to \SIV\ emission ratio measured at the location of the nucleus of IC~4518W would imply a softer radiation field than that at the high-excitation emission line region (or viceversa). This is quite surprising since in closest regions of the AGN the radiation field should be harder than in the surrounding areas. Nevertheless, this can be explained if the Ne abundance (with respect to that of the S) was lower in the high-excitation zone, so that the \NeII/\SIV\ ratio in this region was as high as that in the nucleus not because the radiation field was so hard as in it but because there was no Ne to excite, thus making the ratio to diminish towards this region.
%\textit{Explanations? The \NeII\ line has a lower ion. pot. than the \SIV\ line so, in principle, it could (and should) be ionized farther, which is the contrary to what we see... unless there is no Ne that can be ionized in that region}.

\section{-C- NGC~5135: Separated, Spatially Resolved SF and AGN Activity}\label{s:ngc5135sp}

\subsection{From an Isolated Sy2 Nucleus to a Circumnuclear \HII\ Region}\label{ss:ngc5135nucspec}

\begin{figure}%[!h]
\epsscale{0.75}
%\plotone{./figures/ngc5135_trecs_spectra.ps}
\plotone{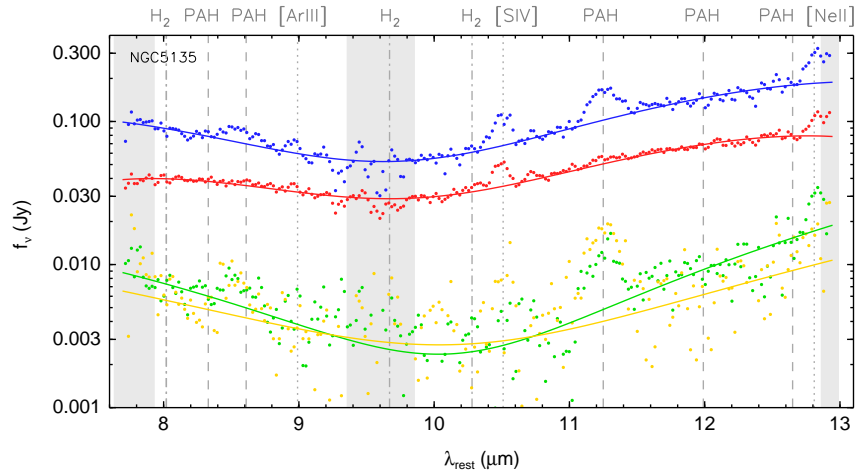}
\vspace{0.25cm}
\caption{\footnotesize Integrated (blue dots) and nuclear (red dots) T-ReCS $N$-band spectra of NGC~5135. The regions of interest of NGC~5135 are an \HII\ region (green dots) located at $\sim\,$2.6\arcsec\ from the nucleus, and the diffuse medium (orange dots) between this region and the nucleus, respectively. The integrated spectrum was extracted with a fixed aperture of 3.6\arcsec\ in length centered at the nucleus of the galaxy, while the nuclear spectrum was extracted with a fixed aperture of 0.36\arcsec. The spectrum of the \HII\ region was extracted with the same aperture as the nucleus while the spectrum of the diffuse medium was extracted with an aperture of 0.72\arcsec. The labels are as in Figure~\ref{f:ngc3256spatpos}.
[\textit{See the electronic edition of the Journal
for a color version of this figure}].}\label{f:ngc5135intnucroispecs} 
\vspace{0.25cm}
\end{figure}

The Seyfert 2 nuclear spectrum of NGC~5135 shows intense \SIV\
line emission (see Figure~\ref{f:ngc5135intnucroispecs}),
suggesting a relatively high ionized medium in agreement with its
classification as Seyfert. There is also faint \NeII\ line
emission probably associated with the central AGN as well. Although
quantitatively speaking there is no PAH emission, there seems to be
a little bump and a peak on the spectra at the positions where the
\PAHa\ should be. Following the evidence
above, the PAHs are supposed to be depleted in the vicinity of an AGN
but in some cases, the molecules can survive to the radiation field
if they are shielded, for example, by the star formation itself
(\citealt{Voit92}; \citealt{Mason07}). If this was the case,
we could not definitively affirm that there is no star formation
in the nuclear region (central $\sim\,100\,$pc) of NGC~5135. In any case,
the AGN effectively dominates the MIR luminosity of the nucleus
and the star formation would account for a very small fraction of it,
and if there is any, it must be at a very low level compared with
that taking place in the circumnuclear region (see below).
%In fact, it would not be the first time that star formation is found very close to the central engine of an active galaxy (NGC~7469, MORE REFERENCES)
%\subsection{Integrated Spectrum}\label{ss:ngc5135intspec}
Contrarily to NGC~3256 and IC~4518W, the integrated spectrum of
NGC~5135 shows significant differences when compared to its nuclear
spectrum. In this LIRG the MIR AGN emission is clearly separated
from that of the star formation (see \citealt{DS08}).
%(see Appendix \ref{ss:detailed} for a detailed description of the morphology of this LIRG).
The integrated spectrum includes not only the emission from the
nucleus but also the emission arising from an \HII\ region located at about
$\sim\,2.6\arcsec$ south-west, and from the diffuse region between both
(see Figure~\ref{f:slits}).
%This is the reason why there is intense 8.6 and \PAHa\ emission in the integrated spectrum but not in the nuclear one. Moreover, we can glimpse the presence of the \ArIII\ emission line, also associated to star-forming regions (\citealt{Snijders07}).
%Thus, the spatial resolution is crucial when looking at the very nuclear regions of LIRGs.
%Therefore, in NGC~5135 we have been able to disentangle the emission produced by the AGN from that of the surrounding star formation; the latter dominates the integrated MIR spectrum of the central region of this galaxy.

%\subsection{Extra-Nuclear Regions}\label{ss:ngc5135regofintspec}

%The \HII\ region of NGC~5135 shows an interesting spectrum when compared to that of the diffuse medium located between the \HII\ region and the nucleus
Both the \HII\ and the diffuse emission region of NGC~5135 (the two
regions of interest of this galaxy; see Figure~\ref{f:slits} and
Figure~\ref{f:ngc5135intnucroispecs}) exhibit strong \PAHa\ emission.
%feature with almost the same peak of emission.
The MIR continuum near the edges of the spectrum is however
significantly brighter in the \HII\ region.
%(between $\sim\,$20\% and 80\% at 8 and 13$\,\micron$, respectively, and even more if both spectra are normalized to their minimum flux density  at $\sim\,$10$\,\micron$).
Thus, the diffuse ISM of NGC~5135
is characterized by intense PAH emission but by a fainter continuum
emission from hot dust when compared with that of the \HII\ region.
The \PAHa\ EW at the diffuse region is 2 times higher
($\sim\,$1$\,\micron$) than that measured at \HII\ region
($\sim\,$0.5$\,\micron$).
This is expected if the typical stellar populations in the region
of diffuse emission are not extremely young ($\lesssim\,$10\,Myr)
neither very massive, which is in agreement with the presence
of weak \Paalpha\ emission.
%betwen the nucleus of the galaxy and the position of the \HII\ region.
Therefore, the lack of a hot dust continuum (associated with the
\Paalpha\ emission) in the diffuse region would make its \PAHa\ EW
larger than in the \HII\ region, supporting the PAH \textit{dilution}
scenario in star-forming regions
(see discussion in \S\ref{ss:denseffect}).
% As a consequence, the \HII\ and diffuse region strongly differ in their \PAHa\ EWs.
% Figure~\ref{f:ngc5135spatprofews} shows that the

%which would be in agreement if the emitting dust is being heated by very young stars, as in a star-forming region.
On the other hand, the extinction could play a role in this situation.
If the PAH and continuum emissions are decoupled, that is, arise from different
regions, they can be obscured by different amounts of cold dust, thus
leading to different extinctions. There is a correlation between the
\Paalpha\ LSD of star-forming regions and their obscuration (brighter
regions are affected by higher extinctions; \citealt{Calzetti07}; and
\citealt{DS08}, their figure~9). Therefore, if the \HII\ region (that shows
bright \Paalpha\ emission) is more absorbed than the diffuse medium, the
strength and shape
%(see Figure~\ref{f:extcurve})
of the \Siabs\ absorption feature would make the continuum of the \HII\
spectrum to be deeper at $\sim\,10\,\micron$ when compared to the
emission at 8 or 13$\,\micron$. This is in agreement with what we see
in Figure~\ref{f:ngc5135intnucroispecs} where the diffuse region shows
a shallower spectrum. However, this is difficult to test since our
data do not enable us to probe the geometry of the regions in detail
or how the dust is distributed.
% (see Figure~\ref{} \textit{pending of insert}), which is also in agreement with what we see in Figure~\ref{f:intnucroispecs}.
% \textit{On the other hand, if PAH and continuum emission suffered from approximately the same extinction, the PAH EW would allow us to differentiate between effectively young star-forming regions and a diffuse medium.} \textit{Moreover, could be the PAH EWs related with the age of the stellar population???}
%Pues podria, porque viendo los perfiles espaciales de la EW de los PAHs, esta es menor en las regiones mas nucleares, donde se espera que las poblaciones estelares sean mas jovenes!

\subsection{Spatial Profiles: Tracers of AGN and Star Formation Activity}\label{ss:ngc5135spatprof}

\begin{figure}%[!h]
\epsscale{0.37}
%\plotone{./figures/ngc5135_ap4pix_fix2_cal_flux_spat_dist_panel_02.ps}\hspace{0.2cm}\plotone{./figures/ngc5135_ap4pix_fix2_cal_flux_spat_dist_panel_01.ps}\hspace{0.2cm}\plotone{./figures/ngc5135_ap4pix_fix2_cal_flux_spat_dist_panel_11.ps}
\plotone{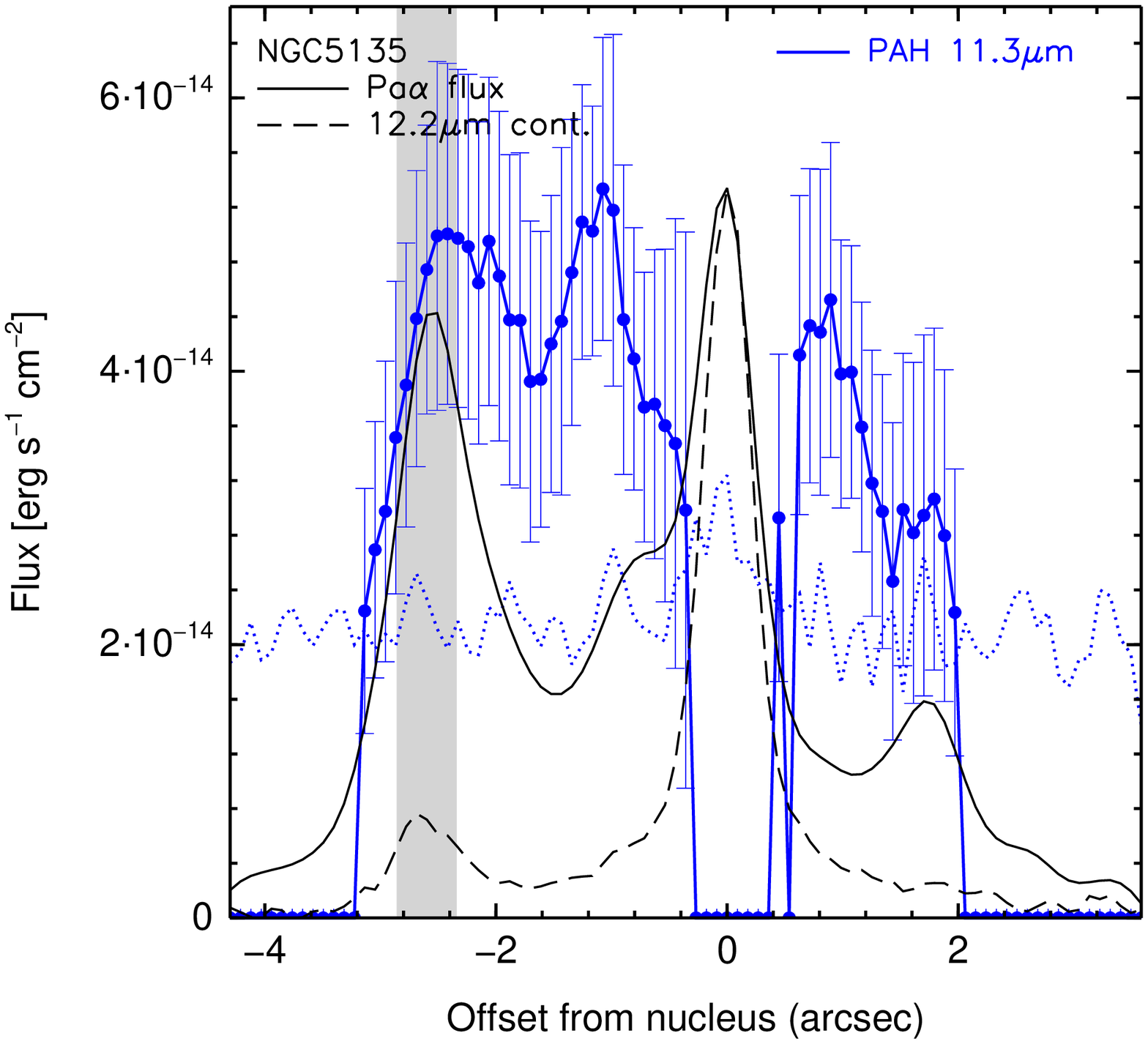}\hspace{0.2cm}\plotone{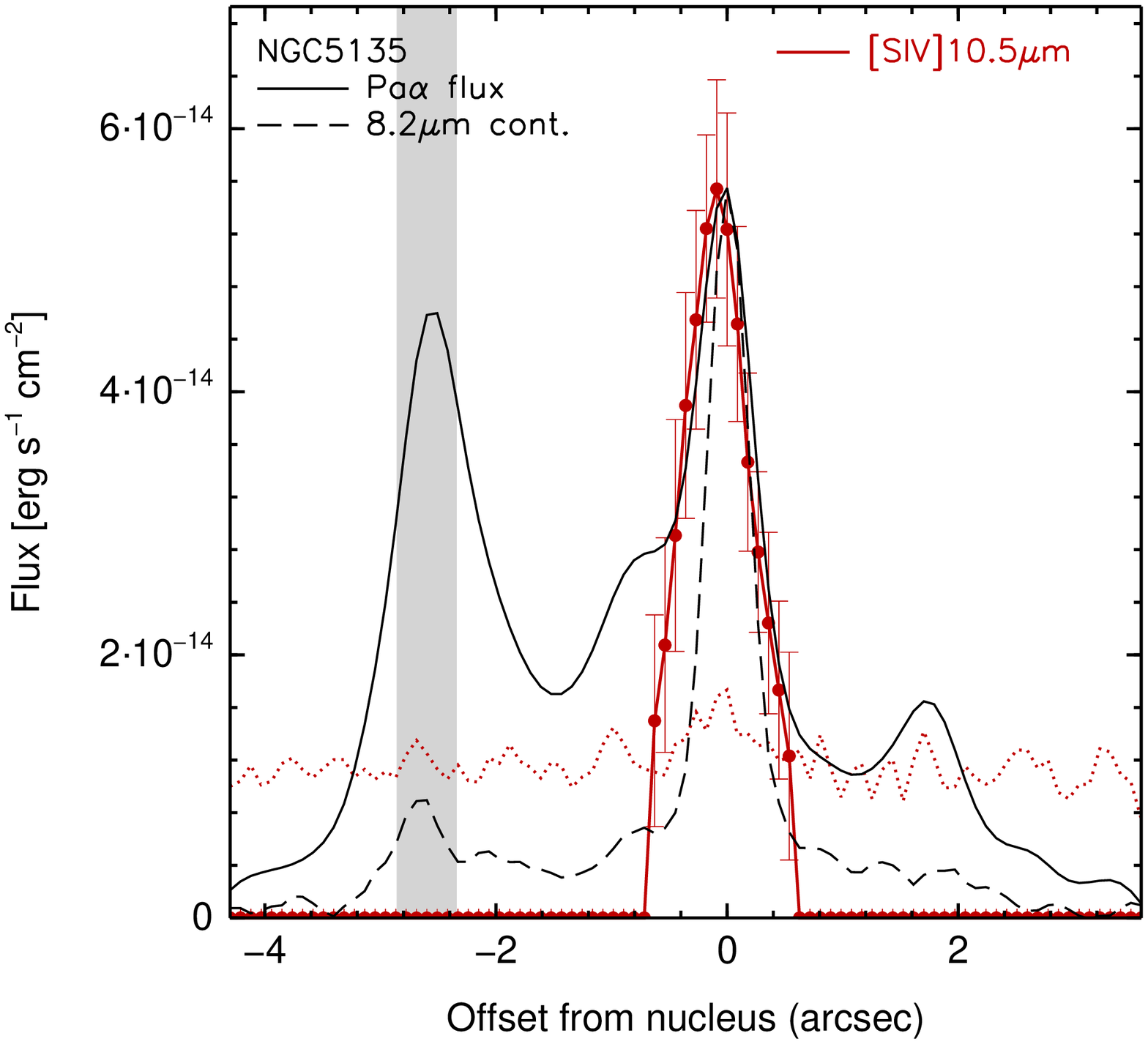}\hspace{0.2cm}\plotone{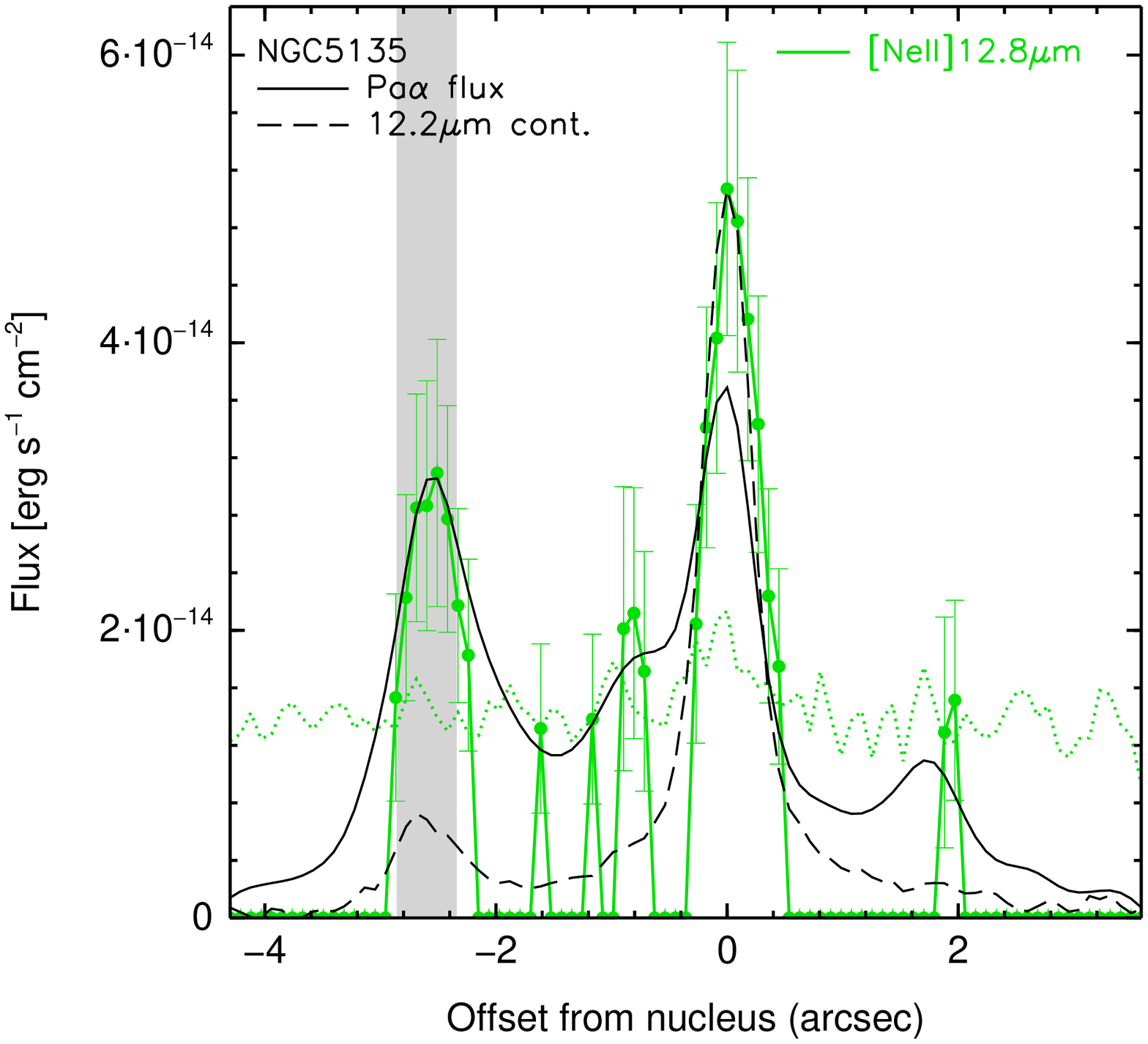}
\vspace{0.25cm}
\caption{\footnotesize Same as Figure~\ref{f:ngc3256spatproffluxes} but for NGC~5135. In this case, the flux spatial profiles of the \PAHa\ (top left), and the \SIV\ (top right) and \NeII\ (bottom) emission lines are shown. No \PAHb\ emission was detected.
[\textit{See the electronic edition of the Journal
for a color version of this figure}].}\label{f:ngc5135spatproffluxes}
\vspace{0.25cm}
\end{figure}

NGC~5135 is a clear example where star formation and AGN activity
are well separated (see Figure~\ref{f:slits}) with our T-ReCS spatially
resolved spectroscopy. While we find \PAHa\ emission
outside ($\gtrsim 0.5\arcsec$) the nucleus but not within it,
the \SIV\ emission line is only detected in the central region
($\lesssim\,0.7\arcsec$), with little overlap between both features
(see Figure~\ref{f:ngc5135spatproffluxes}).

As for NGC~3256, the spatial profile of the \NeII\ emission line
follows well that of the \Paalpha\ line.
%is very well correlated with the \Paalpha\ line.
There is \NeII\ emission not only in the nucleus and in the \HII\
region located at 2.6\arcsec\ south-west (negative offset),
but it is also detected in the \Paalpha\ bump at $\sim\,$0.8\arcsec.
%It is even detected tentatively at $\sim\,$2\arcsec\ to the north, although the detection limit of the \NeII\ line is slightly above the \Paalpha\ emission.
However, unlike NGC~3256, which is classified as \HII, the nuclear
\NeII\ emission of NGC~5135 should be mainly associated to
%(and be produced primarily by) its Sy2
the AGN. Therefore we have scaled
the \Paalpha\ profile to the \NeII\ emission at the
location of the \HII\ region as we found that, for star-forming
dominated regions (as in NGC~3256), both emission lines correlate
quite well (see \S\ref{s:neii}).
Taking this into account, we can see that the nucleus
shows a slightly higher \NeII/\Paalpha\ ratio than the \HII\ region.
%(see Figure~\ref{f:ngc5135spatprofneiipaafluxratio}).
This excess might be attributed to the extinction
%since it is rather constant along the slit
as the \NeII\ line is less affected by obscuration than \Paalpha.

Fig~\ref{f:ngc5135spatproffluxes} shows that the \PAHa\ emission is
not only produced by the youngest stellar populations as traced by the
\Paalpha\ line emission, but it is also detected in the diffuse ISM.
This confirms that the PAH emission is associated not only to the
ionizing stars but also to the more evolved stellar populations
and their UV flux (see also, e.g., \citealt{Peeters04}; \citealt{TG05}).
I.e., the spatial profile of the PAH
emission profile does not resemble that of the \Paalpha\
line (see also \S\ref{s:pah}), and the diffuse region shows a
\PAHa\ intensity similar to that
of the \HII\ region. On the other hand, the \NeII\ emission is
fainter in the diffuse region and bright where there is an
enhancement of the \Paalpha\ line.

\section{-D- NGC~7130: Mixed Star Formation and AGN Activity}\label{s:ngc7130sp}

\subsection{AGN and Star Formation Emissions Together within $\leq\,$100\,pc}\label{ss:ngc7130nucspec}

\begin{figure}%[!h]
\epsscale{0.75}
%\plotone{./figures/ngc7130_trecs_spectra.ps}
\plotone{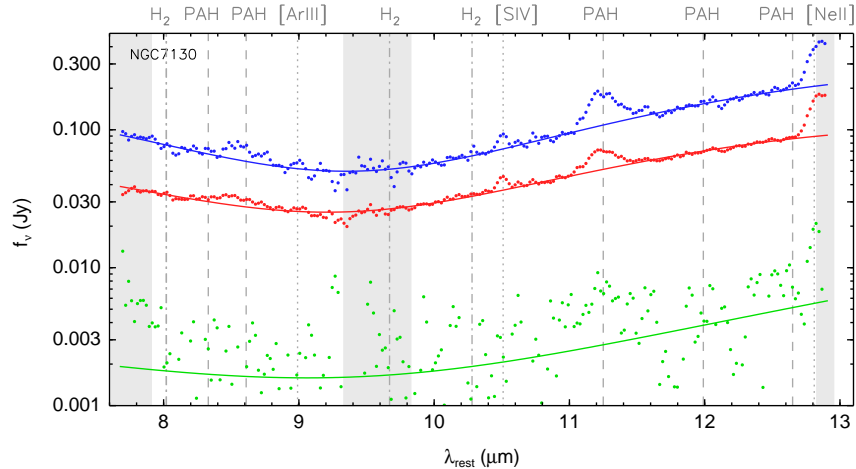}
\vspace{0.25cm}
\caption{\footnotesize Integrated (blue dots) and nuclear (red dots) T-ReCS $N$-band spectra of NGC~7130. The region of interest (green dots) of NGC~7130 is an \HII\ region located at 9.5\arcsec\ to the north of the nucleus. The integrated spectrum was extracted with a fixed aperture of 3.6\arcsec\ in length centered at the nucleus of the galaxy, while the nuclear spectrum was extracted with a fixed aperture of 0.36\arcsec\ and the spectrum of the \HII\ region with an aperture of 0.72\arcsec. The labels are as in Figure~\ref{f:ngc3256spatpos}.
[\textit{See the electronic edition of the Journal
for a color version of this figure}].}\label{f:ngc7130intnucroispecs} 
\vspace{0.25cm}
\end{figure}

The MIR emission of the nucleus of NGC~7130 is resolved (\citealt{DS08})
and shows signatures of both star formation and AGN activity. Its optical
classification as LINER/Sy is in agreement with the detection of the \NeV\
emission line in the \textit{Spitzer} IRS spectrum of the galaxy
(\citealt{AAH09b}; \citealt{PS09b}), supporting the existence of an AGN.
The T-ReCS nuclear
spectrum reveals clear 8.6 and \PAHa\ emission together with the \SIV\
emission line (see Figure~\ref{f:ngc7130intnucroispecs}).
%Besides, the 12$\,\micron$ PAH is marginally detected too.
This is a clear example of the coexistence of PAH emission and an AGN within
less than $\sim\,$100\,pc. There are two explanations for this:
(a) the PAH molecules are far away enough from the AGN so that they
cannot be destroyed by its radiation field; (b) the PAH molecules are
being shielded in some manner.
%, as might be in NGC~5135 (see below).
%\subsection{Integrated Spectrum}\label{ss:ngc7130intspec}
The integrated T-ReCS spectrum shows the same features as
the nuclear spectrum since the nucleus of NGC~7130 is quite
compact and outside $\sim\,500\,$pc there is no emission
(neither continuum nor feature emission).
%The PAH EWs are similar in the nuclear and the integrated spectra (see Table~\ref{t:intews}). The \ArIII\ emission line is also barely seen in the integrated spectra, as in NGC~5135. However, unlike for the nucleus of NGC~5135, and even with the high spatial resolution achieved by T-ReCS, we can see that the star formation and the AGN activity are still mixed in the nucleus of NGC~7130.

%\subsection{Extra-Nuclear Regions}\label{ss:ngc7130regofintspec}

%Finally, as an example, Figure~\ref{f:ngc7130intnucroispecs} shows the spectrum of an \HII\ region of NGC~7130 (see Figure~\ref{f:slits}; the region of interest of this galaxy). Unfortunately, this is too noisy to infer any conclusion from it. Still, it seems to show a hint of the \PAHa\ emission.

\subsection{Spatial Profiles: Star-Formation Surrounding a Sy2 Nucleus}\label{ss:ngc7130spatprof}

\begin{figure}%[!h]
\epsscale{0.37}
%\plotone{./figures/ngc7130_ap4pix_fix2_cal_flux_spat_dist_panel_02.ps}\hspace{0.2cm}\plotone{./figures/ngc7130_ap4pix_fix2_cal_flux_spat_dist_panel_12.ps}\hspace{0.2cm}\plotone{./figures/ngc7130_ap4pix_fix2_cal_flux_spat_dist_panel_01.ps}\plotone{./figures/ngc7130_ap4pix_fix2_cal_flux_spat_dist_panel_11.ps}
\plotone{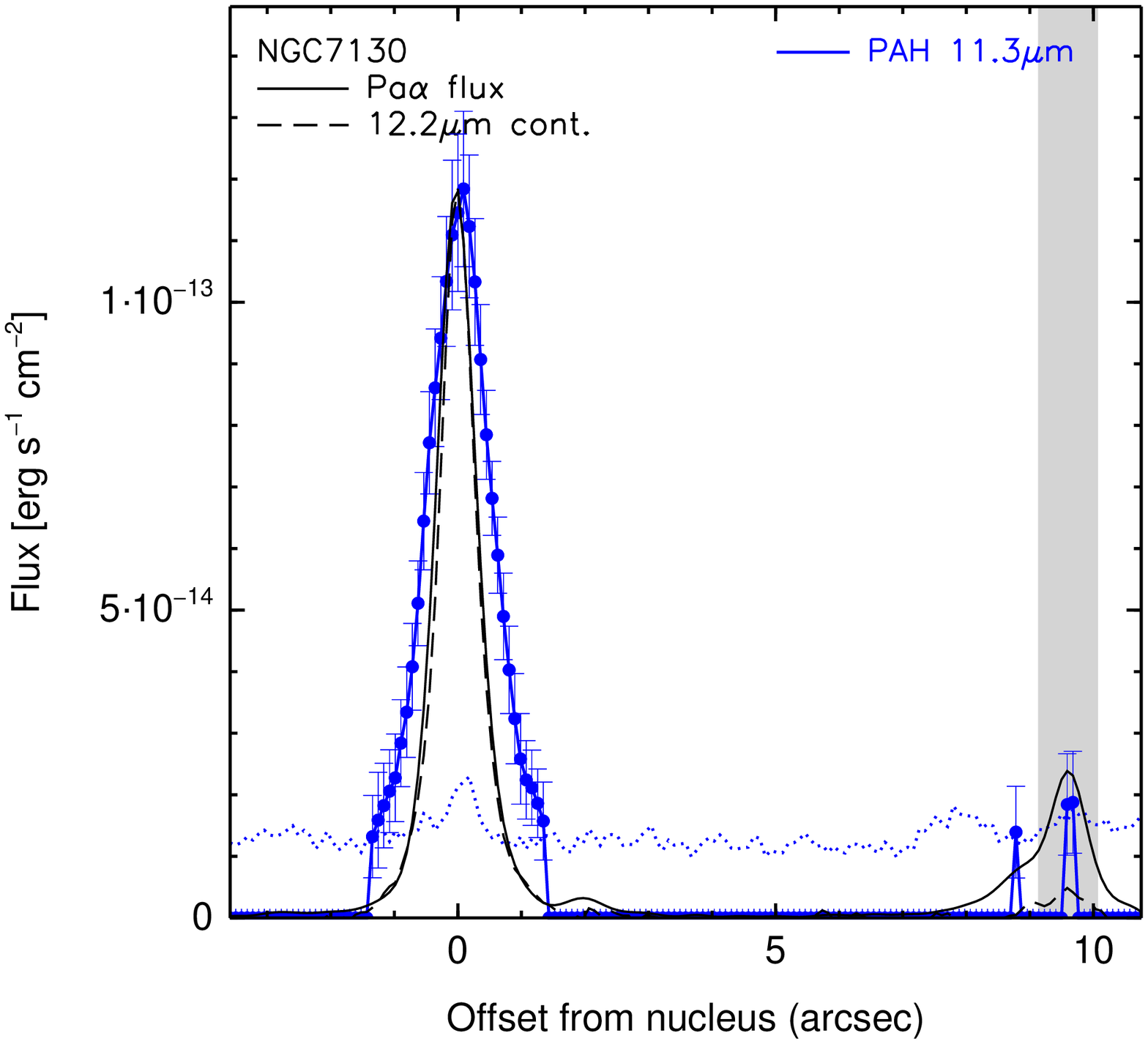}\hspace{0.2cm}\plotone{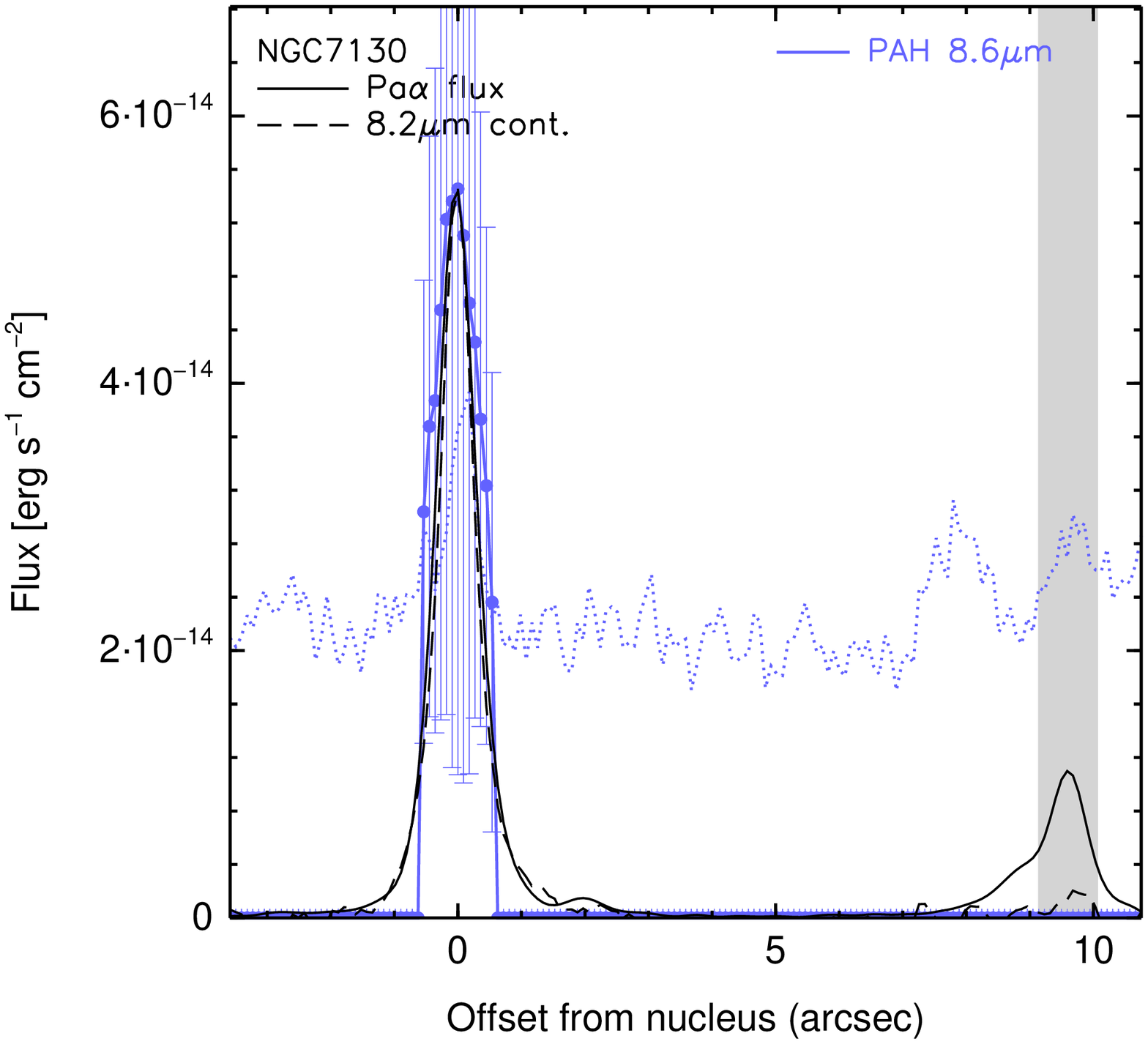}\hspace{0.2cm}\plotone{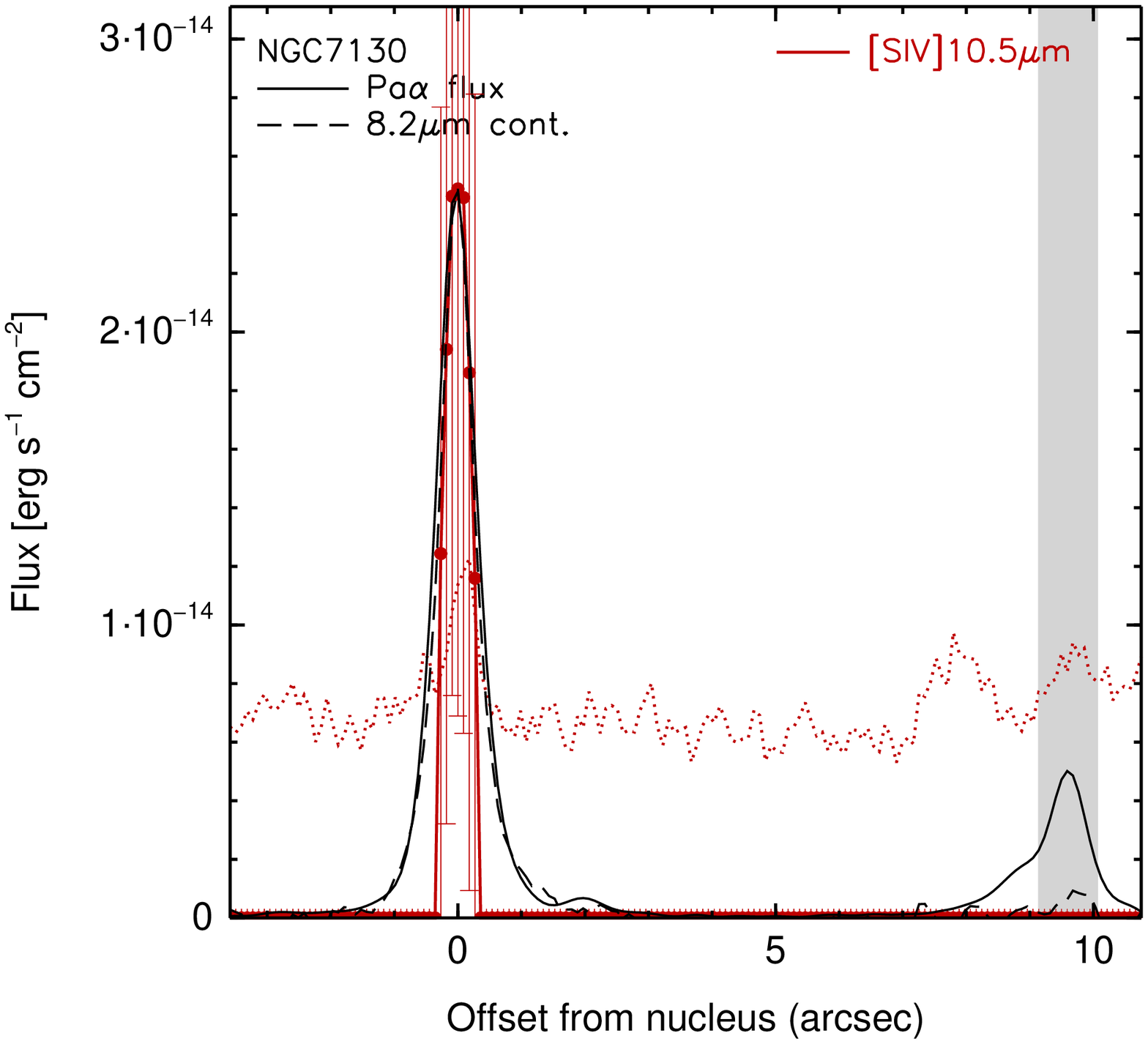}\hspace{0.2cm}\plotone{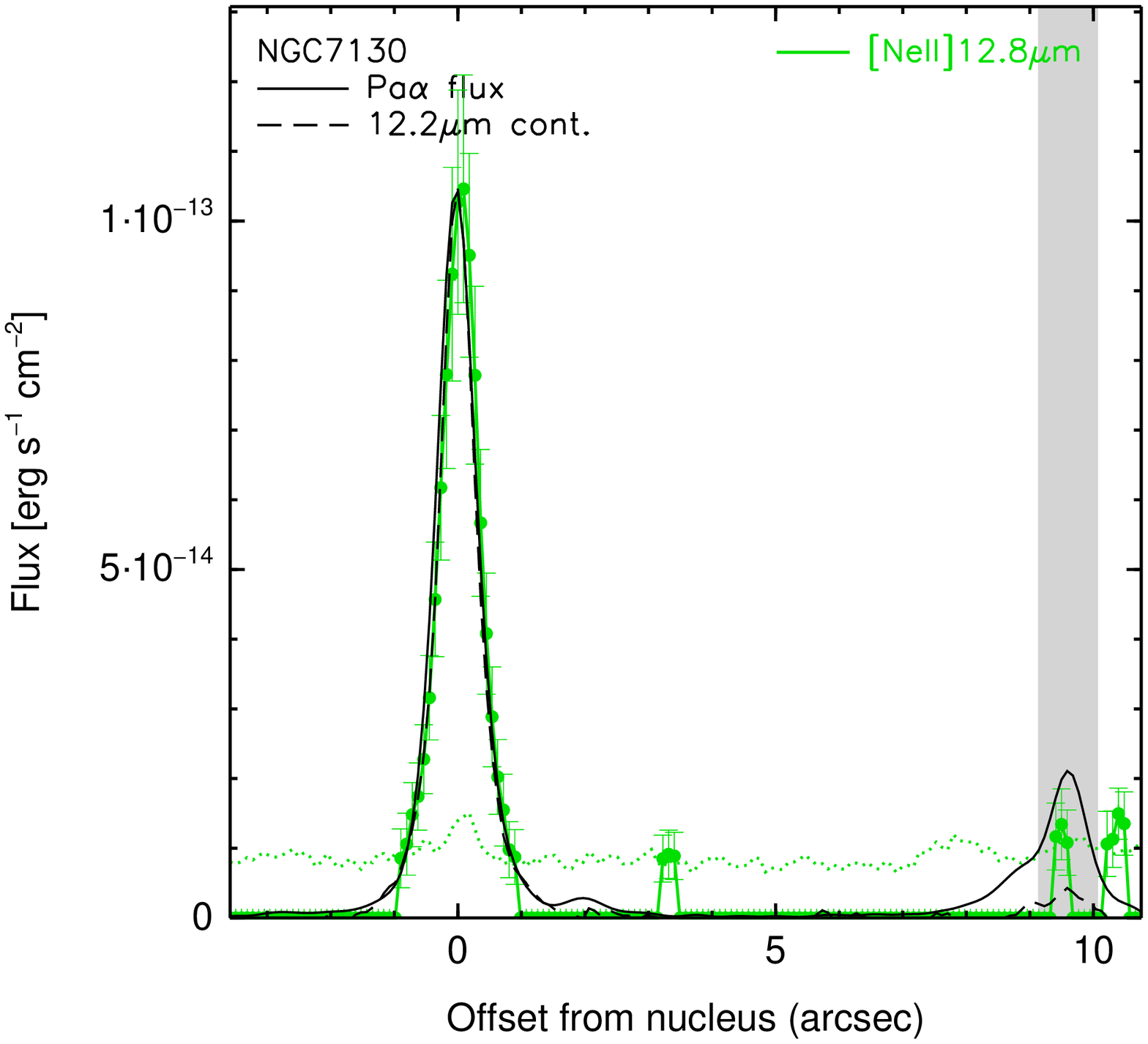}
\vspace{0.25cm}
\caption{\footnotesize Same as Figure~\ref{f:ngc3256spatproffluxes} but for NGC~7130. In this case, all flux spatial profiles are shown: the 11.3$\,\micron$ (top left) and \PAHb\ (top right) features, and the \SIV\ (bottom left) and \NeII\ (bottom right) emission lines.
[\textit{See the electronic edition of the Journal
for a color version of this figure}].}\label{f:ngc7130spatproffluxes}
\vspace{0.25cm}
\end{figure}

The nucleus of NGC~7130 is a clear example where star formation
and AGN activity manifest their characteristic spectral features
within the same region (see Figure~\ref{f:ngc7130spatproffluxes}).
%; \citealt{GD01}).
We find that
%\NeII\ line emission but, unlike the PAHs, it is as compact as the \Paalpha\ profile. In fact,
the \NeII\ and \Paalpha\ profiles are as compact as the (resolved)
continuum emission. We also detect 8.6 and \PAHa\ emission,
but extended within the central 1.5\arcsec\ of the galaxy.
Both, the PAH features, and the \NeII\ and \Paalpha\ emission lines
clearly indicate that there is star formation within the nuclear region.
Moreover, Fig~\ref{f:ngc7130spatproffluxes} shows that in
NGC~7130 the PAH molecules can survive within a distance of less
than $\sim\,$100\,pc from the AGN without being destroyed.
%, whether it be because they really can survive to the radiation field, or because they are being shielded from it.
%are signatures of star formation, the \NeII\ line traces the \textit{ionizing} stars, which means that the youngest populations are rather concentrated to the nucleus, while the extended PAH emission is tracing the more diffuse UV radiation field. On the other hand,
The intrinsic hard ($2-10\,$keV) X-ray luminosities
($L_{\rm 2-10\,keV}$) of the AGNs of NGC~7130 and NGC~5135
are very similar ($\approx\,$1$\,\times\,$10$^{43}$\,erg\,s$^{-1}$;
\citealt{Levenson05}, and \citealt{Levenson04}, respectively).
Therefore, the existence of PAH emission in the nucleus of NGC~7130
but not in the nucleus of NGC~5135 suggests that the PAH carriers
are not being destroyed in the latter but the absence of PAHs is due to
the fact that there is no nuclear star formation (or at least it is
very weak). The \SIV\ line is also detected (although with a high uncertainty)
as in the nucleus of IC~4518W and NGC~5135, and is as compact as the
continuum emission.
% (see Figures~\ref{f:ngc7130spatproffluxes} and \ref{f:ngc7130spatprofconts}).

%As discussed above, the PAH molecules are not destroyed in the inner regions of NGC~7130, suggesting that they are being shielded from the radiation field of the active nucleus. An alternative explanation could be that we are not able to resolve the structure within the nucleus. For example, if the star-forming region is arranged in a circumnuclear ring-like pattern with a diameter $\lesssim 0.5\arcsec\,\simeq 150\,$pc, (as seen at other wavelengths; \citealt{GD01}; \citealt{Dopita02}), lower than our resolution, we would not be able to fully resolve the structure and we would be seeing PAH emission in the very center of the nucleus when it really would not exist.

The \NeII/\Paalpha\ ratio is almost constant in the nuclear region of
the galaxy (between $3-5$ within the inner 1\arcsec$\,\simeq\,$340\,pc).
%Figure~\ref{f:ngc7130spatprofratios} shows that
The 11.3/\PAHb\ ratio is also almost constant along the central
$\sim\,$200\,pc with a mean value of $\simeq\,2.3$. This value
is slightly higher than that seen in NGC~3256 ($\simeq\,1.7$),
but both are in agreement with the median ratio found by \cite{Smith07}
for the SINGS galaxy sample (1.5$\pm^{1.4}_{0.3}$).
%However, the difference between the ratios of the nuclear regions of our two LIRGs suggests that in NGC~7130 we would be looking at a more neutral medium. This disagrees with the fact that in the nucleus of NGC~7130 there is an AGN which would tend to increase the hardness of the radiation field and the ionization level in the area thus decreasing (not enhancing) the 11.3/\PAHb\ ratio. Alternatively, this simply may indicate variations of this ratio from galaxy to galaxy.
% \textit{And this why!!!??? Maybe the ionized PAHs are not more ionized but directly destroyed by the AGN thus making the \PAHb\ intensity to diminish and then increasing the ratio...}
The spatial profile of the \PAHa/\NeII\ ratio has his minimum at the
nucleus of the galaxy with a value of $\simeq\,$1.1 (similar to that
of the nucleus of NGC~3256) and increases outwards from the central region.
%At about 1\arcsec\ ($\simeq\,$340\,pc) away from the nucleus this ratio reaches a value between $\simeq\,2-3$, similar (or slightly higher) to those found in the external star-forming regions of NGC~3256 and NGC~5135. This suggests that this value may be typical of more evolved (a few Myr) stellar populations and regions of diffuse \Paalpha\ emission).
% and seems to stabilize at a distance of $\sim\,1\arcsec\,$ at a value of $\simeq 0.3-0.4$ (almost equal to the values found for the external regions of NGC~3256 and NGC~5135, and that could be typical of a diffuse ISM with intermediate-age stellar populations). The values of this ratio at the nucleus and around are equal to those found at the nucleus and the surrounding regions of NGC~3256, suggesting presence of intense star formation in the nucleus of NGC~7130.
%The datapoint measured in the \HII\ region located $\sim\,9.5$\arcsec\ to the north of the nucleus also shows a value of $\simeq\,$2 (although with a large uncertainty).
%, similar to that of the \HII\ region seen in NGC~5135.

%The \SIV/\PAHa\ and the \NeII/\SIV\ ratios (as well as the 11.3/\PAHb\ ratio) are subject to large uncertainties.
%(see Figure~\ref{f:ngc7130spatprofratios}).
%Bearing this in mind, we can see that
The \SIV/\PAHa\ ratio peaks at the nucleus, while the
%profile of the
\NeII/\SIV\ ratio has the minimum there.
The former is due to the fact that the \PAHa\ emission is more
extended than the \SIV\ line emission. In the
same manner, the \SIV\ emission seems to be more concentrated
than the \NeII\ emission as the \NeII/\SIV\ ratio
increases outwards from the nucleus. The nucleus of NGC~7130
is resolved in continuum emission so these variations are indeed real.
% and attest internal structures.
In fact, this would be in agreement with the \SIV\ line emission being
the signature of the unresolved AGN emission, and the \NeII\ line emission
being mostly associated to the surrounding star-forming ring seen
in the UV (\citealt{GD01}; also detected in PAH emission).
%\textbf{In summary, young stars imply PAH emission but not the contrary.}
Because the 8.6 and \PAHa\ emissions are more extended than the continuum
and line emissions, the spatial profiles of their EWs show again
their minima in the nucleus, increasing outwards. In contrast,
the \SIV\ EW is approximately constant (within the large uncertainties)
%peaks at the center,
confirming that the region from where the \SIV\ line
emission arises is very compact (at least as compact as the
continuum).
%The \NeII\ EW shows an interesting spatial profile. The maximum and minimum of the undulating shape ($\pm\,$0.5\arcsec) approximately coincide with the location of the circumnuclear star-forming ring seen in the nucleus of the galaxy (\citealt{GD01}). Such a gradient in the \NeII\ EW, although small ($\Delta\,$EW$\,\sim\,$\,0.2$\,\micron$), implies that the flux of the line is effectively changing across the ring and is not related to changes in the continuum emission which is very symmetric with respect to the position of the nucleus.
% (see Fig~\ref{f:ngc7130spatprofconts}).
%The physical interpretation of this change is difficult though. The variation of the \NeII\ EW can be due to a change in the ionization conditions of the region, that might be caused by the AGN emission, or to a variety in the ages of the stellar populations in the circumnuclear ring of star formation. The differential extinction again can also be playing a role in these variations of the EW.

\bibliographystyle{/home/tanio/mypapers/apj}%este estilo nombra en la lista de referencia solo a los tres primeros  autores
%\bibliographystyle{./apj}%este estilo nombra en la lista de referencia solo a los tres primeros  autores
%\bibliography{/home/tanio/mypapers/bib}{}
%\bibliography{./bib}{}

\end{document}